\documentclass[rmp,twocolumn,superscriptaddress,showpacs,floatfix]{revtex4}

\usepackage{epsfig}
\usepackage{graphicx}

\newcommand{\be}{\begin{equation}}
\newcommand{\ee}{  \end{equation}}
\newcommand{\ba}{\begin{eqnarray}}
\newcommand{\ea}{  \end{eqnarray}}
\newcommand{\ve}{\varepsilon}

\begin{document}

\title{Random Matrices and Chaos in Nuclear Physics \\
Part 1: Nuclear Structure}

\author{H. A. Weidenm\"uller}
\affiliation{Max-Planck-Institut f\"ur Kernphysik, D-69029 Heidelberg,
Germany}

\author{G. E. Mitchell}
\affiliation{North Carolina State University, Raleigh, North
Carolina 27695 \\
Triangle Universities Nuclear Laboratory, Durham, North Carolina 27706}

\begin{abstract}The authors review the evidence for the applicability
of random--matrix theory to nuclear spectra. In analogy to systems
with few degrees of freedom, one speaks of chaos (more accurately:
quantum chaos) in nuclei whenever random--matrix predictions are
fulfilled. An introduction into the basic concepts of random--matrix
theory is followed by a survey over the extant experimental
information on spectral fluctuations, including a discussion of the
violation of a symmetry or invariance property.  Chaos in nuclear
models is discussed for the spherical shell model, for the deformed
shell model, and for the interacting boson model. Evidence for chaos
also comes from random--matrix ensembles patterned after the shell
model such as the embedded two--body ensemble, the two--body random
ensemble, and the constrained ensembles. All this evidence points to
the fact that chaos is a generic property of nuclear spectra, except
for the ground--state regions of strongly deformed nuclei.
\end{abstract}

\pacs{ 24.60.Lz,25.70.Gh,21.60.Cs,21.60Ev} 

\maketitle

\section{Introduction}

In the 1970s chaos became a household word of physicists. Chaos had
been known to mathematicians, astrophysicists and other specialists
since the early years of the century. It took the advent of the
personal computer to make chaos generally known: The exponential
divergence of the trajectories of a chaotic system with Hamiltonian
dynamics could easily be simulated. Many Hamiltonian systems,
especially those with few degrees of freedom, have been analyzed
since. The subject has been reviewed, for instance, by
Gutzwiller (1990).

In the wake of this development, physicists in several fields became
interested in quantum manifestations of classical chaos (``quantum
chaos''). Again, this caused a flurry of activity, especially in the
study of systems with few degrees of freedom, see McDonald and Kaufman
(1979), Casati {\it et al.}  (1980), and Berry (1981). The work
culminated in the Bohigas--Giannoni--Schmit (1984) conjecture.  The
conjecture connects the spectral fluctuation properties of quantum
systems which are chaotic in the classical limit with predictions of
random--matrix theory (RMT). A summary of these developments is given
by Haake (2001).

Independently of these developments and preceding them, RMT had been
developed in the framework of nuclear physics by Wigner and Dyson --
see papers in the compilation by Porter (1965).  Data accumulated in
the 1960s and analyzed in the 1980s provided evidence that nuclear
spectra follow RMT predictions.  The wide interest enjoyed by RMT in
the 1980s among practitioners of quantum chaos had repercussions on
nuclear physics: Nuclei are many--body systems, and chaos manifests
itself here in ways different from those of few--degrees--of--freedom
systems. At the same time, nuclei are paradigmatic for the
applicability of RMT to fermionic many--body systems and for the
occurrence of chaos in such systems.  Thus much work was done to
establish RMT and to analyze and interpret quantum chaos in nuclei.

The present review is the first part of a two--part series dealing
with RMT and chaos in nuclear physics. It focuses on spectral
properties of nuclei, while the second part will deal with RMT and
chaos in nuclear reactions. Topics such as compound--nucleus
scattering, Ericson fluctuations, isobaric analog resonances, and
parity violation in epithermal neutron scattering are not dealt with
here. The paper is intended as an introductory review to the field. It
is aimed mainly at two groups of physicists, those who work on quantum
chaos in fields different from nuclear physics, and nuclear physicists
who wish to learn about RMT and chaos in nuclei. We accordingly assume
no prior knowledge of RMT and chaos nor do we assume any detailed
knowledge of nuclear physics. We aim at giving a comprehensive survey
which is focused on concepts and illustrative examples while
derivations and formulas are kept to a minimum. The last comprehensive
review of the field has been given in Reviews of Modern Physics over
25 years ago by Brody {\it et al.} (1981); a short review was later
published by Bohigas and Weidenm\"{u}ller (1988). Wherever possible
we have avoided giving a large number of references in favor of citing
review articles: Readability of the article was our primary concern,
followed by completeness.

In Section~\ref{RMT} we motivate RMT and introduce those concepts of
RMT which are frequently used in nuclear physics. We pay particular
attention to the Gaussian orthogonal ensemble of random matrices
(GOE). We establish the connection between RMT and quantum chaos. In
Section~\ref{applRMT} we describe those applications of RMT to nuclear
spectra which do not make use of specific nuclear--structure
concepts. The comparison between RMT predictions and spectroscopic
data is used to establish the domain of applicability of RMT to
nuclear spectra. Special attention is devoted to the breaking of
isospin symmetry, and to a test of time--reversal invariance. The role
of RMT and chaos in nuclear models is described in
Section~\ref{models}. We focus attention on the two most important
nuclear--structure models, the spherical shell model and the
collective model in two of its versions, but also mention a number of
other applications of RMT. Nuclei (and other fermionic many--body
systems) are governed by the mean field (in nuclei: the shell model)
and a residual interaction dominated by two--body forces. In such
systems, the structure of the Hamiltonian is very different from that
of a typical GOE matrix. That difference has given rise to a number of
random--matrix ensembles which are closer in structure to the
mean--field approach than is the GOE. These are treated in
Section~\ref{concepts}. Section~\ref{concl} contains a summary and
conclusions.

\section{Random Matrices}
\label{RMT}

\subsection{Why Random Matrices?}
\label{why}

Random Matrices were introduced into Nuclear Physics in the 1960s by
Wigner (See Wigner's papers in Porter (1965)). That step was
preceded (and, in our view, probably motivated) by Bohr's insight that
nuclei are systems of great complexity. It is useful to recall the
arguments that led Bohr (1936) to this insight.

\begin{figure}[h]
\vspace{5 mm}
\includegraphics[width=\linewidth]{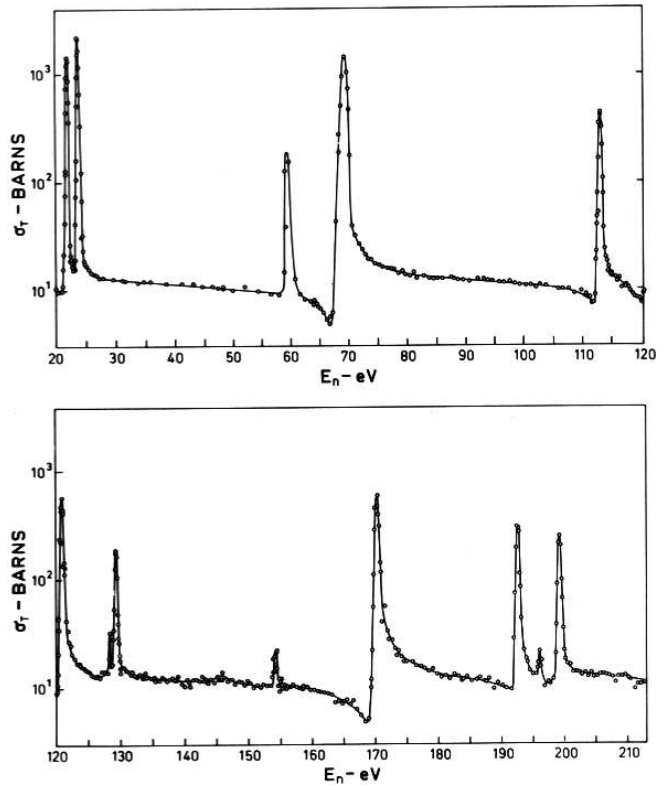}
\vspace{3 mm}
\caption{The total neutron cross section on $^{232}$Th versus neutron
energy $E_n$ in eV. From Ref.~\cite{Neu64} as reproduced in Bohr and
Mottelson (1969), Vol. 1, p.178.}
\label{fig1}
\end{figure}

Experiments in the 1930s, especially by Fermi and his group in Rome on
neutron scattering by light nuclei, had revealed the existence of
numerous narrow resonances (Fermi {\it et al.}, 1934, 1935).  We show
in Fig.~\ref{fig1} data of a similar type taken in the 1950s by
Rainwater and his group at Columbia University (Garg {\it et al.},
1964).  That group used time--of--flight spectroscopy of slow neutrons
to measure the total neutron cross section on a number of heavy
even--even nuclei (nuclei with even numbers of protons and
neutrons). The cross section versus neutron energy $E_n$ shown in the
panels of Fig.~\ref{fig1} for the target nucleus $^{232}$Th displays
narrow resonances with widths $<$ 1 eV and spacings of about 20
eV. The target nucleus $^{232}$Th has spin zero and positive parity;
the incident slow neutrons carry zero angular momentum and have spin
$1/2$. Therefore the resonances all have spin/parity $1/2^+$. These
resonances correspond to excited states of the ``compound nucleus''
$^{233}$Th with an excitation energy slightly above the neutron
separation energy of $4.786$ MeV (the ``neutron threshold''). The
number of resonances observed in each compound nucleus was limited by
the resolution of the spectrometer and was never much larger than
$200$. Similar data on proton resonances at the Coulomb barrier in
lighter nuclei were later taken by the Triangle Universities group
(Wilson {\it et al.}, 1975).  Together these data form what has been
called the ``nuclear data ensemble'' by Haq {\it et al.} (1982) and
Bohigas {\it et al.} (1983).

\begin{figure}[h]
\vspace{5 mm}
\includegraphics[width=\linewidth]{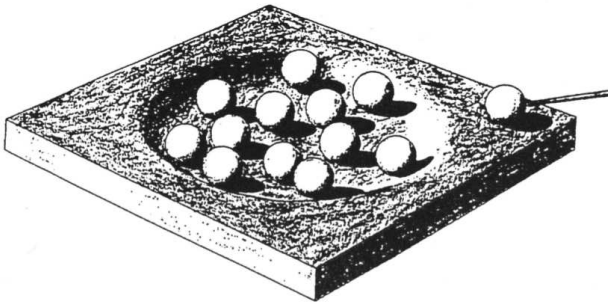}
\vspace{3 mm}
\caption{Bohr's wooden toy model of the compound nucleus. From Nature
(1936).}
\label{fig2}
\end{figure}

Bohr argued that the existence of numerous narrowly spaced and narrow
compound--nucleus resonances was incompatible with
independent--particle motion and was due to strong nucleon--nucleon
interactions. Indeed, assuming an independent--particle model with a
nuclear radius of about 5 fm and a potential well depth of several ten
MeV, one finds that the single--particle states have a typical spacing
of several hundred keV and widths of the order of ten keV or larger,
in complete disagreement with the data. To account qualitatively for
the data, Bohr proposed his ``compound--nucleus model''
(Fig.~\ref{fig2}): The incident nucleon carries kinetic energy (as
indicated by the billiard cue), collides with the nucleons in the
target and shares its energy with many nucleons. In units of the time
for passage of the nucleon through the nuclear interior, it takes the
system a long time until one of its constituent nucleons acquires
sufficient energy to be re--emitted from the system.

\begin{figure}[h]
\vspace{5 mm}
\includegraphics[width=\linewidth]{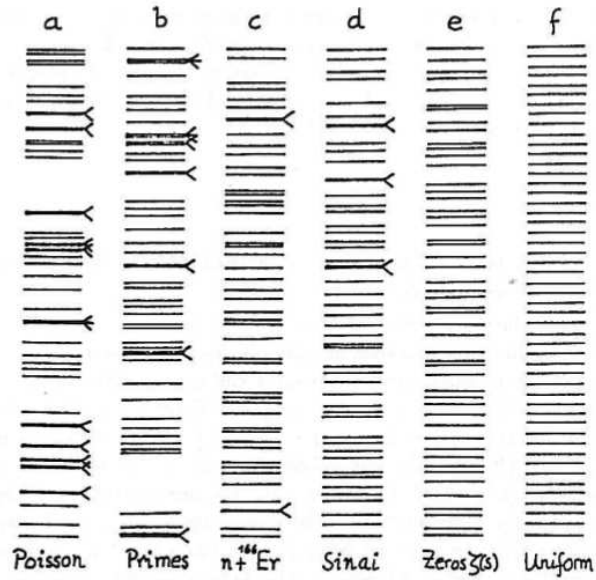}
\vspace{3 mm}
\caption{Six spectra with 50 levels each and the same mean level
spacing. From right to left: The one--dimensional harmonic oscillator,
a sequence of zeros of the Riemann Zeta--function, a sequence of
eigenvalues of the Sinai billiard (see Section~\ref{RMTchaos}), a
sequence of resonances seen in neutron scattering on $^{166}$Er, a
sequence of prime numbers, and a set of eigenvalues obeying Poisson
statistics (see section~\ref{RMTchaos}). From Bohigas and Giannoni
(1984).}
\label{fig3}
\end{figure}

Bohr's idea that the nucleus is a complex, strongly interacting system
was adopted by the community and held sway until the discovery of the
nuclear shell model in 1949. Bohr's idea almost certainly motivated
Wigner to introduce random matrices. To explain the spirit of the
approach, we focus attention on nuclear levels with the same quantum
numbers (total spin $J$, parity $\Pi$, and, at least in light nuclei,
total isospin $T$) and ask: Can we identify generic spectral
properties of a system with strong interactions? Fig.~\ref{fig3} shows
six spectra, all having the same total number of levels, and spanning the
same total energy interval, and therefore having the same average
level spacing. The spectra differ only in the way the spacings between
neighboring levels are distributed. For the one--dimensional harmonic
oscillator (the rightmost spectrum), all spacings are identical. The
spacing distributions differ more and more from a delta function as we
go ever more to the left. The random--matrix approach characterizes
spectra by their fluctuation properties: The distribution of spacings
of nearest neighbors is the first and obvious measure for spectral
fluctuations. It is referred to as the nearest--neighbor spacing (NNS)
distribution. There are other measures such as the correlation between
nearest spacings, between next--nearest spacings, etc. Some of these
are introduced below.

To implement this approach, we need to develop a statistical theory of
spectra. Random matrices provide the tool to do so. Instead of
considering the actual nuclear Hamiltonian (which was not known in the
1950s) we consider an ensemble of Hamiltonians (each given in matrix
form). The ensemble is defined in terms of some probability
distribution for the matrix elements, hence the name random matrices.
The ensemble is chosen in such a way that the member Hamiltonians
incorporate generic features. The spectral distribution functions are
calculated as averages over the ensemble and are compared with the
actual fluctuation properties of nuclear spectra.

Canonical random--matrix theory (RMT) as developed by Wigner and
Dyson -- see the compilation by Porter (1965) --  classifies systems by their symmetry properties.
Nuclei are invariant under time reversal. The matrix representation of
the nuclear Hamiltonian can accordingly be chosen real and symmetric.
The random--matrix ensemble which is considered almost exclusively in
the sequel is therefore an ensemble of real and symmetric matrices.

The random--matrix approach does not aim at calculating individual
spectra and at comparing them with data. Rather, one determines the
joint probability distribution of the eigenvalues and from here
calculates certain spectral fluctuation measures such as the NNS
distribution as averages over the ensemble. RMT contains one (or, in
the general case, a number of) input parameter(s). In the case of
spectral fluctuations, that input parameter is the average nuclear
level spacing. The fluctuation measures predicted by RMT are scaled by
the average level spacing and, thus, parameter--free. If the observed
spectral fluctuation properties agree with RMT predictions, and if no
further information on the system is available, one concludes that the
system is generic. This implies that no information beyond the average
nuclear level spacing can be deduced from the available spectral
information. If, on the other hand, the data do not agree with RMT
predictions, this indicates that the spectrum is not generic and that
the available spectral information may be used to deduce further
properties of the system. The harmonic oscillator in one dimension is
a case in point.

The random--matrix approach to spectral fluctuations (and to other
properties of complex systems) has some similarity to classical
thermodynamics. There one is also interested in a generic description
of systems in terms of a few parameters. These parameters (specific
heat, magnetic susceptibility, etc.) are system--specific, but within
the framework of classical thermodynamics, need not be determined from
the system's Hamiltonian. In that sense, classical thermodynamics and
random--matrix theory are phenomenological theories that do not refer
to an underlying system--specific Hamiltonian. The random--matrix
approach differs fundamentally from the dynamical approach used in
most fields of physics where one integrates the equations of motion
and fits a few parameters of an (otherwise known) Hamiltonian to the
data. Similar to classical thermodynamics, the random--matrix approach
has been applied to many systems beyond nuclear physics (Guhr {\it et
  al.}, 1998).

\subsection{The Gaussian Orthogonal Ensemble}
\label{GOE}

In Section~\ref{GOE} and in Section~\ref{propGOE} we define the
Gaussian orthogonal ensemble (GOE) and collect and interpret a number
of results for this ensemble. We also introduce the Gaussian unitary
ensemble (GUE). Proofs may be found in the books by
Porter (1965)  and Mehta (2004). For the GOE we consider
real and symmetric Hamiltonian matrices $H$ in a Hilbert space of
dimension $N$. With $\mu, \nu = 1, \ldots, N$ the matrix elements obey
$H_{\mu \nu} = H_{\nu \mu} = H^*_{\mu \nu}$. For realistic systems
Hilbert space is infinite--dimensional, so we consider the limit $N
\to \infty$ in the sequel. The ensemble is defined in terms of an
integration over matrix elements. The volume element in matrix space
\be
{\rm d} [ H ] = \prod_{\mu \leq \nu} {\rm d} H_{\mu \nu}
\label{1}
\ee
is the product of the differentials ${\rm d} H_{\mu \nu}$ of the
independent matrix elements (i.e., of the matrix elements not connected
by symmetry). The ensemble is defined by the probability density
${\cal P}(H)$ of the matrices $H$,
\be
{\cal P}(H) \ {\rm d} [ H ] = {\cal N}_0 \exp \bigg\{ - \frac{N}{4
\lambda^2} {\rm Trace} ( H^2 ) \bigg\} \ {\rm d} [ H ] \ .
\label{2}
\ee
Here ${\cal N}_0$ is a normalization factor and $\lambda$ a parameter.
This parameter defines the average level density (see Eq.~(\ref{4})
below). In applications of the GOE to data, $\lambda$ is determined by
the empirical average level density. The spectral fluctuation
properties of the GOE are then predicted in a parameter--free fashion.

The Gaussian weight factor is a cutoff that ensures convergence of the
ensemble averages of observables for large values of the integration
variables. We use the symmetry of the matrices to write the trace in
the exponent as $\sum_{\mu < \nu} 2 H^2_{\mu \nu} + \sum_\mu H^2_{\mu
\mu}$. Then the probability density
\ba
{\cal P}(H) \ {\rm d} [ H ] &=& {\cal N}_0 \prod_\mu \exp \{ - \frac{N}
{4 \lambda^2} H^2_{\mu \mu} \} {\rm d} H_{\mu \mu} \nonumber \\
&& \qquad \times \prod_{\rho < \sigma} \exp \{ - \frac{N}{2 \lambda^2}
H^2_{\rho \sigma} \} {\rm d} H_{\rho \sigma}
\label{2b}
\ea
is a product of terms each of which depends only on a single matrix
element. Therefore the GOE has the following properties: The independent
matrix elements are uncorrelated Gaussian--distributed random variables
with zero mean value and a second moment given by
\be
\overline{H_{\mu \nu} H_{\rho \sigma}} = \frac{\lambda^2}{N} \big(
\delta_{\mu \rho} \delta_{\nu \sigma} + \delta_{\mu \sigma} \delta_{\nu
\rho} \big) \ .
\label{2d}
\ee
Here the overbar denotes the ensemble average. Defining the GOE by
these properties is equivalent to the definition~(\ref{2}).

While the form of the probability measure in Eq.~(\ref{2}) is fixed by
symmetry requirements, the Gaussian cutoff in that equation seems
completely arbitrary. However, using plausible assumptions one can
actually derive that factor. Rosenzweig and Porter (1960) have shown
that the distribution~(\ref{2}) is obtained when one assumes that the
ensemble is orthogonally invariant, and that matrix elements not
connected by symmetry are statistically independent. And Balian (1968)
has derived the distribution~(\ref{2}) from a maximum entropy
principle.

In the GOE every state in Hilbert space is connected to itself and to
every other state by a matrix element of $H$. Since all non--diagonal
matrix elements have the same first and second moments, every state is
coupled to all other states with equal average strength. This results
in level repulsion between any pair of levels, and in a complete
mixing of states in Hilbert space. The importance of such coupling is
seen when we consider a more general ensemble with probability density
\ba
{\cal P}_\alpha(H) {\rm d} [H] &=& \tilde{{\cal N}}_0 \prod_\mu
\exp \{ - \frac{N} {4 \lambda^2} H^2_{\mu \mu} \} {\rm d} H_{\mu \mu}
\nonumber \\ && \qquad \times \prod_{\rho < \sigma} \exp \{ -
\frac{N}{2 \alpha \lambda^2} H^2_{\rho \sigma} \} {\rm d} H_{\rho
\sigma}
\label{2f}
\ea
where the positive parameter $\alpha$ ranges from zero to one. For
$\alpha = 0$ all non--diagonal elements vanish, and the
ensemble~(\ref{2f}) consists of diagonal matrices with independent,
Gaussian--distributed diagonal elements. The shape of the average
spectrum is Gaussian, there is no level repulsion, and the spectral
fluctuations are Poissonian (see Section~\ref{RMTchaos}). For $\alpha =
1$, the ensemble coincides with the GOE. For values of $\alpha$
between these two limits, the shape of the spectrum and the spectral
fluctuations interpolate between those two limiting cases. Significant
mixing between levels occurs when the mean--square mixing matrix
element $\overline{H^2_{\mu \nu}}$ with $\mu \neq \nu$ is roughly
equal to the square of the mean level spacing. Taking for the latter
the GOE value $d = \pi \lambda / N$ at the center of the semicircle
(see Eq.~(\ref{4}) below), we find that significant mixing occurs when
$\alpha$ is of order $1 / \sqrt{N}$. We see that for $N \to \infty$,
mixing sets in as soon as $\alpha$ differs from zero. These
observations are used in Sections~\ref{isos} and \ref{trsb}.

Reality and symmetry of the matrices $H_{\mu \nu}$ are preserved under
orthogonal transformations of the basis. The ensemble~(\ref{2}) is
accordingly chosen in such a way that it is invariant under such
transformations: With each matrix $H$ belonging to the ensemble, all
matrices obtained from $H$ by orthogonal transformations also belong
to the ensemble. As a consequence, there does not exist a preferred
direction in Hilbert space, and the ensemble is generic. Because of
that invariance and the Gaussian cutoff, the ensemble is referred to as
the Gaussian orthogonal ensemble of random matrices.

Instead of the $N(N+1)/2$ integration variables $H_{\mu \nu}$ with
$\mu \leq \nu$ used in Eq.~(\ref{2}), we may use the $N$ eigenvalues
$E_\mu$ of the matrices $H$ and the $N(N-1)/2$ generators of the
orthogonal transformation ${\cal O}$ which diagonalizes $H$. Then the
volume element ${\rm d} H$ takes the form
\be
{\rm d} H = {\rm d} {\cal O} \prod_{\mu < \nu} |E_\mu - E_\nu|
\prod_\rho {\rm d} E_\rho \ .
\label{1a}
\ee
The factor ${\rm d} {\cal O}$ stands for the Haar measure of the
orthogonal group in $N$ dimensions. (The Haar measure is the unique
invariant measure that can be assigned to every compact group and that
is used to define integrals over that group (Conway, 1990)). The
probability density ${\cal P}(H)$ takes the form
\ba
{\cal P}(H) {\rm d} [ H ] &=& {\cal N}_0 \ {\rm d} {\cal O} \exp
\bigg\{ - \frac{N}{4 \lambda^2} \sum_\mu E^2_\mu \bigg\} \nonumber \\
&& \times  \prod_{\rho < \sigma} |E_\rho - E_\sigma| \prod_\nu {\rm
d} E_\nu \ .
\label{2a}
\ea
The right--hand side of Eq.~(\ref{2a}) is the product of two factors.
One factor depends only on the eigenvalues and the other, only on the
diagonalizing matrices. It follows that the eigenvalues and the
eigenvectors of the matrices $H$ are uncorrelated random variables.
In the limit $N \to \infty$, the projections of the eigenvectors onto
an arbitrary direction in Hilbert space have a Gaussian distribution.
Some properties of the eigenvalue distribution can be read off
directly from Eq.~(\ref{2a}). The factor $\prod_{\mu < \nu} |E_\mu -
E_\nu|$ stems from the volume element in matrix space and reflects the
orthogonal invariance of the ensemble. It causes the probability
density for the eigenvalues to go to zero as two eigenvalues approach
each other. This is a manifestation of level repulsion, a basic
feature of quantum mechanics.

Another property of the GOE is displayed when we write the probability
density in the form
\ba
&&{\cal P}(H) {\rm d} [ H ] = {\cal N}_0 \ {\rm d} {\cal O} \prod_\mu
{\rm d} E_\mu \nonumber \\
&& \qquad \times \exp \bigg\{ - \frac{N}{4 \lambda^2} \sum_\nu E^2_\nu
+ \sum_{\rho < \sigma} \ln |E_\rho - E_\sigma| \bigg\} \ .
\label{2e}
\ea
For the interpretation of the eigenvalue distribution in
Eq.~(\ref{2e}) we use an analogy to classical statistical mechanics. We
consider the eigenvalues $E_\mu$ as position coordinates of $N$
particles in one dimension. The probability density of the eigenvalues
in Eq.~(\ref{2e}) then has the form of the canonical partition
function (integrated over the momentum variables) of a gas of $N$
classical point particles with repulsive two--body interactions
(``Coulomb gas'') moving in a common harmonic oscillator potential at
inverse temperature $\beta = 1$. The particles will tend to keep apart
as much as is consistent with the overall harmonic oscillator
potential. This property leads to the ``spectral stiffness'' of the
GOE discussed below.

Besides the GOE there exist two more canonical random--matrix
ensembles: The Gaussian unitary ensemble (GUE) and the Gaussian
symplectic ensemble (GSE). The GUE is the Gaussian ensemble of
Hermitian (but not necessarily real) matrices. This ensemble plays a
role for systems which are not time--reversal invariant. If that
invariance does not hold, the Hamiltonian matrix is Hermitian but
cannot in general be chosen real and symmetric. In nuclear physics,
the GUE is used for tests of time--reversal invariance. It is also
used as a theoretical testing ground because the calculation of
ensemble averages over observables is typically simpler for the GUE
than for the GOE. For these reasons, we briefly introduce the GUE in
the next paragraphs. The GSE applies to systems with half--integer
spin which are invariant under time reversal but which are not
rotationally invariant. The GSE does not apply to nuclei directly
(see, however, Lombardi {\it et al.}, 1994) and is not discussed in
this review. In addition to the three canonical ensembles introduced
by Dyson (see Porter, 1965) and often distinguished by the label
$\beta$ with $\beta = 1$ for the GUE, $\beta = 2$ for the GOE, $\beta
= 4$ for the GSE, there exist seven more random--matrix ensembles
which are defined in terms of invariance requirements (Altland and
Zirnbauer, 1997) ; their construction is based upon Cartan's
classification of Lie groups (see Chevalley, 1946). Some of these
ensembles play a role in the low--energy behavior of quantum field
theories and find application in lattice QCD, the discretized form of
quantum chromodynamics. Others relate to the scattering of electrons
off the interphase between a normal conductor and a superconductor.

For the GUE the independent variables are the real and the imaginary
parts of the complex elements $H^{\rm GUE}_{\mu \nu}$ of the
Hamiltonian (a matrix of dimension $N$). The invariant measure has the
form
\be
{\rm d} [ H^{\rm GUE} ] = \prod_{\mu < \nu} {\rm d} [ \Re
H^{\rm GUE}_{\mu \nu} ] {\rm d} [  \Im H^{\rm GUE}_{\mu \nu} ]
\prod_\sigma {\rm d} H^{\rm GUE}_{\sigma \sigma} \ .
\label{18a}
\ee
With this definition, the equation for the probability density of the
GUE is similar to Eq.~(\ref{2}) and reads
\ba
&& {\cal P}(H^{\rm GUE}) \ {\rm d} [ H^{\rm GUE} ] \nonumber \\
&& = {\cal N}_0 \exp \bigg\{ - \frac{N}{2 \lambda^2} {\rm Trace}
(H^{\rm GUE})^2 \bigg\} \ {\rm d} [ H^{\rm GUE} ] \ .
\label{2g}
\ea
The GUE is invariant under unitary transformations of Hilbert space.
The real and imaginary parts of the matrix elements are uncorrelated
random variables with equal Gaussian probability distributions
centered at zero. The factors in the exponent are chosen in such a way
that the second moments have the values
\be
\overline{H^{\rm GUE}_{\mu \nu} H^{\rm GUE}_{\rho \sigma}} =
\frac{\lambda^2}{N} \delta_{\mu \sigma} \delta_{\nu \rho} \ .
\label{2h}
\ee

Level repulsion in the GOE is linear, see Eqs.~(\ref{2a}) and
(\ref{7}). This is a consequence of orthogonal invariance. In the GUE,
the transformation to eigenvectors and eigenvalues as new integration
variables involves a unitary transformation ${\cal U}$ and yields
\ba
{\cal P}(H^{\rm GUE}) {\rm d} [ H^{\rm GUE} ] &=& {\cal N}_0 \ {\rm d}
{\cal U} \exp \bigg\{ - \frac{N}{2 \lambda^2} \sum_\mu E^2_\mu \bigg\}
\nonumber \\ 
&& \times  \prod_{\rho < \sigma} (E_\rho - E_\sigma)^2 \prod_\nu {\rm
d} E_\nu \ .
\label{2i}
\ea
Here ${\rm d} {\cal U}$ denotes the Haar measure of the unitary group
in $N$ dimensions. Instead of the factor $|E_\rho - E_\sigma|$
occurring in Eq.~(\ref{2a}), Eq.~(\ref{2i}) contains the factor
$(E_\rho - E_\sigma)^2$. As a result, level repulsion for the GUE is
quadratic. That difference between GOE and GUE is easily understood:
In the GOE, the coupling of any pair of levels is described by a
single parameter, the real coupling matrix element. For two levels to
have a small spacing, the value of that parameter must be small. In
the GUE, the coupling is described by two parameters, the real and the
imaginary parts of the coupling matrix element. For two levels to have
a small spacing, both parameters must be small, and the probability of
small spacings is reduced accordingly.

As concerns the GUE analog of Eq.~(\ref{2e}), the form of the volume
element in matrix space leads to a different inverse temperature
$\beta = 2$, while for the GSE we have $\beta = 4$.  The ``Dyson
parameter'' $\beta$ with $\beta = 1,2,4$ is often used to label the
three canonical random--matrix ensembles GUE, GOE, GSE.

\subsection{Properties of the GOE}
\label{propGOE}

\subsubsection{Average Level Density}
\label{avrho}

A central property of the GOE is the mean level density $\rho(E)$, a
function of the energy $E$. It is defined as
\be
\rho(E) = \overline{ \sum_\mu \delta(E - E_\mu)}
\label{3}
\ee
and, for $N \to \infty$, given by
\be
\rho(E) = \frac{N}{\pi \lambda} \sqrt{1 - \bigg( \frac{E}{2 \lambda}
\bigg)^2} \ .
\label{4}
\ee
The average spectrum extends from $- 2 \lambda$ to $+ 2 \lambda$.
This confinement of the spectrum to a finite stretch of the energy
axis is a consequence of the Gaussian cutoff (or, for that matter, of
any other sufficiently strong cutoff factor), and of the factor $N$ in
the exponent of Eq.~(\ref{2}). When plotted versus $E / (4 \lambda)$,
$\rho(E)$ has the shape of a semicircle. (That shape is specific for
the Gaussian cutoff). That is why Eq.~(\ref{4}) is often referred to
as ``Wigner's semicircle law''. The factor $N$ on the right--hand side
of Eq.~(\ref{4}) ensures that $\rho(E)$ is normalized to the total
number of levels. The mean level spacing $d(E)$ is defined by
\be
d(E) = \rho^{-1}(E)
\label{4a}
\ee
and tends to zero as $N \to \infty$ because we fit a spectrum of $N$
eigenvalues into a finite energy interval of length $4 \lambda$. At
the center of the spectrum, we have $d(0) = \pi \lambda / N$.

\subsubsection{Universality}
\label{univ}

The form of the spectrum is due to the Gaussian cutoff factor. That
form is obviously totally unrealistic: Hardly any real physical system
possesses such a spectrum. While reality and symmetry of the matrices
$H_{\mu \nu}$ reflect time--reversal invariance and are thus a
consequence of quantum theory, the Gaussian cutoff is not, although
the arguments of Rosenzweig and Porter (1960) and of Balian (1968)
lend some plausibility to its use. The Gaussian cutoff is preferred
from a practical point of view, of course, because of the ease with
which Gaussian integrals can be performed. But the GOE is physically
interesting only if it furnishes information which is independent of
the form of the cutoff factor. That property is guaranteed by the
universality of the GOE.

In using the GOE we are usually not interested in the overall shape
of the spectrum. Interest rather focuses on local spectral fluctuation
properties such as the NNS distribution or correlations between level
spacings. These are predicted in a parameter--free fashion. That means
that all local spectral fluctuation properties are functions of a
dimensionless parameter $s$, the ratio of the actual level spacing and
the mean level spacing. Local spectral fluctuations characterize
properties of the spectrum on an energy scale which in the limit $N
\to \infty$ is negligibly small compared to the length $4 \lambda$ of
the spectrum. On that scale, the spectral fluctuations are universal:
As functions of the parameter $s$ they have the same form for both the
GOE and all non--Gaussian cutoff factors, as long as the latter are
orthogonally invariant and confine the spectrum to a finite
singly--connected piece of the energy axis (Hackenbroich and
Weidenm\"{u}ller, 1995).

Non--Gaussian cutoffs which obey that proviso modify the overall shape
of the spectrum. In fact, for any given form of the spectrum it is
always possible to find a cutoff factor such that the resulting
random--matrix ensemble has an average spectrum of that form. The
local fluctuation properties are unaffected by such a choice: In the
limit $N \to \infty$, the local fluctuation properties separate from
the global spectral properties and become universal.

\subsubsection{Ergodicity}
\label{ergod}

Theoretical predictions of the GOE are obtained as averages over the
ensemble. How can we compare such predictions in a meaningful way with
data which, after all, are taken from a physical system with a single
Hamiltonian (and not from an ensemble of Hamiltonians)? That question
is answered by the property of ergodicity of the GOE. Spectral data on
a given system can be used to calculate spectral measures such as the
mean level spacing or the NNS distribution as running averages over
the spectrum. We denote such a running average by angular brackets.
We would like to ascertain that $\overline{O} = \langle O \rangle$
holds true for all members of the ensemble and for all observables $O$
that  describe local spectral properties. That equation cannot be
proved in general because there is no way to evaluate $\langle O
\rangle$ in the framework of the GOE. It is possible, however, to
prove the slightly weaker statement (Brody {\it et al.}, 1981)
\be
\overline{\big(\overline{O} - \langle O \rangle\big)^2} = 0 \ .
\label{5}
\ee
The proof is made possible because all terms on the left--hand side
are ensemble averages. The statement says that for almost all members
of the ensemble (with the exception of a set of measure zero; the
measure being defined in Eq.~(\ref{1})) the running average of an
observable $O$ (calculated for a single member of the ensemble) is
equal to the ensemble average of the observable. That property is
referred to as ergodicity. The name derives from the formal similarity
of the statement with ergodicity in classical statistical mechanics
(equality of phase--space average and time--average along a single
trajectory).

\subsubsection{Information Content of GOE Spectra}
\label{inform}

Eq.~(\ref{2b}) shows that in the GOE, every state in Hilbert space is
coupled to every other one by a Gaussian--distributed random matrix
element: In the GOE all states in Hilbert space are completely mixed
with each other. Choosing the parameters $N$ and $\lambda$ and drawing
all independent matrix elements from the resulting Gaussian
distribution generates a random GOE matrix. Diagonalizing that matrix
yields a GOE spectrum. By construction, that spectrum contains no
information beyond the input parameters $N$ and $\lambda$. In
particular, the spectral fluctuations are void of physical
information. If the spectral fluctuations of an experimental spectrum
agree with GOE predictions, and if there is no further information on
that system, then the spectral data alone cannot be used to extract any
physical information on the system beyond the mean level density.

That conclusion is also reached when we ask: How many pieces of
spectral data are needed to determine the underlying Hamiltonian $H$?
In the case of a GOE spectrum, counting shows that we need all $N$
eigenvalues and all $N$ orthonormal eigenfunctions to determine the
$N(N+1)/2$ independent matrix elements of $H$. This must be compared
with the usual dynamical approach to physical systems where the
Hamiltonian is given in terms of a few (say $n$) parameters. Then $n$
pieces of data suffice to determine the Hamiltonian. Further data can
be used to check the consistency of the underlying theory.

\subsection{GOE Fluctuation measures}
\label{fluct}

\subsubsection{Porter--Thomas distribution}
\label{porter}

We recall that in the GOE eigenvalues and eigenfunctions are
uncorrelated random variables. For $N \to \infty$, the projections of
the eigenfunctions onto an arbitrary vector in Hilbert space have a
Gaussian distribution centered at zero. Therefore the squares
$\psi^2$ of such projections have a $\chi^2$--distribution with one
degree of freedom. We introduce the variable
\be
y = \psi^2 / \ \overline{\psi^2} \ .
\label{6a}
\ee
The resulting distribution is also known as the Porter--Thomas
distribution and has the form
\be
P(y) = \frac{1}{\sqrt{2 \pi y}} \exp ( - y / 2 ) \ .  
\label{6}
\ee
The function $P(y)$ is given in terms of the mean value $\Gamma =
\overline{\psi^2}$. That parameter is an input parameter which is not
predicted by random--matrix theory. The distribution can be checked
experimentally: Transition probabilities of nuclear levels to a fixed
final state and decay widths to a fixed channel are proportional to
squares of matrix elements containing the nuclear wave functions.
These matrix elements can be read as projections of the wave functions
onto a particular vector in Hilbert space.

It may happen that the mean value $\Gamma$ undergoes a secular
variation.  This is the case, for instance, for doorway states. Then
it is necessary to ``unfold'' the fluctuations by scaling the
intensities properly, see the end of Section~\ref{door}.

\subsubsection{Nearest--Neighbor--Spacing Distribution and
$\Delta_3$--Statistic}
\label{nns}

It takes substantial theoretical effort to work out the spectral
fluctuation measures in the GOE. That is not described here. We
confine ourselves to introducing two fluctuation measures that have
found wide application in the analysis of experimental data: The
nearest--neighbor--spacing (NNS) distribution $P(s)$ and the
$\Delta_3$ statistic due to Dyson and Mehta. These are obtained in the
limit $N \to \infty$. Prior to using these measures for data analysis,
it is neccessary to ``unfold'' the experimental spectra, see
Section~\ref{unfo}.

\begin{figure}[h]
\vspace{5 mm}
\includegraphics[height=\linewidth,angle=270]{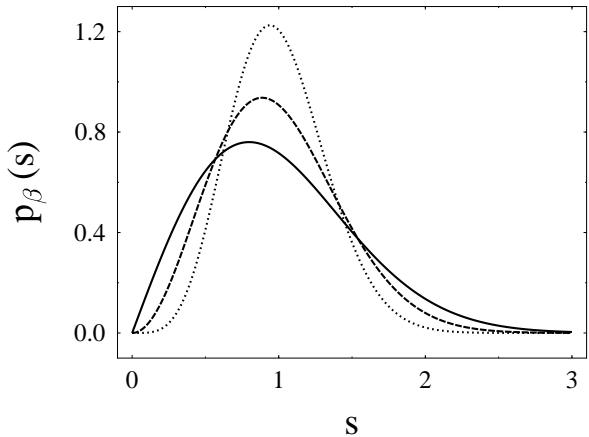}
\vspace{3 mm}
\caption{The nearest--neighbor--spacing (NNS) distribution of the GOE
(solid line) versus $s$, the ratio of the actual level spacing and 
the mean level spacing. For the sake of comparison we also show the
NNS distributions for the GUE (dashed line) and the GSE (dotted line).
The parameter $\beta$ is the Dyson index with $\beta = 1,2,4$ for GUE,
GOE, GSE, respectively.}
\label{fig4}
\end{figure}

The NNS distribution $P(s)$ depends on $s$, the ratio of the actual
level spacing and the mean level spacing $d$.  It cannot be given in
closed form. An excellent approximation due to Wigner is known as the
Wigner surmise,
\be
P(s) = \frac{\pi}{2} s \exp ( - \pi s^2 / 4 ) \ .
\label{7}
\ee
The linear increase with $s$ for small $s$ is due to GOE level
repulsion as displayed in Eq.~(\ref{2a}). Universality shows that the
Gaussian falloff is not related to the Gaussian cutoff factor defining
the GOE and simply accounts for the fact that very large spacings are
unlikely to occur. The exact expression for $P(s)$ was first derived
by Gaudin (1961). $P(s)$ is displayed in Fig.~\ref{fig4}.

\begin{figure}[h]
\vspace{5 mm}
\includegraphics[height=\linewidth,angle=270]{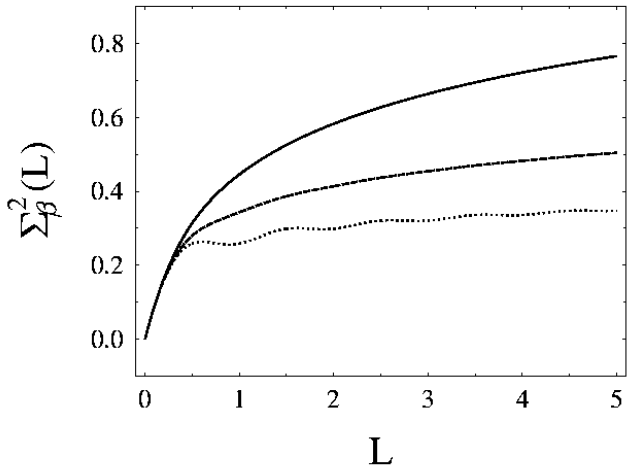}
\vspace{3 mm}
\caption{The number variance versus the length $L$ of the interval
($L$ is in units of the mean level spacing) for the three canonical
ensembles. Top curve: GOE; middle curve: GUE; bottom curve: GSE. The
parameter $\beta$ is the Dyson index, see Fig.~\ref{fig4}. From
Guhr {\it et al.} (1998).}
\label{fig5}
\end{figure}

The NNS distribution describes the distribution of level spacings but
does not contain information about their correlations. Such
information is contained in another fluctuation measure, the
$\Delta_3$--statistic. The number staircase function
\be
{\cal N}(E) = \int_{- \infty}^E {\rm d} E' \ \sum_\mu \delta(E' -
E_\mu)
\label{8}
\ee
counts the number of eigenvalues below energy $E$. With increasing
energy, it increases by unity as $E$ passes a (non--degenerate)
eigenvalue and is otherwise constant. The number of eigenvalues in the
energy interval $[E_0, E_0 + L]$ is given by $n(E_0, L) = {\cal N}(E_0
+ L) - {\cal N}(E_0)$. By definition of the mean level spacing $d(E)$,
we have $\overline{n(E_0, L)} = L / d(E_0)$. (We use that for $N \to
\infty$, $d(E)$ is constant (independent of $E$) in any energy
interval containing a finite number of levels). The number variance
$\Sigma^2_\beta(L) =$ $\overline{n^2(E_0, L)} -$ $(\overline{n(E_0,
L)})^2$ is a fluctuation measure which contains information about
correlations between level spacings. Suppose, for instance, that actual
GOE spectra can be constructed by drawing spacings at random from the
NNS distribution. In this case, $\Sigma^2_\beta(L)$ would grow linearly
with $L$. In actual fact $\Sigma^2_\beta(L)$ is, for large $L$,
proportional to $\ln L$. The slow growth indicates that large spacings
and small spacings do not follow each other at random but almost
alternate and reflects the ``stiffness'' of GOE spectra, see the text
below Eq.~(\ref{2e}). For the three canonical ensembles, the number
variance is shown in Fig.~\ref{fig5}. The number variance is seldom
used in nuclear physics because it fluctuates too strongly, and one
uses the $\Delta_3$--statistic by Dyson and Mehta instead. The latter
is defined by
\ba
&&\Delta_3(L) = \\
&& {\rm min}_{a,b} \frac{1}{L} \bigg\langle \int_{E_0}^{E_0 + L} {\rm
d} E' \ \overline{ \bigg( {\cal N}(E') - a - b E' \bigg)^2 } \bigg
\rangle_{E_0} \ . \nonumber
\label{9}
\ea
We integrate the ensemble average of the square of the difference
between the number staircase function and the straight line $(a + b
E')$ over an energy interval, divide by the length $L$ of that
interval, and minimize the result with respect to the parameters $a$
and $b$ of the straight line. The angular brackets denote an average
over the initial point $E_0$. It can be shown that $\Delta_3(L)$ can
be written as an integral over the number variance
$\Sigma^2_\beta(L)$. Therefore $\Delta_3$ is much smoother than
$\Sigma^2_\beta(L)$ and is better suited for data analysis. Similar to
$\Sigma^2_\beta(L)$, $\Delta_3(L)$ grows logarithmically with $L$. For
large $L$,
\be
\Delta_3(L) \approx \frac{1}{\pi^2} \{ \ln L - 0.0687 \} \ .
\label{10}
\ee
Similar to $\Sigma^2$, the $\Delta_3$--statistic reflects the
stiffness of GOE spectra and is often referred to as ``spectral
stiffness''. Fig.~\ref{fig7} shows $\Delta_3(L)$ versus $L$ for the
GOE.

\subsubsection{Unfolding of Spectra. Purity and Completeness.}
\label{unfo}

In the limit $N \to \infty$, the average level density $\rho(E)$ of
the GOE in Eq.~(\ref{4}) is constant in every energy interval
containing a finite number of levels, and the same is true of the
average level spacing $d$. In nuclei the situation differs: The level
density grows nearly exponentially with energy. In many cases, even a
fairly short stretch of levels displays this fact: The spacings of the
lowest--lying levels are consistently larger than those of the
highest--lying ones. That fact distorts the spectral fluctuation
measures and must be taken into account prior to comparing data with
GOE predictions. This is done by ``unfolding'' the spectra: The actual
spectrum is modified such that the average level spacing is
constant. GOE predictions relate to spectra consisting of levels with
identical quantum numbers.  Spectra obtained experimentally may be
incomplete (i.e., miss levels (especially those with small or very
large widths)), or not be pure (i.e., may contain levels with
uncertain or incorrect quantum number assignments). It is important to
know how lack of completeness and/or purity affects the comparison of
data with GOE predictions.

Unfolding requires the knowledge of the average level density
$\rho(E)$ for the data at hand. The situation is easy if a theoretical
prediction for the average level density is available. This is the
case, for instance, in billiards (where a point particle moving in two
dimensions is scattered elastically on some surface). Here the Weyl
formula (see Baltes and Hilf, 1976) gives the average level density in
closed form in terms of the area enclosed by the surface and the
length of the boundaries of that surface. Given $\rho(E)$, the
spectrum (or the spectra) are subsequently unfolded by mapping the
eigenvalues $E_\mu$ onto new eigenvalues $\ve_\mu$ by the prescription
\be
\ve_\mu = \int_{- \infty}^{E_\mu} {\rm d} E \ \rho(E) \ .
\label{11}
\ee
By construction, the new eigenvalues are dimensionless and have an
average level spacing equal to unity. The $\ve_\mu$ can be used to
construct the NNS distribution and the $\Delta_3$--statistic. We
observe that the right--hand side of Eq.~(\ref{11}) is the average of
the staircase function defined in Eq.~(\ref{8}). The unfolded
eigenvalues $\ve_\mu$ are the values of that function taken at
$E_\mu$. Usually, however, the exact form of the average level density
is not known. If the data are obtained by numerical simulation of an
ensemble (diagonalization of many matrices), the average level density
is best found by numerically averaging over the ensemble. If we deal
with an empirical spectrum of, say, several tens of levels, it is
advantegeous to use the data to construct the staircase function
rather than the level density (the representation of the latter in the
form of a histogram depends on the bin width chosen), and to fit a
low--order polynomial to that function. The unfolded eigenvalues are
again given by the values of the fitted staircase function taken at
the original eigenvalues $E_\mu$.

How does the omission of levels affect spectral fluctuation
properties? If a fraction $f$ of levels is removed at random from a
complete sequence (no missing levels, no levels with wrong quantum
numbers) then the resulting correlation functions can easily be
related to the correlation functions of the complete sequence~(Bohigas
and Pato, 2004). As a consequence, in the limit of large level numbers
the information on spectral correlations is fully preserved provided
$f$ is known. In nuclei, levels with small widths are hard to detect
and easily missed. But widths are theoretically predicted to be
uncorrelated with the positions of eigenvalues. The removal of levels
with small widths is thus a random process, and the analysis of
Bohigas and Pato (2004) applies. The fraction $f$ can be estimated
using the Porter--Thomas distribution and the experimental detection
efficiency. The admixture to a complete sequence of levels of levels
with wrong quantum numbers can be treated similarly (Bohigas and Pato,
2004).

\subsection{Discussion}
\label{discGOE}

The random--matrix approach described above is based on a few general
principles: Invariance of the system under time--reversal (which leads
to an ensemble of real and symmetric matrices), absence of a preferred
direction in Hilbert space (which makes the ensemble generic and
implies orthogonal invariance), and confinement of the spectrum to a
singly--connected finite stretch of the energy--axis (which is
realized for the Gaussian cutoff as well as for many other cutoff
factors).

The GOE is universal: Cutoff factors different from the Gaussian
cutoff but subject to the conditions formulated in Section~\ref{univ}
lead to different forms of the average spectrum but to identical
predictions for the local spectral fluctuation measures. GOE
predictions of local spectral fluctuation measures are useful for the
analysis of data because the ensemble is ergodic: For almost all
members of the ensemble, the fluctuation measure calculated as an
average over the ensemble equals the result obtained by taking a
running average over the spectrum of that member.

It is rather amazing that the few principles just mentioned lead to
parameter--free quantitative predictions for the local spectral
fluctuation measures.  Eq.~(\ref{6}) gives the distribution law
relevant for transition matrix elements and decay
widths. Eqs.~(\ref{7}) and (\ref{10}) present the two measures which
are most frequently used for the analysis of spectral data. Needless
to say, other fluctuation measures have also been worked
out (see Brody {\it et al.}, 1981).

Random--matrix theory predicts fluctuation properties in terms of mean
values. In the examples treated so far, the input mean value has been
the average of the squares of the projected wave functions as in
Eq.~(\ref{6}), or the local average level density $\rho(E)$ (or the
average level spacing $d$) as in Eqs.~(\ref{7}) and (\ref{10}). Both
parameters must be determined from the data. Then the theory predicts
the distribution of the relevant observables. In case the level
density changes significantly over the length of the given spectrum,
an unfolding of the spectrum must precede the comparison with GOE
predictions.

\subsection{Random Matrices and Chaos}
\label{RMTchaos}

There exists a close connection between random matrices and quantum
chaos. The latter term refers to quantum systems that are chaotic in
the classical limit. In classical mechanics, chaos is a dynamical
property characterized by the exponential divergence in time of
trajectories starting in close--lying points of phase space. The phase
space of a fully chaotic system is filled with such chaotic
trajectories and is void of islands with regular dynamics. The
analysis of chaos in classical conservative systems uses the existence
of periodic orbits and was pioneered, among others, by Gutzwiller. The
results are summarized in his book (Gutzwiller, 1995). Periodic--orbit
theory has been successfully applied to many systems with few degrees
of freedom (including systems that are not fully chaotic). To the best
of our knowledge, a similarly complete understanding of chaos in
classical many--body systems with their high--dimensional phase space
does not exist. Atomic nuclei pose the additional difficulty that the
matter density is very high. Even in the classical limit it is not
possible to neglect the fact that neutrons and protons are fermions.
This fact leads to complications regarding periodic--orbit theory
(see, for instance,   Sommermann and Weidenm\"{u}ller, 1993;
Weidenm\"{u}ller, 1993; Sakhr and Whelan, 2003).

\begin{figure}[h]
\vspace{5 mm}
\includegraphics[height=0.7\linewidth,angle=270]{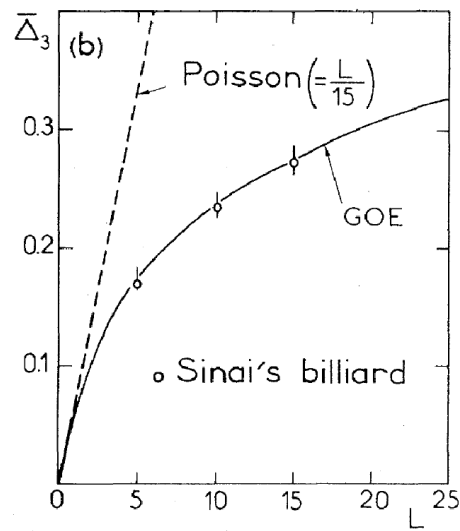}
\vspace{3 mm}
\caption{The $\Delta_3$--statistic for the Sinai billiard (open
circles), the GOE prediction (solid line), and the Poisson result
(dashed line). From Bohigas {\it et al.} (1984).}
\label{fig6}
\end{figure}

\begin{figure}[h]
\vspace{5 mm}
\includegraphics[width=\linewidth]{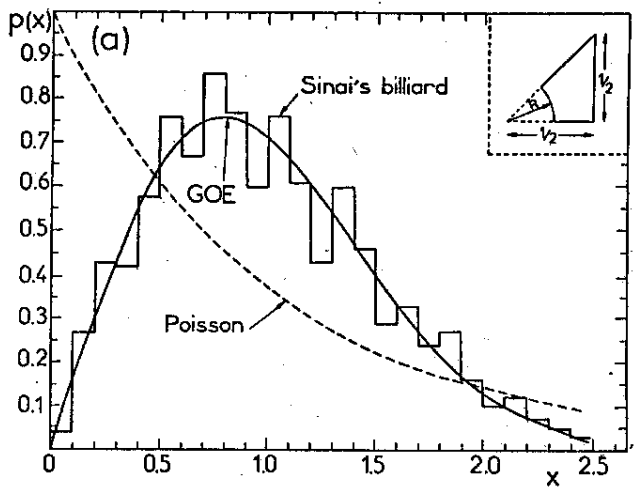}
\vspace{3 mm}
\caption{The NNS distribution for the Sinai billiard (histogram),
the GOE prediction (solid line), and the Poisson result (dashed
line). While the theoretical literature commonly uses the label $s$
for the actual level spacing in units of the mean level spacing, that
quantity is usually referred to as $x$ when RMT predictions are
compared with data. We follow that usage here. Insert: The eigenvalues
must be selected according to the symmetries of the
eigenfunctions. From Bohigas {\it et al.} (1984).}
\label{fig7}
\end{figure}

Since the late 1970s much effort has been devoted to identifying the
dynamical properties of quantum systems that are fully chaotic in the
classical limit (see Gutzwiller, 1990; Haake, 2001). There were two
lines of development: Some authors looked for signals of classical
chaos in the time--evolution of wave packets, others focused attention
on the fluctuation properties of eigenvalues and eigenfunctions of
closed systems (which possess a discrete spectrum). Here we confine
ourselves to the latter development which has had repercussions in
nuclear physics. Mounting numerical evidence from classically chaotic
few--degrees--of--freedom systems due to McDonald and Kaufman (1979),
Casati (1980), Berry (1981), and others culminated in the work of the
Orsay group (Bohigas {\it et al.}, 1984) on the Sinai billiard: A
point particle moving in two dimensions is scattered elastically by
the interior surface of a square and by the exterior surface of a
circle inscribed into the square. That system is fully chaotic and
invariant under time reversal. Solving the Schr\"odinger equation for
the Sinai billiard numerically, the Orsay group accumulated a sequence
of about 1000 consecutive eigenvalues belonging to eigenfunctions of
the same symmetry class. For a meaningful evaluation of the spectral
fluctuation measures, their numerical accuracy had to be much better
than the average level spacing. It was also important to make sure
that no eigenvalue was missed. The Weyl formula (see Baltes and Hilf,
1976) was used to unfold the spectrum. The number of eigenvalues was
large enough to evaluate for the first time the $\Delta_3$--statistic
(in addition to the NNS distribution which had been used before). The
results are shown in Figs.~\ref{fig6} and \ref{fig7}.

In the figures the data on the Sinai billiard are compared with GOE
predictions and with the Poisson distribution. The latter has an
exponential form, is typical for regular (or integrable)
systems~(Berry and Tabor, 1977), and is briefly explained below. The
figures show excellent agreement between the results for the Sinai
billiard and the GOE predictions. This agreement led Bohigas {\it et
al.} (1984) to formulate a conjecture (the ``Bohigas--Giannoni--Schmit
(BGS) conjecture''): The spectral fluctuation properties of a quantum
system which is fully chaotic in the classical limit coincide with
those of the canonical random--matrix ensemble having the same
symmetry. Massive numerical evidence on other
few--degrees--of--freedom--systems has given extensive support to the
conjecture. Attempts to prove the conjecture analytically have partly
been based on periodic--orbit theory (which has been very important
all along in the understanding of classical chaos) and the
semiclassical approximation. Berry (1985) has established a connection
between classical chaos and the $\Delta_3$ statistic. More recently
and for a special fluctuation measure (the two--point function, a
quantity intimately related to the $\Delta_3$--statistic), the
conjecture has been proved or, to use a mathematically less demanding
term, demonstrated in Heusler {\it et al.} (2007). The argument uses
generic properties of periodic orbits in classically chaotic systems
with few degrees of freedom.

The semiclassical arguments used in Heusler {\it et al.} (2007) do not
apply directly to chaos in classical or quantum many--body systems.
This point was made at the beginning of this Section and is taken up
again in Section~\ref{npres}. For the time being, we adopt the BGS
conjecture also for many--body systems. Thus we speak of quantum chaos
(or, briefly, of chaos) whenever in a many--body system such as the
nucleus the spectral fluctuation measures introduced above agree with
GOE predictions. We have to keep in mind that the NNS distribution is
a less safe indicator of quantum chaos than the
$\Delta_3$--statistic. To see this, we consider~(Rosenzweig and Porter
(1960)) a real symmetric matrix with diagonal entries that are
eigenvalues of a regular system with average level spacing $d$ and
with non--diagonal elements of typical strength $v$ (see the example
in Eq.~(\ref{2f})).  The parameter which rules the regular--to--chaos
transition is $v / d$. As we increase that parameter from very small
values, neighboring levels begin to repel, and the NNS distribution of
the GOE is approached. This happens before the long--range stiffness
of the spectrum as manifest in the $\Delta_3$--statistic is attained.

In classical mechanics, the case of complete chaos is a limiting case.
Another limiting case is that of completely regular motion. In quantum
mechanics, regular motion corresponds to the existence of a complete
set of quantum numbers that label every state. There is no level
repulsion, and no correlation between levels. As shown by Berry and
Tabor (1977), this case generically yields a spectrum where the
spacings have an exponential distribution (``Poisson spectrum''), see
Fig.~\ref{fig7}. The general case in classical mechanics is the one
where phase space consists of islands of regular trajectories
separated by domains filled with chaotic trajectories. There does not
seem to exist a generic description of the distribution of level
spacings for that case. The Brody distribution defined in
Eq.~(\ref{14a}) below is a purely heuristic (and not the only)
interpolation formula between the Poisson distribution and the Wigner
surmise.

\subsection{Doorway States}
\label{door}

As explained in Section~\ref{GOE} above, the GOE yields the generic
description of the spectra of bound quantum systems. It offers the
best first guess of spectral properties if we have no specific
knowledge of the system  except for the fact that it is
invariant under time reversal. There are cases, however, where we
possess some limited additional dynamical information. We may then
look for a theoretical description which takes that information fully
into account but is otherwise generic. Doorway states are a case in
point. Other examples for this type of approach are described in
Section~\ref{viol} below.

We consider a particular mode of excitation of the system. To be
specific, we take the electric dipole operator $D$ acting on the
ground state $| g \rangle$ of an even--even nucleus in the
long--wavelength limit. With $H$ the Hamiltonian, we shift for
simplicity the energy such that the ground state has energy zero, $H |
g \rangle = 0$. We choose the normalization of the dipole mode $| 0
\rangle = D | g \rangle$ such that $\langle 0 | 0 \rangle = 1$. Then
the expectation value $E_0 = \langle 0 | H | 0 \rangle$ gives the mean
excitation energy of the dipole mode. In general the dipole mode is
not an eigenstate of $H$. Therefore the variance of $H$ with respect
to the dipole mode, i.e., the expression $\Delta H^2 = \langle 0 | H^2
| 0 \rangle - E^2_0 = \sum_{\mu \geq 1} (H_{0 \mu})^2$, does not
vanish. (Here $\mu$ labels states with the same quantum numbers as,
but orthogonal to the dipole mode). As a consequence the cross section
for dipole absorption possesses a large number of sharp lines, each
occurring at an eigenstate of $H$. (Here we disregard the fact that
for most nuclei, $E_0$ is greater than the threshold for particle
emission, so that the sharp lines actually become more or less broad
resonances). We observe that $(1/N)\Delta H^2 = (1/N) \sum_{\mu \geq
1} (H_{0 \mu})^2$ represents the mean coupling strength of the dipole
mode with the other states of the system. The ratio of that expression
and the mean level spacing (taken at $E_0$) is a measure of the length
of the energy interval over which the dipole mode is strongly mixed
with other states. Within that interval, the peak heights of the
dipole absorption lines are enhanced, and an average of the cross
section for dipole excitation (taken with a Lorentzian weight function
whose width is larger than the mean level spacing) displays a
resonance at or near the energy $E_0$. This is the giant dipole
resonance. Dipole absorption may be viewed as a two--step process:
First the dipole mode is formed. The ground states of even--even
nuclei have spin zero and positive parity, and the dipole operator is
the $z$--component of a vector. Therefore in these nuclei the dipole
mode has spin one and negative parity. Subsequently that mode decays
into the eigenstates of $H$ with the same quantum numbers. It is
obvious that that same picture applies to many other nuclear
reactions: neutrons impinging on a nucleus give rise to a
single--particle mode, protons generate isobaric analog modes,
etc. The picture can likewise be used to describe the distribution of
simple configurations like one--particle one--hole states over the
eigenstates of the system. Early summaries of the doorway state idea
may be found in Bohr and Mottelson (1969) and Mahaux and
Weidenm\"{u}ller (1969). For a recent discussion which goes beyond the
standard model described in the next paragraph, see Zelevinsky {\it et
al.} (1996) and De Pace {\it et al.} (2007).

We turn to the standard description of a doorway state within
random--matrix theory. To this end, we define the following extension
of the GOE. With $E_0$ the mean excitation energy of the doorway
state, and with $H_{0 \mu}$ (where $\mu = 1, \ldots, N$) the coupling
matrix elements of the doorway state with the other states of the
system, the Hamiltonian matrix $H$ for the doorway--state model
takes the form
\ba
H = \left( \matrix{ E_0 & H_{0 \nu} \cr H_{\mu 0} & H_{\mu \nu} \cr}
\right) \ .
\label{12}
\ea
Here $H_{\mu \nu}$ is a GOE matrix of dimension $N$. With $| 0
\rangle$ the doorway state, we obviously have $ \langle 0 | H | 0
\rangle = E_0$ and $(1/N) \Delta H^2 = (1/N) \sum_\mu (H_{0
\mu})^2$. Because of the distinct role of the doorway state, the
doorway model of Eq.~(\ref{12}) is not orthogonally invariant. It does
possess that invariance, however, in the subspace of states carrying
labels $\mu \geq 1$. That statement implies that ensemble averages of
observables cannot depend on the individual coupling matrix elements
$H_{0 \mu}$ but depend only on the orthogonal invariant $(1/N)
\sum_\mu (H_{0 \mu})^2$, the mean square coupling matrix element.

Every random--matrix model that incorporates additional dynamical
information does so at the expense of complete orthogonal invariance.
That invariance is extremely helpful in working out the spectral
properties of the GOE. Therefore non--invariant extensions of the GOE
are often very difficult to handle. This is not the case for the model
of Eq.~(\ref{12}) because we add only a single state to the GOE. As a
consequence, in the limit $N \to \infty$ both the spectral statistics
and the Porter--Thomas distribution of the Hamiltonian~(\ref{12})
coincide with those of the GOE. The only distinct feature of the model
is the strength function for the doorway state. It is defined as the
ensemble average of $\sum_\tau |\langle 0 | \tau \rangle|^2 \delta(E -
E_\tau)$ and gives the probability per unit energy interval to find
the doorway state admixed to the eigenstates $| \tau \rangle$ of $H$
with eigenvalues $E_\tau$. Put differently, the strength function is
the average value of the level density weighted with the square of the
overlap matrix element. In form, the strength function is closely
related to the local density of states used in condensed--matter
physics. Because of the ensemble average, the strength function is a
smooth function of energy $E$ and has the form
\be
\overline{\sum_\tau |\langle 0 | \tau \rangle|^2 \delta(E - E_\tau)}
= \frac{1}{2 \pi} \frac{\Gamma^{\downarrow}}{(E - E_0)^2 + (1/4)
(\Gamma^\downarrow)^2} \ .
\label{13}
\ee
Here
\be
\Gamma^\downarrow = 2 \pi \bigg( (1/N) \sum_\mu (H_{0 \mu})^2 \bigg)
\rho(E) 
\label{13a}
\ee
is the spreading width of the doorway state, with $\rho(E) = (1 / d)$
the average level spacing of the GOE at energy $E$. Since $(1/N)
\Delta H^2 / d = \Gamma^\downarrow / 2 \pi$, the spreading width
measures the length of the energy interval within which the
eigenstates of $H$ carry significant admixtures of the doorway
state. The notation for the spreading width with a downarrow is a
reminder of the fact that the spreading width does not account for a
decay process into some open channel (with the ensuing probability
flux of particles at large distance), but for the mixing of a
particular mode with other bound states.

The Lorentzian form of the strength function applies approximately for
$\Gamma^\downarrow \ll \lambda$ where $2 \lambda$ is the radius of the
semicircle. For small coupling, Eq.~(\ref{13a}) has the form of
Fermi's Golden Rule. In the framework of random--matrix theory, the
result~(\ref{13a}) is valid beyond the perturbative regime, however,
and holds even if $\Gamma^\downarrow \rho(E) \gg 1$. Integrating the
left--hand side of Eq.~(\ref{13}) and using completeness we obtain
unity. That same statement applies to the right--hand side, and the
strength function is properly normalized. The Lorentzian
form~(\ref{13}) does not apply when the energy $E_0$ of the doorway
state is close to one of the end points of the semicircle, or when the
spreading width becomes very large (i.e., comparable with the radius
of the semicircle), see Kota (2001), De Pace {{\it et al.} (2007), and
numerical examples given, for instance, in Zelevinsky {\it et al.} 
(1996).

The spreading width $\Gamma^{\downarrow}$ has a remarkable property
which makes it a useful measure for the spreading of a doorway state.
We ask: How does the mean square matrix element $(1/N) \sum_\mu (H_{0
\mu})^2$ change with the average level density of the GOE states? That
question arises in the nuclear context because doorway phenomena are
encountered at various excitation energies and in nuclei with widely
different mass numbers for which the nuclear level density differs
markedly. An intuitive answer is obtained by noting that a significant
increase of the level density implies a significant increase in the
complexity of the wave functions making up the GOE states. Therefore,
each of the matrix elements $H_{0 \mu}$ connecting the doorway state
with the GOE states is strongly reduced, and so is the mean square
matrix element. But in the expression~(\ref{13a}) for the spreading
width this strong reduction of the mean square matrix element is
essentially compensated by the increase of the average level
density. That compensation is exact in simple models (Brody {\it et
al.}, 1981) and is expected to hold to a good degree of approximation
in realistic cases. An experimental verification of this expectation
comes from investigations of the spreading width for isospin mixing
(Harney {\it et al.}, 1986) and will be discussed in Part 2 of this
review. We conclude that in contrast to the exponential dependence of
the average level density on excitation energy and mass number, the
spreading width is expected to be a slowly varying function of these
parameters and is thus a useful measure for the spreading of a doorway
state.

If the model~(\ref{12}) applies in reality, a doorway state has hardly
any influence on spectral properties of the system: The average level
density of the states that carry the same quantum numbers as the
doorway state is unchanged, their spectral fluctuation properties are
the same as for the GOE, and their partial widths for decays different
from electric dipole decay to the ground state have the same
Porter--Thomas distribution as for the GOE. The only difference to the
pure GOE case is the Lorentzian enhancement~(\ref{13}) of the strength
function for dipole absorption. This is the only trace left of the
doorway state after we take account of its mixing with the complicated
states. Dividing the partial widths for dipole absorption of the
eigenstates of $H$ by the value of the strength function (taken at the
corresponding eigenvalue) removes that trace and should yield
quantities that have a pure Porter--Thomas distribution.

\section{Application of RMT to Nuclear Spectra}
\label{applRMT}

\subsection{General Remarks}
\label{gen}

Nuclear energy levels are characterized by quantum numbers that
reflect the symmetries of the nuclear Hamiltonian: Total spin ($J$)
reflects rotational symmetry, parity ($\Pi$) reflects invariance under
mirror reflection, isospin ($T$) reflects proton--neutron symmetry. We
exhibit the consequences of such symmetries for the application of RMT
to nuclear spectra.

The total Hilbert space is spanned by many--body wave functions that
carry the quantum numbers $J, \Pi, T$. These can be arranged in such a
way that the matrix representation of the nuclear Hamiltonian has
block structure,
\be
H = \left( \matrix{ H^{J_1 \Pi_1 T_1} &  0     & 0 & \cdots \cr
                     0     & H^{J_2 \Pi_2 T_2} & 0 & \cdots \cr
                   \vdots & \vdots & \ddots & \cdots \cr} \right) \ .
\label{14}
\ee
Here $\{ J_1 \Pi_1 T_1 \} \neq \{J_2 \Pi_2 T_2 \} \neq \ldots$, and
each of the matrices $H^{J \Pi T}$ couples only many--body states
which carry the same quantum numbers. If in addition the nuclear
dynamics is chaotic, then the BGS conjecture (see
Section~\ref{RMTchaos}) implies that each of the matrices $H^{J \Pi
T}$ is a member of a random--matrix ensemble. Since nuclei obey
time--reversal invariance, the suitable ensemble is the GOE. In the
framework of RMT, Hamiltonian matrices referring to different sets of
quantum numbers are assumed to be uncorrelated.

To compare RMT predictions on spectral fluctuations with data on
nuclear energy levels, sequences of levels carrying the same quantum
numbers are needed. The data are subject to three requirements: (i)
The sequence(s) should be as long as possible, (ii) the sequence(s)
should be pure (i.e., should not contain levels carrying quantum
numbers which differ from those of the rest), and (iii) the
sequence(s) should be complete (i.e., there should not be any levels
that were not detected). The first requirement is needed to ensure
that the running average over the actual spectrum is as close as
possible to the running average over the complete spectrum, the latter
by ergodicity (Section~\ref{ergod}) being equal to the GOE ensemble
average. The two other requirements guarantee that the statistical
predictions of RMT can meaningfully be applied to the data, see
Section~\ref{unfo} and the discussion in Section~\ref{viol}.
Unfortunately, the number of nuclear data sets of sufficient quality
to provide detailed tests of RMT is fairly limited. This is primarily
due to the requirements of purity and completeness imposed by the
sensitivity of the standard fluctuation measures.

In many cases nuclear levels are observed as narrow particle-unstable
resonances, see Fig.~\ref{fig1}. Then a multi--level $R$--matrix fit
(Lane and Thomas, 1958) is used to determine the positions the levels
would have if they were stable under particle decay. These positions
are used to calculate level spacings and to test GOE predictions. (In
R--matrix theory, the nucleus is thought to be enclosed by a
ficticious boundary that lies some distance beyond the nuclear radius.
Boundary conditions on that surface and the nuclear Hamiltonian
jointly define a set of discrete states within the boundary. These
states appear as resonances in the scattering matrix $S$.
Approximations to the resulting formal expression for $S$ serve as the
basis of the fits to data).

In comparing nuclear data with GOE predictions for the NNS
distribution and/or the $\Delta_3$--statistic, one faces a difficulty:
Both distributions are parameter free, and it is difficult to assess
the significance of the usual tests for goodness of fit such as the
$\chi^2$--test when one is far from these limiting cases. Therefore,
one uses measures which interpolate between the GOE prediction and the
case of a totally regular system. These do have free parameters, and
the goodness--of--fit tests are easily interpreted. The NNS
distribution has the form of the Wigner surmise~(\ref{7}) for the GOE
and is proportional to $\exp ( - s)$ (Poisson distribution) for
regular systems (Berry and Tabor, 1977). An expression that
interpolates between both is the Brody distribution (Brody {\it et
al.} (1981).  It depends on a single parameter $\omega$ and is given
by
\be
P_\omega(s) = (1 + \omega) \alpha s^\omega \exp ( - \alpha s^{1 +
\omega} ) 
\label{14a}
\ee
where $s$ is the actual level spacing in units of the mean level
spacing and the constant $\alpha = [ \Gamma \{ ( 2 + \omega ) / ( 1 +
\omega ) \} ]^{1 + \omega}$ is fixed by normalization. For $\omega =
0 \ (\omega = 1)$, the Brody distribution equals the Poisson
distribution (the Wigner distribution), respectively. For all $\omega
> 0$, the Brody distribution vanishes at $s = 0$. The Brody formula is
only one of several formulas that interpolate between the Poisson and
the Wigner distribution. Another example is the Berry--Robnik
distribution (Berry and Robnik, 1984)    .

For the GOE, the $\Delta_3$--statistic has the logarithmic dependence
on the length $L$ of the energy interval shown in Eq.~(\ref{9}), while
for a regular system it is linear in $L$. A parameter--dependent
measure for deviations from the GOE is obtained by considering a
spectrum which is a superposition of $k$ independent GOE spectra. For
$k \gg 1$, the $\Delta_3$--statistic approaches the linear dependence
of the regular case. (Apparently this was first noticed by Gurevich
and Pevsner (1956)). The deviation from the GOE prediction is
significant already for $k = 2$, see Fig.~\ref{fig16} below.

The predictions of RMT on fluctuation properties of nuclear wave
functions can only be tested in terms of the distribution of matrix
elements (either for decay into open channels or for electromagnetic
transitions, weak interaction matrix elements not being numerous
enough for such a test). Here the size of the sample again is
important. In addition there is usually an experimental cutoff for
small matrix elements so that only part of the Porter--Thomas
distribution can be tested.

Two properties of nuclei are central for tests of RMT. (i) In every
nucleus, the average level density increases roughly exponentially
with excitation energy. Thus, while typical level spacings near the
ground state are several hundred keV, spacings of levels having the
same spin and parity (a subset of all levels!) at neutron threshold in
heavy nuclei are typically $10$ eV, see Fig.~\ref{fig1}. For fixed
excitation energy the level density increases likewise with mass
number $A$ (save for corrections due to nuclear shell structure, see
Section~\ref{shell}). The requirements on experimental energy
resolution obviously increase with increasing level density and, in
general, limit nuclear spectroscopy except for very fortuitous
situations such as those leading to the data in Fig.~\ref{fig1}. (ii)
Levels below the threshold for particle emission have only small
widths (in comparison with the mean level spacing) due to beta-- or
gamma--decay. Above particle threshold, the total widths of the
nuclear resonances increase rapidly with increasing excitation
energy. This is because the number of open channels for particle decay
increases rapidly (the number of states available for decay in the
daughter nuclei increases roughly exponentially with excitation energy
in these nuclei). As a consequence, isolated resonances as shown in
Fig.~\ref{fig1} are observed only just above the lowest particle
threshold. A few hundred keV above that threshold, resonances begin to
overlap (the mean level spacing decreases, the average total width
increases), and it is no longer possible to investigate spectral
fluctuations. Rather, this is the domain of statistical nuclear
reaction theory (see Part 2 of this review). Thus tests of GOE
predictions in nuclear spectra are limited to the energy interval
between the ground state and an energy somewhat above the first
particle threshold.

In describing the application of RMT to nuclear data, we first discuss
the experimental methods that have been used to obtain the relevant
spectral information (Section~\ref{exp}). We then review the results
of comparing the data with GOE predictions on spectral fluctuations
(Section~\ref{tests}). RMT can be extended to deal with violations of
symmetry or invariance. This is described in Section~\ref{viol}.

\subsection{ Experimental Methods}
\label{exp}

\subsubsection{Neutron Resonances}

The early tests of RMT involved neutron resonances,  as shown in
Fig.~\ref{fig1}. The scattering of slow neutrons enables the study of
individual resonances in a narrow window of energies at high
excitation energy in the compound nucleus -- typically 5 to 7 MeV. At
these energies the level density in medium--weight and heavy nuclei is
very large. However, the angular momentum barrier for the incident
neutron severely restricts the neutron's orbital angular momentum and
thus the spins $J$ of the compound nuclear resonances which
contribute to the scattering. For slow neutrons, only $s$-- and
$p$--wave resonances are usually observed.

The experimental method of choice is a time--of--flight measurement.
Longer flight paths allow for better energy resolution (essential to
resolve the resonances), but reduce the counting rate because of the
smaller detector solid angle. Thus high--intensity neutron sources are
required. Today the most intense neutron beams are produced at
spallation sources.

The most common experiment is a transmission measurement. The
transmission of the neutron beam through a target with nuclei of mass
number $A$ determines the total cross section for the $n$ + $A$
reaction. Neutron capture followed by $\gamma$ emission is also very
helpful in determining the resonance parameters. Analysis of the
resonance data is normally performed with the Lane and Thomas version
of the Wigner--Eisenbud R--matrix formalism (Lane and Thomas,
1958). The classic monograph on neutron resonance reactions is by Lynn
(1968).

For a comparison with RMT predictions the levels in a sequence must
have the same quantum numbers. Thus one key issue is to determine the
spin $J$ and parity $\Pi$ of each resonance. This is done using the
angular momentum $\ell$ of the scattered neutron. For spin--zero
targets all $s$--wave resonances have $J^\Pi$ = 1/2$^{+}$. For slow
neutrons, the difference in penetrabilities for ${\ell}$ = 0 and
${\ell}$ = 1 is so large that at first sight $\ell$ can be assigned by
inspection -- strong resonances are $s$--wave and weak resonances are
$p$--wave. One normally formalizes this with a Bayesian analysis
(Bollinger and Thomas, 1968), but this approach is not reliable in the
gray area between weak $s$--wave and strong $p$--wave resonances. Of
course other experiments can be used to improve the spin and parity
assignments. We mention neutron capture with high resolution
${\gamma}$-ray spectroscopy or with calorimeters (where for each
capture reaction the total number and individual energies of the
emitted gamma rays are registered).  However, these are very time
consuming.

In addition to the issue of spurious resonances (incorrect spin or
parity assignments), the other major problem is missing levels. The
missing levels are expected to be the weakest levels, hence the focus
on signal--to--noise ratios and energy resolution in resonance
measurements. The Porter--Thomas distribution predicts many weak
resonances. Assuming that distribution one can estimate the fraction
of missing levels. Unfortunately this nearly universally used
correction method can be misleading if non--statistical effects are
present. An alternative approach utilizing the NNS distribution was
developed only recently (Agvaanluvsan {\it et al.}, 2003). Bohigas and
Pato (2004) extended the investigation of the effects of missing
levels to other level fluctuation measures.

Almost all of the neutron resonance data suitable for detailed
comparison with RMT are obtained for spin--zero targets. Most of the
neutron resonance data used in the early evaluation of RMT were
obtained by Rainwater's group at Columbia (Liou {\it et al.}, 1972a
and 1972b) for a number of nuclei with mass numbers $A > 110$. Due to
experimental limitations, the number of resonances in each nucleus was
never significantly larger than 200.

\subsubsection{Proton Resonances}

Due to the Coulomb barrier proton resonances cannot be studied near
zero bombarding energy as neutron resonances are. High--resolution
proton resonance measurements are typically taken at bombarding
energies corresponding to 60--70{\%} of the Coulomb barrier. This has
two major advantages. First, the Coulomb barrier penetrability serves
to narrow the proton widths and makes possible the resolution of
individual resonances in rather dense spectra. Second, the addition of
Coulomb and nuclear resonance scattering amplitudes leads to striking
interference patterns that are used to identify the spin and parity of
each resonance. The parity assignment is normally apparent by
inspection, since the interference patterns for even and odd orbital
angular momenta are quite different. This is important because the
proton penetration factors for different orbital angular momenta do
not differ as much as for the neutron resonances. As a result one
usually observes $s$--, $p$--, and $d$--wave and sometimes even $f$--
and $g$--wave resonances. The primary difficulty consists in
determining the $J$ value of the proton resonances. Additional
experiments (inelastic scattering or capture) can resolve this
problem, but as in the neutron case, these experiments are very time
consuming and therefore are rarely performed. Although most of the
high--quality proton resonance data are also for $s$--wave sequences,
there are (in contrast to the neutron case) a few $p$--wave sequences
that are considered pure and complete. With very few exceptions the
level density becomes too great for this method to work much beyond mass
number $A = 60$. Here the typical number of resonances of the
same spin and parity is  50 or so. Thus the proton resonance
data complement the neutron resonance data; the best results for each
set are obtained in quite different mass regions.  Almost all of the
data used to compare with RMT are for spin--zero targets and from the
Triangle Universities Nuclear Laboratory (TUNL) (Wilson {\it et al.},
1975; Watson {\it et al.}, 1981).

In order to observe (nearly) all of the resonances one needs very good
beam--energy resolution and high beam intensity. These requirements
seem contradictory. One approach is to accept the time--dependent
energy fluctuations intrinsic to the accelerator and make a correction
later. The most successful correction method solves the
resolution--intensity impasse by using two beams. One
(high--intensity) beam is used to perform the experiment; the other
beam is used to generate a feedback signal that follows the beam
energy fluctuations; this signal generates a voltage difference which
is applied to the target. Thus the time--dependent energy fluctuations
are canceled. The method works well for a Van de Graaff accelerator
where most of the fluctuations have low frequency. The details are
given by Bilpuch {\it et al.} (1976).

\subsubsection{Low--lying Levels}
\label{low}

The spectroscopy of low--lying levels (excitation energies below 2 MeV
or so) has always been a primary object of study in nuclear physics.
Many different approaches have been used: various nuclear reactions
including inelastic scattering, pickup, and transfer reactions,
${\gamma}$-ray spectroscopy following ${\beta}$ or ${\alpha}$ decay,
etc. Almost all of these processes are quite selective. Therefore, one
needs to use many different approaches to ensure that all levels (in
some energy interval) are observed. One very powerful technique uses
the neutron capture reaction. Neutron capture on nucleus $A$ is
followed by sequences of $\gamma$ transitions which finally populate
the ground state of nucleus $A + 1$. The average neutron resonance
capture technique (Bollinger and Thomas, 1968) effectively averages over many
neutron resonances and is non--selective; it also averages over
Porter--Thomas fluctuations, increasing the probability of observing
weak transitions. Every low--lying state within some spin range is
expected to be populated. The combination of neutron capture and
direct reactions has led to a number of complete level schemes at low
energies (von Egidy {\it et al.}, 1986).

\subsubsection{High--spin States}

The ground states of even--even nuclei have spin zero, those of
even--odd, odd--even and odd--odd nuclei have small spin values. The
excitation energy $E(J)$ of the lowest state with given spin $J$
generically increases with $J$; the function $E(J)$ defines the
``yrast line'', see Fig.~\ref{fig8}. High--spin states located near
the yrast line are typically investigated via the collision of two
heavy nuclei (``heavy--ion collisions''). At non--zero impact
parameter, the two colliding nuclei with mass numbers $A_1$ and $A_2$
typically carry a large angular momentum of relative motion. The
high--spin intermediate complex formed by the collision (with spin
values as large as $60 \hbar$ or so) may decay by a sequence of
$\gamma$ transitions (perhaps with intermittent neutron evaporation)
to the ground state of a nucleus whose mass number is smaller than but
close to $A_1 + A_2$. The $\gamma$ rays emitted in this process are
analyzed using large--scale gamma--ray detection arrays. This approach
has generated a very large amount of data. These comprise many
rotational bands (with many states in each) for a range of heavy
nuclei. (Rotational bands are typical for deformed nuclei and are
dealt with in Section~\ref{coll}). The observed states have relatively
high spin and rather large excitation energies, but are not far above
the yrast line.  Some aspects of the method of analysis are summarized
by D{\o}ssing {\it et al.} (1996), where further references may be
found.

Unfortunately, the difficulties in obtaining suitable data sets for
comparison with RMT are many. One often sees many rotational bands,
but of course each band has only one state of a specific spin and
parity. Thus one is forced to combine results from many bands and
nuclei. Another serious issue is the problem of quantum number
assignments. Within a given band with a well--known bandhead the
assignments are reliable; assignments based on interband transitions
are more problematic. Until now, levels up to a few $100$ keV above
the yrast line have been analyzed (Garrett {\it et al.}, 1997);
higher--lying rotational bands cannot be individually resolved. The
evidence here points to regular motion, see
Section~\ref{anahss}. Theoretical expectations are that at around
$800$ keV above the yrast line, the spectral fluctuations become
chaotic, see Section~\ref{chaosyrast}.  It is to be hoped that with
improved resolution (perhaps attainable with the next generation of
large--scale detectors), spectroscopic data in that interesting energy
region will become available.

\subsubsection{Complete Level Schemes}
\label{cls}

The ideal is a complete scheme which begins at the ground state and
extends into the neutron or proton resonance region, with perfect
quantum number assignments to each level. Obtaining such a complete
level scheme is exceptionally difficult at best, and impossible in
medium--weight to heavy nuclei. The level densities are simply too
great. For very light nuclei, on the other hand, the level density is
small and the total number of states is not sufficient for a detailed
statistical analysis. Only for nuclei in the mass range between $20$
and $40$ or so does the level density have suitable values. These are
essentially the nuclei belonging to the $2s1d$--shell, see
Section~\ref{shell} below.

It might seem that complete spectra might best be measured by using a
variety of reactions as done for the spectroscopy of low--lying
states. However, this approach meets practical difficulties. Most
reactions are not only selective, but also provide information only in
a limited energy range. In the approach that was successfully used for
two nuclei -- $^{26}$Al and $^{30}$P -- the properties (quantum
numbers, positions, widths) of a number of proton resonances were
determined. The proton capture reaction was then measured for these
resonances which had different quantum numbers. The method essentially
guarantees that all the levels below the proton separation energy are
observed. Quantum numbers are assigned to the observed levels using
high--resolution gamma--ray spectroscopy, including angular
distributions of primary and secondary gamma rays. The experimental
procedure for $^{30}$P is described in detail by Grossmann {\it et
al.} (2000). The general approach is summarized by Mitchell and
Shriner (2001). With the help of the neutron capture reaction, a
``nearly complete'' level scheme below $4.3$ MeV excitation energy was
measured in $^{116}$Sn (Raman {\it et al.}, 1991).

The analysis of the complete spectra must allow for isospin--symmetry
breaking and is dealt with in Section~\ref{isos}.

\subsubsection{Low--lying Modes of Excitation}

The emphasis here is not on complete level schemes in a restricted
energy range (as in Section~\ref{low}), but rather on phenomena that
relate to levels of fixed spin and parity and seem linked to the
concept of a doorway state, see Section~\ref{door}. The primary
example of a doorway state is the giant electric dipole resonance
which manifests itself in a large resonance--like structure in the
absorption cross section for $\gamma$ rays. The isobaric analog
resonances provide another classic example of a doorway--state
phenomenon. The detailed analysis of both types of resonances involves
nuclear reaction theory and is dealt with in Part 2 of this review.

There are several other interesting excitation modes at lower energy
-- including the low--lying isovector magnetic orbital dipole or
scissors mode with $J^{\Pi} = 1^+$ (Bohle {\it et al.}, 1984; Richter,
1995) and the electric pygmy dipole resonance with $J^{\Pi} = 1^-$
(both spin assignments applying to even--even nuclei). Nuclear
resonance fluorescence measurements have been used to generate
extensive data sets of 1$^{+}$ and 1$^{-}$ states; the method provides
a unique $J$ value of 1 and a probable parity assignment, while
measurements with a polarized photon beam provide a definitive parity
assignment. Work at Darmstadt has been focused on both the scissors
mode (Enders {\it et al.}, 2000) and  the electric pygmy dipole
resonance (Enders {\it et al.}, 2004).

\subsubsection{Summary}

Fig.~\ref{fig9} summarizes in a qualitative fashion the domains of
excitation energy $E_x$ and mass number $A$ where spectral
fluctuations have been investigated. The figure is largely
self--explanatory; suffice it to say that the proton resonances around
$A = 50$ are measured above threshold but below the Coulomb
barrier. The figure shows that high--spin states are measured at
comparatively high excitation energies. This statement has to be taken
with a grain of salt, however. For an even--even nucleus with mass
number $A$ around $160$ or so, Fig.~\ref{fig8} shows schematically
the excitation energy of the lowest state with spin $J$ versus $J$
(the ``yrast'' line). The high--spin states that were analyzed so far
lie up to a few $100$ keV above the yrast line and, for that value of
$J$, represent low--lying excited states. Sequential decay of a
rotational band close to the yrast line by repeated emission of
$\gamma$ rays reduces both $J$ and the overall excitation energy while
the distance to the yrast line remains essentially the same.

\begin{figure}[h]
\vspace{5 mm}
\includegraphics[height=0.7\linewidth]{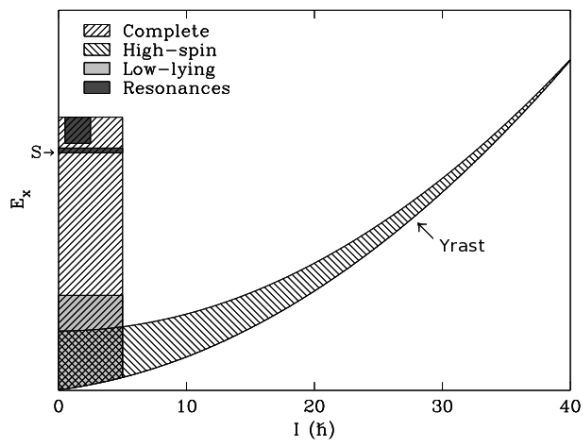}
\vspace{3 mm}
\caption{The yrast line, i.e., the excitation energy $E(J)$ of the
lowest state with spin $J$ versus $J$ for an even--even nucleus with
mass number around $160$ (schematic). The shaded area indicates the
domain where spectroscopic information is available, see D{\o}ssing
{\it et al.} (1996). The letter $S$ denotes the particle threshold.
Depending on $A$, $S$ typically lies between 5 and 8 MeV.}
\label{fig8}
\end{figure}

\begin{figure}[h]
\vspace{5 mm}
\includegraphics[height=0.7\linewidth]{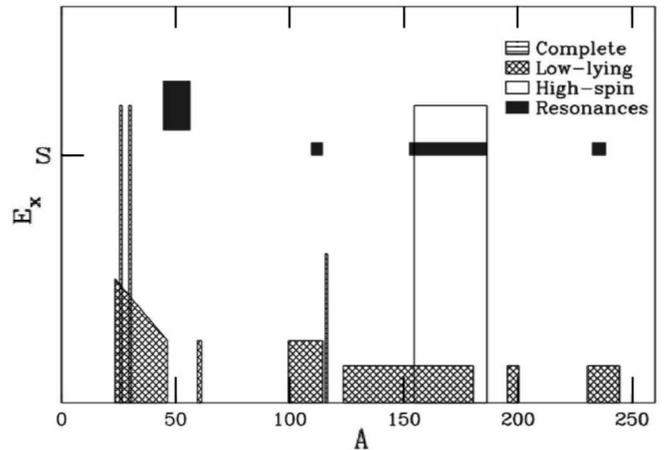}
\vspace{3 mm}
\caption{The domains of excitation energy $E_x$ and mass number $A$
where spectral fluctuations have been investigated are shown for four
classes of states (resonances, low--lying states, high--spin states,
and states that belong to a complete spectrum. Complete spectra are
known for three nuclei only, $^{26}$Al, $^{30}$P, and $^{116}$Sn).
The letter $S$ denotes the particle threshold.}
\label{fig9}
\end{figure}

\subsection{Tests of Fluctuation Measures}
\label{tests}

The fluctuation measures described in Section~\ref{fluct} have been
applied to nuclear data obtained with the experimental methods
summarized in Section~\ref{exp}. We emphasize again that due to the
sensitivity of these measures, the quality of the data sequence (the
degree of purity and completeness) is of paramount importance; 
only a very small fraction of all nuclear data can be used for such
tests.

We mention in passing another measure, the ``correlation hole''. As a
test for spectral fluctuations of the GOE type, it has been applied
much more widely in molecular physics (Leviandier {\it et al.} 1986,
Guhr and Weidenm{\"u}ller 1990b, Lombardi {\it et al.} 1994) than in
nuclear physics (Alhassid and Whelan 1993). Let $1 - Y_2(b)$ denote
the probability of finding two levels at a distance $b$. For
completely uncorrelated (Poissonian) spectra one has $Y_2(b) = 0$ for
all $b$ while GOE level repulsion implies $Y_2(0) = 1$. The Fourier
transform of the spectral autocorrelation function (a function of time
$t$) depends on the Fourier transform of $Y_2$ and is sensitive to the
difference between regular and chaotic motion. For chaotic motion it
displays a ``correlation hole'' at $t = 0$.

\subsubsection{Neutron and Proton Resonances}
\label{npres}

Although the fluctuation measures described in Section~\ref{fluct}
were proposed to describe neutron resonances in the 1950s, even in the
early 1960s there were no neutron data of sufficient quality to
provide an adequate test of RMT. For example, Dyson and
Mehta (1963) considered the best available neutron resonance
data and concluded the data were such that the RMT ``model'' was
neither proved nor disproved. They exhorted experimentalists to
improve the data quality.

By the early 1970s the high--quality neutron resonance data from the
Columbia group( Liou {\it et al.}, 1972a, 1972b) were available and
seemed to confirm the predictions of the GOE version of RMT. However,
due to the limited number of resonances (of order 100) for each
nucleus, these results were considered suggestive but not
definitive. Haq {\it et al.} (1982) and Bohigas {\it et al.} (1983)
combined neutron resonance data sets from a number of nuclei. This was
made possible by scaling level spacings in units of the mean level
spacing (the GOE fluctuation measures depend on that scaled parameter
only). The authors also included some of the proton resonance data in
their analysis. A complication (relative to the neutron data) was here
the much larger energy range needed in order to obtain a reasonable
sample size. This larger energy range required a correction
(unfolding) of the experimental data in order to transform to a new
set of levels with constant mean level spacing, see
Section~\ref{unfo}. The analysis included also suitable spectra for
other than $s$--wave proton resonances. Using all these data, the
authors obtained a set of 1407 levels that they labeled the nuclear
data ensemble (NDE). The analysis of the NDE is done by combining an
energy average (for every nucleus) with an ensemble average (over all
nuclei that are included in the NDE). Both the NNS and the
${\Delta_{3}}$ statistic for the NDE agreed very well with the GOE
predictions, see Figs.~\ref{fig10} and \ref{fig11}. A number of later
tests with other measures also agreed well with GOE predictions (see,
for instance, Lombardi {\it al.}, 1994) even though the presence of
non--statistical effects can never be excluded (Koehler {\it et al.},
2007).

\begin{figure}[h]
\vspace{5 mm}
\includegraphics[width=\linewidth]{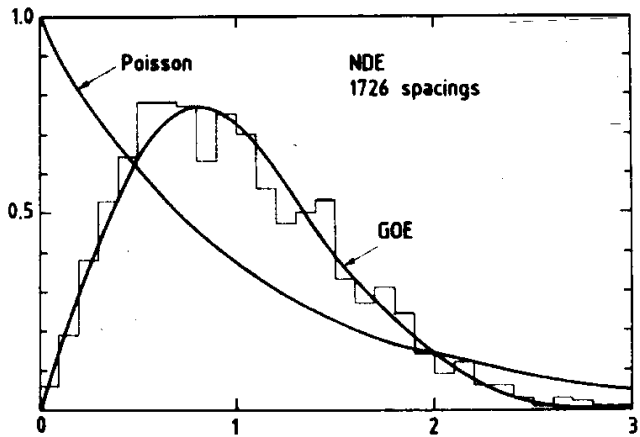}
\vspace{3 mm}
\caption{The NNS distribution for the nuclear data ensemble
(histogram) and the GOE prediction (solid line). From Bohigas {\it et
  al.}, 1983. (By the time that paper was published, the
nuclear data ensemble had grown to 1726 spacings).}
\label{fig10}
\end{figure}

\begin{figure}[h]
\vspace{5 mm}
\includegraphics[height=\linewidth,angle=270]{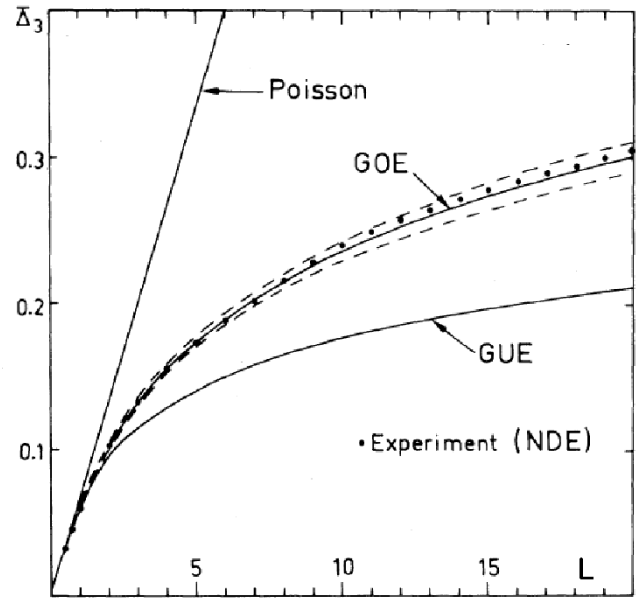}
\vspace{3 mm}
\caption{The $\Delta_3$--statistic for the nuclear data ensemble
(data points) and the GOE and GUE predictions (solid lines). The
dashed lines estimate the finite--range--of--data errors. From
Haq {\it et al.} (1982).}
\label{fig11}
\end{figure}

The analysis by Haq {\it et al.} (1982) and by Bohigas {\it et al.}
(1983), and in some of the subsequent papers marks a turning point in
the history of applications of RMT to nuclear spectra. As a result of
these analyses it became generally accepted that proton and neutron
resonances in medium--weight and heavy nuclei agree with GOE
predictions. With the later recognition of the connection between
spectral RMT fluctuations and quantum chaos (see
Section~\ref{RMTchaos}), the term chaos began to be used by nuclear
physicists.

As mentioned in Section~\ref{RMTchaos}, chaos in classical many--body
systems has not been investigated as thoroughly in terms of
periodic--orbit theory as in classical few--degrees--of--freedom
systems, not to speak of the complications due to the exclusion
principle. Therefore, the connection between classical chaos and RMT
is less well established and the use of the term ``chaos'' is somewhat
more tentative in nuclei. By the same token, the use of semiclassical
periodic--orbit theory in nuclei has been basically limited to
independent--particle motion. In that domain, it has been very
successful. We mention early applications by Strutinsky (1966, 1967,
1968) whose ``shell--correction method'' is reviewed in
Section~\ref{sdb}, by Balian and Bloch, 1970, and recent work by
Bohigas and Leboeuf, 2002; Leboeuf and Roccia, 2006; Roccia and
Leboeuf, 2007.

According to RMT, the eigenvalues and eigenvectors are uncorrelated
random variables, see Eq.~(\ref{1a}). In nuclei, this prediction was
tested by Bohigas {\it et al.} (1983). The correlation coefficient was
found to be $0.017 \pm 0.029$, the error reflecting the finite number
of data points. In microwave billiards, the test yields $0.02 \pm
0.05$ (Alt {\it et al.}, 1995). Another test (also giving agreement
with the GOE) has been reported in molecules by Lombardi and Seligman
(1993).

\subsubsection{Low--lying Levels}
\label{analow}

After the success of RMT in describing the fluctuation properties of
highly excited (resonance) states, it was natural to attempt to extend
such analyses to low--lying states. Although there is an enormous
amount of experimental information available for states near the
ground state, for most nuclides the quantum numbers are known for only
a very limited number of low--lying states. In particular, complete
and pure sequences of levels with the same spin and parity are
typically very short. Therefore data from several or many nuclei must
be combined to generate a sufficiently large ensemble. Moreover, the
shortness of the available sequences precludes the study of other
fluctuation measures than the NNS distribution. An extensive data set
was compiled by von Egidy {\it et al.} (1986). An initial analysis of a
subset of these data was performed by Abul-Magd and Weidenm\"{u}ller
(1985). A more extensive analysis of this same data set was performed
by Shriner {\it et al.} (1991). The spacing distributions and their
cumulative sums are shown in Fig.~\ref{fig12}. The nuclei are grouped
into classes according to mass number $A$. The size of each class was
determined by the data available. Obviously, the cumulative sums have
smaller fluctuations.  While the agreement with the Wigner
distribution looks satisfactory in most cases, clear deviations occur
for $150 < A \leq 180$ (rare earth nuclei) and for $230 < A$ (very
heavy nuclei). In both ranges of mass numbers, sizable nuclear
deformations occur and cause rotational motion, see
Section~\ref{coll}. The rotational model is integrable and the motion
therefore regular. The same statement holds for other forms of
so--called collective motion, see Section~\ref{collmod}.

The spacing distributions were fit with the Brody distribution (Brody
{\it et al.}, 1981). The overall trend of the Brody parameter
${\omega}$ was to decrease with increasing mass number $A$ -- for the
lightest region ($A$ = 25 -- 50) the average value of ${\omega}$ was
about 0.7, while for the heaviest mass region ($A$ = 225 -- 250) the
value of ${\omega}$ was 0.2. Various theoretical works have attempted
to explain this behavior, including Bae {\it et al.} (1992) and
Yoshinaga {\it et al.} (1993). To exhibit the connection of the NNS
distributions with the degree of collectivity, attention was focused
on the behavior of the 2$^{+}$ and 4$^{+}$ states because these states
play a prominent role in collective rotations and vibrations of the
nucleus. Two types of nuclei were considered: nuclei with
approximately spherical ground states and nuclei with strongly
deformed ground states, see Section~\ref{collmod}. The transition
between both classes was studied. Depending on the model chosen, the
motion is chaotic or regular in one but not in the other limit.
Fig.~\ref{fig13} shows that the experimental results are striking.
2$^{+}$ and 4$^{+}$ states in strongly deformed nuclei have NNS
distributions that agree with the Poisson distribution, while the
corresponding states in spherical nuclei have spacing distributions
that agree with the Wigner distribution. Unfortunately, the limited
amount of data and the corresponding large uncertainties preclude a
more detailed assessment of the effects of collectivity.

Another approach by Abul-Magd {\it et al.} (2004) focuses on specific
states (the lowest 2$^{+}$ states in even--even nuclei) and collects
all complete sequences of low--lying $2^+$ states from the nuclear
data tables. The sequences are short in most cases. Nuclei are
classified by the ratio $R_{4/2}$ of the excitation energy of the
lowest $4^+$ state over that of the lowest $2^+$ state. That ratio is
a well--known measure of collectivity. The size of each class is
chosen such that it contains a sufficient number of sequences for a
meaningful statistical analysis.  The NNS distributions are analyzed
using a measure different from the Brody distribution and obtained by
superposing a number of uncorrelated GOE sequences, each of mean
fractional level density $f$.  The parameter $f$ serves as a fit
parameter and is referred to as ``chaoticity parameter''. In its
dependence on $R_{4/2}$, this parameter displays deep minima when
$R_{4/2}$ equals $2.0$, $2.5$, and $3.3$. These values correspond to
the dynamical symmetries of a specific collective model, the
interacting boson model described in Section~\ref{coll}. Whenever one
of these symmetries prevails, the motion of the nucleus is integrable
and thus regular.

In summary, there is evidence that the nuclear dynamics in the
ground--state region is partly chaotic and partly regular. The regular
features are dominant whenever collective motion with a high degree of
symmetry applies.


\begin{figure}[t]
\vspace{5 mm}
\includegraphics[width=\linewidth]{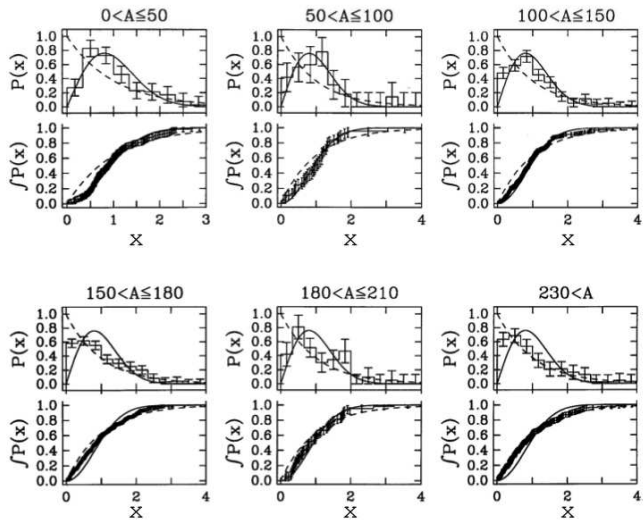}
\vspace{3 mm}
\caption{Upper panels: NNS distributions (histograms) are compared
with the Wigner distribution (solid lines) and the Poisson distribution
(dashed lines) for several ranges of mass numbers. Lower panels: same
for the cumulative distributions (number of spacings smaller than $x$).
(Cf. the caption of Fig.~\ref{fig7}). The approximate number of levels
in each region is: A = 0-50 (N = 150), A = 50-100 (N = 50), A =
100-150 (N = 270), A = 150-180 (N = 450), A = 180-210 (N = 60), A$>$
230  (N = 190). From Shriner {\it et al.} (1991).}
\label{fig12}
\end{figure}


\begin{figure}[h]
\vspace{5 mm}
\includegraphics[width=\linewidth]{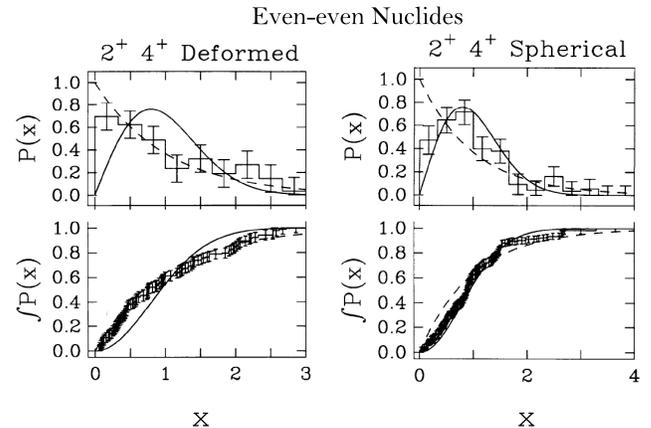}
\vspace{3 mm} 
\caption{Comparison of the NNS distributions versus $x$, the level
spacing in units of the mean level spacing, for $2^+$ and $4^+$
states in strongly deformed (left panel) and in spherical nuclei
(right panel). Adapted  from Shriner {\it et al.} (1991).}
\label{fig13}
\end{figure}

\subsubsection{High--spin States}
\label{anahss}

The only extensive analysis of the statistical properties of
high--spin states was performed by Garrett {\it et
al.} (1997). The data set comprised energy levels in deformed
nuclei in the range of proton numbers $Z$ = 62 -- 75 and mass numbers
$A$ = 155 -- 185. The spin values ranged up to $J = 40$. The levels
were at high excitation energies, but only up to several hundred  keV
above the yrast line, see Fig.~\ref{fig8}. The authors found that the
NNS distribution agreed best with the Poisson distribution. This is
consistent with the results for low--lying states in deformed nuclei,
see Section~\ref{chaosyrast}.

An observed deficiency of small spacings is not well understood. It is
possible that right above the yrast line a symmetry related to the $K$
quantum number is partly broken. This quantum number measures the
projection of the nuclear spin onto the body--fixed symmetry axis, see
Section~\ref{coll}. This might lead to level repulsion at small
distances. Hopefully the analysis of similar data will shed more light
on that question. We return to the general issue of symmetry breaking
and its influence on spectral fluctuation properties in
Section~\ref{isos} below.

\subsubsection{Analysis of Low--Lying Modes of Excitation}

At low excitation energies, one observes several modes of excitation.
Statistical measures have been used in order to identify the character
of the mode in the case of the scissors mode (Enders {\it et al.},
2000) and of the pygmy dipole resonance (Enders {\it et al.}, 2004).

For the scissors mode (a low--lying isovector magnetic dipole mode)
data were generated by nuclear resonance fluorescence
measurements (Enders {\it et al.}, 2000). The spectra of $13$ heavy deformed
even--even nuclei with neutron numbers in the 82 - 126 range
(corresponding to a major shell) were used to generate an ensemble of
152 scissors--mode states with spin/parity $1^+$, all in a range of
excitation energy between $2.5$ and $4.0$ MeV. In each nucleus, the
sequence of states used in the analysis was required to contain a
minimum number of $8$ states. After unfolding, the ensemble was
analyzed with the standard RMT fluctuation measures. The data agreed
very well with Poisson statistics. The authors examined carefully the
effects of missing levels on the spacing and width distributions and
concluded that missing levels can be ruled out as a cause of this
behavior. They concluded that the levels of the scissors mode are
excited by a common mechanism. The levels are collective but it is not
possible to identify a common doorway. It seems that the underlying
microscopic mechanism is not yet fully understood.

The electric pygmy dipole resonance is so named because of its small
strength relative to the giant electric dipole resonance. In heavy
nuclei, the pygmy resonance is located at excitation energies around
$5$ to $7$ MeV. Enders {\it et al.} (2004)  studied the
statistical properties of this mode in four isotones, all with neutron
number 82. They created an ensemble of 184 $1^{-}$ states in the
excitation energy range of 4 to 8 MeV, along with their dipole
transition strengths to the ground state. After unfolding, the
spectral fluctuations (strength and spacing distributions) are close
to Poissonian. Because of a significant number of missing levels the
analysis is rather involved in this case, however, and the authors
conclude that the weak correlations found point to GOE behavior of the
complete spectra. That conclusion is reinforced by an extensive
comparison with spectra calculated using a particular nuclear model,
the quasiparticle phonon model. These agree with the data but yield
GOE behavior for the full spectra (including the levels missing in the
data). The fundamental mode of excitation is collective. Many other
states exist at the excitation energy where it occurs. These fragment
the doorway state and produce a correlated spectrum of GOE type.

In comparing their results for these two modes, the authors conclude
that the key reason for the apparently different statistical behavior
is the difference in excitation energy. The higher--energy mode
(electric pygmy dipole) is in a region of greater level density. This
results in correlated spectra.

\subsubsection{Eigenvector Distribution}

The GOE predicts a Gaussian distribution for the projections of the
eigenvectors and the Porter--Thomas distribution for their squares,
see Section~\ref{porter}. While both the NNS distribution and the
$\Delta_3$--statistic are strongly affected by missing levels and/or
impure sequences, the Porter--Thomas distribution (a probability
density and not a correlation function) is expected to be less
sensitive to missing or wrongly assigned levels. The experimental data
normally used are the reduced widths (for resonances) or the reduced
transition strengths (for electromagnetic transitions). Early neutron
resonance data appeared to agree with the Porter-Thomas distribution;
a frequently quoted example is for neutron resonances on $^{232}$Th
measured by Rainwater's group at Columbia University (see Garg {\it et
al.}, 1964). The analysis by the Orsay group of the nuclear data
ensemble included a test of the Porter--Thomas distribution for the
widths (see Bohigas {\it et al}, 1983), and used a total of $1182$
measured widths. In addition to a direct comparison with the
Porter--Thomas distribution, a search for the best $\chi^2$
distribution was also done. Very good agreement with the GOE
prediction was found.

Further and more detailed attempts to confirm the Porter--Thomas
distribution have run into the following difficulty. For a set of
Gaussian--distributed amplitudes $\{ \gamma_i \}$ the second and the
fourth moments are related by $\left< \gamma^4 \right> = 3 \left<
\gamma^2 \right>^2$ where the angular brackets denote the running
average. According to Harney (1984), the error (square root of
the variance) of the ratio $R = \left< \gamma^4 \right> / 3 \left<
\gamma^2 \right>^2$ is $\sqrt{8/(3k)}$, where $k$ is the number of data
points. This is a rather large value. For the often quoted $^{232}$Th
data, for example, $k = 171$, and the error of $R$ is 0.125. Thus this
excellent data set only confirms the Gaussian nature of the amplitude
distribution at the 12 ${\%}$ level.

To overcome this problem, a larger data set seemed useful. An ensemble
of 1117 reduced widths was formed with TUNL data (Shriner {\it et
al.}, 1987). With $y = \gamma^2 / \langle \gamma^2 \rangle$ and $P(y)$
the Porter--Thomas distribution, the result for $\sqrt{y} P(y)$ is
shown in Fig.~\ref{fig14}. The visual agreement with the GOE
prediction is striking. However, the value of $R$ for this ensemble
turned out to be 1.26. The problem is that a relatively small number
of non--statistical large widths has a major impact on $R$ because $R$
depends on the fourth moment of the amplitudes.

\begin{figure}[h]
\vspace{5 mm}
\includegraphics[width=\linewidth]{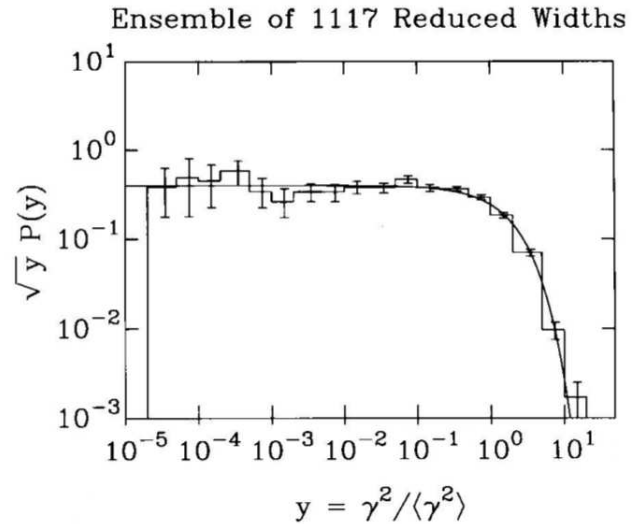}
\vspace{3 mm}
\caption{The distribution $\sqrt{y} P(y)$ for $y = \gamma^2 /
\overline{\gamma^2}$ for 1117 reduced widths. From Shriner {\it et
  al.} (1987).}
\label{fig14}
\end{figure}

To overcome that difficulty, a different measure which is less
sensitive to a few non--statistical amplitudes was needed. The
normalized linear correlation coefficient
\begin{equation}
  \rho(x,x') = \frac{\sum_i \left( x_i - \langle x \rangle \right)
     \left( x'_i - \langle x' \rangle \right)}
     {\left[\sum_i\left( x_i - \langle x \rangle \right)^2
     \sum_i \left( x'_i - \langle x' \rangle \right)^2\right]^{1/2}}
  \label{lcc}
\end{equation}
for two data sets $\{x_i\}$ and $\{x'_i\}$ is expected to be a
sensitive measure of correlations, as it combines information on both
the magnitudes and the phases of the data points. To test whether two
sets of amplitudes $\{a_i\}$ and $\{a'_i\}$ follow the Gaussian
distribution, one calculates the amplitude correlation coefficient
$\rho(a, a')$ and the width correlation coefficient $\rho(w, w')$
where $w$ stands for the square of the amplitude. (Since it is
impossible to measure the absolute sign of an amplitude, $\rho(a, a')$
is calculated with the assumption that $\langle a \rangle = 0 =
\langle a' \rangle$). The Gaussian distribution predicts $\rho^2(a,
a') = \rho(w, w')$ and this is the relation which is tested.

To work out the correlation coefficients it is necessary to measure
partial width amplitudes including their relative phases. That was
done using the inelastic decay of proton resonances. For example, for
a spin--zero target with mass number $A$ and a $2^+$ first excited
state, a proton resonance in the nucleus with mass number $A + 1$ and
spin/parity $3/2^-$ can decay to the $2^+$ state by emitting the
proton with angular momentum/spin $p_{1/2}$ or $p_{3/2}$. The relative
phase of the two decay amplitudes is determined by measuring the
angular distribution of the inelastically scattered protons and the
subsequent de--excitation $\gamma$ rays. Application of this approach
is described in detail in a review by Mitchell {\it et al.} (1985).

The relation $\rho^2(a, a') = \rho(w, w')$ can be checked for each
data set. However, the sample size for each spin value in each nuclide
was small, and it was necessary to combine the data sets. To this end,
an ambiguity in the definition of the correlation coefficient was
used. Instead of writing the amplitudes in terms of the decay channels
$p_{1/2}$ and $p_{3/2}$, any two linear combinations of amplitudes
obtained by orthogonal transformations from the original set can be
used. The correlation coefficient depends on the chosen
representation. There is always a representation in which the
amplitude correlation coefficient is zero. Each of the data sets was
individually transformed to that representation. The resulting width
correlations and their uncertainties were combined for all of the data
sets in order to obtain a final value for the width correlation. The
result was $\overline{\rho(w, w')}$ = -0.01 ${\pm}$
0.03, (Shriner {\it et al.}, 1987, 1989)
, in agreement with the prediction based on the
Gaussian distribution. This is the most sensitive test of the Gaussian
assumption.

\subsection{Violation of Symmetry or Invariance}
\label{viol}

Two of the symmetries mentioned in Section~\ref{gen} hold only
approximately: Isospin symmetry is broken by charge effects, and
parity conservation is violated by the weak interaction. Is it
possible to extend RMT so as to account for such symmetry breaking?
And how do the resulting statistical measures compare with data? We
answer these questions here for the case of isospin symmetry breaking.
Parity violation is a very weak effect that has so far received a
statistical analysis only in the framework of nuclear reaction theory,
see Part 2 of this review. We apply the results to the complete
spectra of $^{26}$Al and $^{30}$P.

We have also assumed that nuclei obey time--reversal invariance. One
of the most precise tests in nuclei of that assumption applies RMT. We
describe how a measure for violation of time--reversal invariance is
derived in the framework of RMT and applied to data.

\subsubsection{Model for Isospin Violation}
\label{isos}

Isospin symmetry is violated in nuclei by charge--dependent effects
such as the Coulomb interaction between protons, the neutron--proton
mass difference, or the mass differences between charged and neutral
pions.  Needless to say, the isospin--violating interaction is small
compared to the strong force. The breaking of isospin symmetry
manifests itself differently in different ranges of mass numbers. In
nuclei with mass numbers around $40$ or more, it leads to the
occurrence of fragmented isobaric analog resonances. The typical
features of this phenomenon relate to nuclear reaction theory and are
not dealt with here (see Part 2 of this review). In some light nuclei
with mass numbers smaller than $40$ or so, the ground state has
isospin $T = 0$ but the density of states with $T = 1$ in the
ground--state region is roughly the same as that of states with $T =
0$. The charge--dependent forces mix states with $T = 0$ and $T =
1$. It is for these nuclei that the following random--matrix model
applies.

In order to account for isospin violation (and for symmetry breaking
in general), the Hamiltonian~(\ref{14}) must be modified. The matrix
elements of the isospin--violating interaction couple states with
different $T$--quantum numbers. For simplicity of presentation we
consider two diagonal blocks only. The isospin--breaking interaction
conserves parity and total spin so these two blocks carry the same
quantum numbers $J$ and $\Pi$ (which we omit) but different isospin
quantum numbers $T_1$ and $T_2$. The Hamiltonian has the form
\be
H = \left( \matrix{ H^{T_1}_{\mu \nu} & V_{\mu \sigma} \cr V_{\rho
\nu} & H^{T_2}_{\rho \sigma} \cr} \right) \ .
\label{15}
\ee
With $N_1$ ($N_2$) the dimensions of the two block--diagonal
matrices, the running indices in the first (second) block are $\mu,
\nu = 1, \ldots, N_1$ ($\rho, \sigma = N_1 + 1, \ldots, N_1 + N_2$),
respectively. We deviate from our earlier systematic notation and
denote the coupling matrix elements connecting the two blocks by $V$.

We use Eq.~(\ref{15}) to define a random--matrix model for symmetry
breaking (Rosenzweig and Porter, 1960). We assume that the matrices
$H^{T_1}$ and $H^{T_2}$ are each members of a GOE and are
uncorrelated. For simplicity we assume that the two GOEs have
identical semicircle radii $2 \lambda_1 = 2 \lambda_2 = 2 \lambda$ and
equal dimensions $N_1 = N_2 = N$ although in practice it is necessary
to take $N_1 \neq N_2$ in order to account for the fact that the level
densities for states with different isospins differ. The (real) matrix
elements of $V$ are assumed to be Gaussian--distributed random
variables with zero mean value and a common second moment
$\overline{V^2}$. They are not correlated with each other or with the
elements of either of the two GOEs. Strictly speaking, the
symmetry--breaking interaction also contributes to the diagonal blocks
$H^{T_1}$ and $H^{T_2}$. However, such contributions can be
incorporated in the random--matrix description of these blocks and
therefore do not appear explicitly in the model.

To define the strength of $\overline{V^2}$ we recall Eq.~(\ref{2d}).
That equation seems to suggest that we put $\overline{V^2} = \alpha^2
\lambda^2 / N$, with $\alpha^2 \ll 1$ to account for the weakness of
the symmetry--breaking interaction. That is not correct, however, and
we must choose $\overline{V^2} = \alpha^2 \lambda^2 / N^2$, with
$\alpha$ a strength parameter which is independent of $N$ in the limit
$N \to \infty$. Indeed, without symmetry breaking, the spectra of
$H^{T_1}$ and $H^{T_2}$ are uncorrelated. A weak symmetry--breaking
interaction induces level repulsion and stiffness among levels with
different isospin. That happens when the matrix elements of $V$ are of
the order of the mean level spacing $d = \pi \lambda / N$, or when
$\overline{V^2}$ is of the order of $\lambda^2 / N^2$. Hence the ratio
of the strength of the symmetry--breaking interaction (average of the
square of the matrix elements) over that of the symmetry--conserving
interaction vanishes asymptotically as $1 / N$ for $N \to \infty$,
since the mean level spacing likewise vanishes in that limit.
Conversely, symmetry violation becomes detectable in spectral
fluctuation measures when the matrix elements of the
symmetry--breaking interaction are of the order of the mean level
spacing. This condition was already mentioned below Eq.~(\ref{2f}) and
is met by the isospin--violating matrix elements in light nuclei, see
Section~\ref{compl} below. The same conclusion is reached when we
consider the violation of time--reversal invariance in
Section~\ref{trsb} below. Apparently Pandey (1981) was the first
author to note the relevance of such small parameters for violations
of symmetry and/or invariance.

The parameter $\overline{V^2}$ suffers from the same shortcoming as
discussed in Section~\ref{door} for the mean--square--matrix element
of a doorway state: $\overline{V^2}$ changes strongly with excitation
energy and/or mass number. A much smoother measure of symmetry
breaking is the spreading width, see Eq.~(\ref{13a}). There are two
possible definitions for $\Gamma^{\downarrow}$,
\ba
\Gamma^{\downarrow T_1} &=& 2 \pi \overline{V^2} \rho^{T_2} \ ,
\nonumber \\
\Gamma^{\downarrow T_2} &=& 2 \pi \overline{V^2} \rho^{T_1} \ ,
\label{17}
\ea
where $\rho^T$ is the level density for the states with isospin $T$.
It is largely a matter of convenience which of these definitions is
used. To interpret the spreading widths, let us consider without loss
of generality the first of Eqs.~(\ref{17}). We use a basis in which
both $H^{T_1}$ and $H^{T_2}$ are diagonal. We first assume that the
mean level spacing $1 / \rho^{T_1}$ of states with isospin $T_1$ is
significantly larger than $\Gamma^{\downarrow T_1}$ and that
$\Gamma^{\downarrow T_1}$ is significantly larger than $1 /
\rho^{T_2}$. Then the arguments of Section~\ref{door} apply, each
state with isospin $T_1$ acts as an isolated doorway state, and
$\Gamma^{\downarrow}$ is the average width of the probability
distribution for finding the doorway state mixed into the eigenstates
of the full system. We expect that that interpretation remains
qualitatively valid also when the inequalities $1 / \rho^{T_1} \gg
\Gamma^{\downarrow} \gg \rho^{T_2}$ are violated. In other words,
$\Gamma^{\downarrow T_1}$ $(\Gamma^{\downarrow T_2})$ is a measure of
the width in energy with which every eigenstate of $H^{T_1}$ (of
$H^{T_1}$, respectively) is spread over the eigenstates of the full
system.

The random--matrix ensemble~(\ref{15}) lacks the overall orthogonal
symmetry of the GOE. By construction, the ensemble is invariant, of
course, under orthogonal transformations of the first $N_1$ (the last
$N_2$) states, respectively. Nonetheless, the analytical treatment of
symmetry violation in RMT is much more difficult than treating a
single GOE.  While analytical results for the ensemble~(\ref{15}) are
not available, replacing each of the two block--diagonal GOEs in
Eq.~(\ref{15}) by a GUE, and considering the complex matrix elements
in the non--diagonal blocks as uncorrelated Gaussian--distributed
random variables with zero mean value and common variance
$\overline{V^2}$, one arrives at a tractable problem~(Guhr and
Weidenm\"uller, 1990a). Although belonging to a different symmetry
class, the resulting ensemble is expected to possess features which
are qualitatively similar to those of the ensemble~(\ref{15}).
Additional information is generated by a numerical analysis of the
ensemble~(\ref{15}).

As $\overline{V^2} / d^2 \propto \alpha^2$ increases from zero, the
spectral fluctuations change from those of two uncorrelated GUEs to
those of a single GUE. From $\overline{V^2} / d^2 = 0$ to
$\overline{V^2} / d^2 > 0$, the change is discontinuous: Level
repulsion amongst states with different $T$ sets in suddenly. For the
local spectral fluctuation measures considered by Guhr and
Weidenm\"uller (1990a) the case of a single GUE is attained when
$\overline{V^2} / d^2 \approx 1$. We expect this statement to apply
likewise to other fluctuation measures, and to hold similarly for the
ensemble in Eq.~(\ref{15}). The expectation is confirmed by numerical
simulations.

\subsubsection{Complete Level Schemes}
\label{compl}

As described in Section~\ref{cls}, complete level schemes were
determined for the nuclides $^{26}$Al and $^{30}$P. These odd-odd $N$
= $Z$ nuclei are particularly interesting because here the densities
of states with $T = 0$ and with $T = 1$ are almost equal, starting
from the ground state, while in most nuclei the states with higher $T$
are shifted toward higher excitation energies. Thus these nuclei are
ideal to study the effect of isospin--symmetry breaking.

\begin{figure}[h]
\vspace{5 mm}
\includegraphics[width=\linewidth]{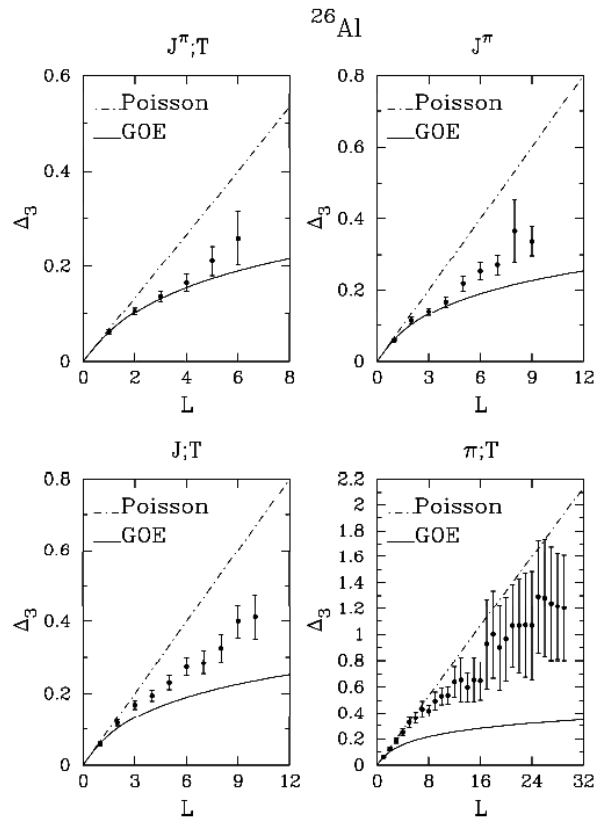}
\vspace{3 mm}
\caption{The four panels show the $\Delta_3$--statistic for the states
in $^{26}$Al obtained by taking into account only the quantum numbers
indicated at the top of each panel.}
\label{fig15}
\end{figure}

A qualitative test for a conserved symmetry is to consider what
happens when that symmetry is neglected. As an example, in
Fig.~\ref{fig15} the ${\Delta_{3}}$--statistic is shown for the
states in $^{26}$Al. Only the quantum numbers shown at the top of each
panel are taken into account in evaluating $\Delta_3$. We note that
ignoring the good quantum number $J$ leads to a major increase in
${\Delta_{3}}$. Ignoring $T$, on the other hand, leads to a very small
change of $\Delta_3$. This seems to suggest that isospin is not a good
quantum number, in apparent contradiction to other evidence that
isospin is only slightly broken (at about the 3 ${\%}$ level) in this
nuclide.

The point is that (as shown in Section~\ref{isos}), the impact of
symmetry breaking on $\Delta_3$ depends on the ratio of the
symmetry--breaking matrix element to the mean level spacing $d$ and is
thus enhanced by small spacings. Hence, a small degree of symmetry
breaking can have a large effect on the statistical measures. Somewhat
fortuitously, the strength of symmetry breaking in $^{26}$Al is such
that the ${\Delta_{3}}$--statistic lies between the values for a
single GOE and for two GOEs, see Fig.~\ref{fig16}. More precisely, the
root--mean--square value of the symmetry--breaking Coulomb matrix
elements in $^{26}$Al is a little smaller than but of the order of the
mean level spacing (Guhr and Weidenm\"{u}ller, 1990a). The results
were consistent with other experimental determinations. Experimental
results on isospin violation for  $^{30}$P  were almost identical with
the $^{26}$Al results  (Shriner   {\bf et al.}, 2000).

\begin{figure}[h]
\vspace{5 mm}
\includegraphics[width=\linewidth]{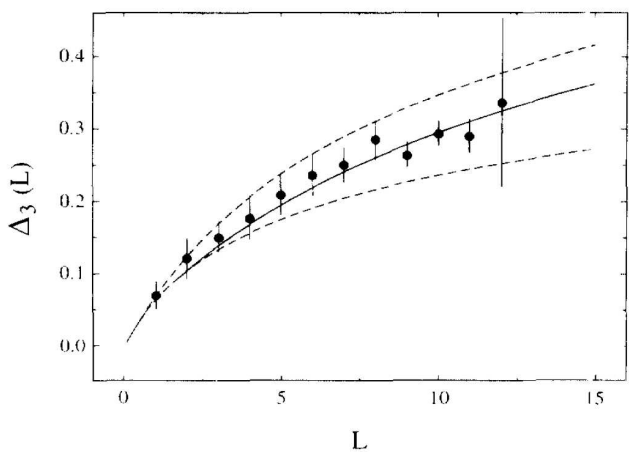}
\vspace{3 mm}
\caption{The spectral rigidity $\Delta_3$ versus $L$ for $75$ levels
with $T = 0$ and $25$ levels with $T = 1$ measured in $^{26}$Al (dots
with error bars). The excitation energies lie between zero and $8$ MeV.
The lower (upper) dashed line is the prediction for a single GOE (for
two uncorrelated GOEs with fractional densities $3/4$ and $1/4$,
respectively). The solid line is the result of a numerical simulation
incorporating symmetry breaking. The strength of the symmetry--breaking
interaction was fitted to the data. From Guhr {\it et al.} (1998).}
\label{fig16}
\end{figure}

Although these results were consistent with theoretical expectations,
they were not considered definitive due to the small sample
sizes. Definitive results on symmetry breaking were provided by
measurements of acoustic resonances in quartz blocks (Ellegaard {\it
et al.}, 1996) and of electromagnetic resonances in coupled microwave
billiards (Alt {\it et al.}, 1998). In these measurements the strength
of the symmetry breaking force could be effectively varied and much
larger sample sizes were obtained. The agreement with the RMT model of
Eq.~(\ref{15}) was excellent.

The third nucleus with a ``nearly complete'' level scheme (for
excitation energies below $4.3$ MeV) is $^{116}$Sn. In that nucleus,
the NNS distribution was studied (Raman {\it et al.}, 1991). Only
sequences with a minimum of five levels with the same spin and parity
assignments were included in the analysis; there were $6$ such
sequences. The histogram for the NNS distribution was fit with the
Brody parametrization~(\ref{14a}). The fit gave $\omega = 0.51 \pm
0.19$, similar to the best fit in $^{26}$Al which gave $\omega = 0.47
\pm 0.14$. In $^{26}$Al the deviation from GOE predictions is due to
isospin symmetry breaking. The cause for the same phenomenon in
$^{116}$Sn is not clear. The nucleus $^{116}$Sn is mentioned in the
present Section only because of its ``nearly complete'' level scheme.

\subsubsection{Transition Strengths}

After the studies of isospin symmetry breaking in the complete spectra
of $^{26}$Al and $^{30}$P (see Section~\ref{compl}), the effect of
symmetry breaking on the eigenvectors in the same systems was also
explored. Although there was no formal proof, heuristic arguments
predicted that the Porter--Thomas distribution would not be changed by
symmetry breaking. The central point of the argument was the
complexity of the wavefunctions of initial and final states of the
transitions.

The TUNL group (Adams {\it et al.}, 1998; Shriner {\it et al.}, 2000)
used the reduced electromagnetic transition probabilities $B$ in these
nuclei as the data set. To eliminate issues of scale, subgroups of the
transitions were considered. The transition probabilities were
classified according to multipolariy (electric dipole E1, electric
quadrupole E2, magnetic dipole M1) and to the isospin difference
$\Delta T = 0$ or $1$ between initial and final states. The parameter
$y$ = $B$ / $\overline{B}$ turns out to have a very large dynamic
range. Therefore it is convenient to use $z = \log_{10}$ $y$. In terms
of that variable, the upper panels of Fig.~\ref{fig17} show the
Porter--Thomas distribution as solid lines. The lower panels give the
integrated distributions. The histograms show the data for
$^{30}$P. The details of the analysis are given by Adams {\it et al.} 
(1998), and Shriner {\it et al.} (2000).  Similar results were
obtained for $^{26}$Al. It is obvious that the data do not agree with
the Porter--Thomas distribution.


\begin{figure}[t]
\vspace{5 mm}
\includegraphics[height=0.8\linewidth]{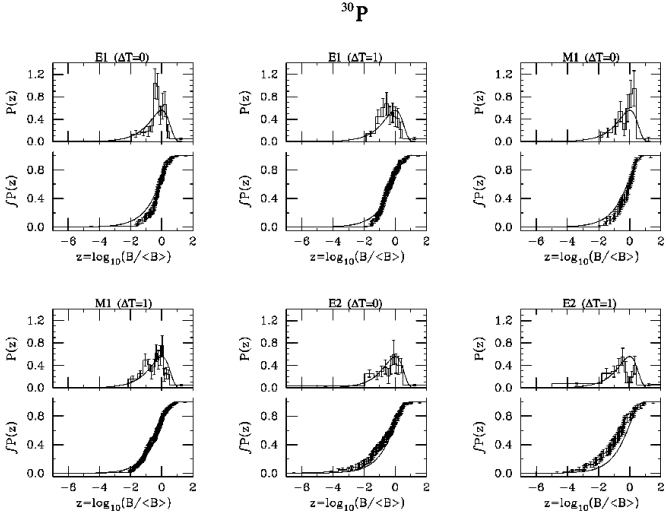}
\vspace{3 mm}
\caption{Comparison between the Porter--Thomas distribution (written
in terms of the variable $z = \log_{10} y$) (solid lines) and the
data (histograms) for several multipole transitions and isospin
differences in $^{30}P$. From Shriner {\it et al.}, 2000.}
\label{fig17}
\end{figure}


It was some years before this somewhat unexpected result was formally
analyzed and explained. Barbosa {\it et al.} (2000) used basically the
same approach as for the description of symmetry breaking on the level
statistics, see Section~\ref{isos}. We only sketch the central
point. We consider electromagnetic transitions of a given
multipolarity (typically M1 or E2). The model~(\ref{15}) has to be
extended because the electromagnetic transition operator does not, in
general, conserve either $T$ or $J$ and connects states with different
spins as well as different isospins. For $\overline{V^2} = 0$ the
eigenfunctions of $H$ are eigenfunctions either of ${\cal H}^{J_1
T_1}$ or of ${\cal H}^{J_2 T_2}$. For fixed $J_1, J_2$ the transition
matrix elements belong to one of three classes: (i) Those coupling two
states with isospin $T_1$; (ii) those coupling two states with isospin
$T_2$; and (iii) those coupling two states with different
isospins. The distribution of matrix elements within each class is
expected to be approximately Gaussian, but the three Gaussian
distributions may have different heights and widths. Moreover,
transitions with different multipolarities behave differently.
Therefore the squares of the transition matrix elements in
Fig.~\ref{fig17} cannot have a simple Porter--Thomas distribution even
for $\overline{V^2} = 0$. As $\overline{V^2}$ increases from zero, the
Gaussian distributions get mixed. Details are only accessible
numerically. With a somewhat different approach Hussein and Pato
(2000) also predicted a deviation from the Porter-Thomas
distribution. Neither group attempted to fit the experimental data in
detail. Such an analysis is still missing. We conclude that while the
effect of isospin symmetry breaking on the level statistics is a
generic phenomenon that can be accounted for completely by a simple
extension of RMT in terms of a single parameter (the spreading width),
the effect of symmetry breaking on the electromagnetic transition
strengths involves additional dynamic elements (classification of the
transition matrix elements).

While the size of the data set in $^{26}$Al and $^{30}$P is limited,
experiments using coupled microwave billiards (Dembowski {\it et al.},
2005) yielded data with much better statistics. The distribution of
transition strengths deviates from the Porter--Thomas
distribution. The theory of Barbosa {\it et al.} (2000) was extended
to this case by Dietz {\it et al.}  (2006).

\subsubsection{Test of Time--reversal Invariance}
\label{trsb}

Because of the anti--unitarity of the  time--reversal operator, the
modeling of a violation of time--reversal invariance in RMT is
fundamentally different from that of a broken symmetry as discussed in
Section~\ref{isos}. As explained in Section~\ref{why}, time--reversal
invariance allows us to choose the Hamiltonian matrix as a real and
symmetric matrix. The matrix ensemble that models such systems is the
GOE. If time--reversal invariance does not hold, the Hamiltonian
matrix is Hermitian but cannot, in general, be chosen real and
symmetric. The matrix ensemble that models such systems is the GUE,
see Section~\ref{GOE}. To describe the violation of time--reversal
invariance in RMT, we need to construct an ensemble that interpolates
between the GOE and the GUE. This is done as follows. Every Hermitian
matrix can be written as the sum of a real symmetric matrix and of $i$
times a real antisymmetric matrix. A stochastic model for a
Hamiltonian with some violation of time--reversal symmetry is then
\be
H = \frac{1}{\sqrt{1 + (1/N) \alpha^2}} \bigg( H^{\rm GOE} + N^{-1/2}
\alpha i A \bigg) \ .
\label{18}
\ee
Here $H^{\rm GOE}$ stands for the GOE, and the elements of the real
antisymmetric matrix $A$ are Gaussian--distributed random variables
with zero mean value and a second moment given by
\be
\overline{A_{\mu \nu} A_{\rho \sigma}} = \frac{\lambda^2}{N} \big(
\delta_{\mu \rho} \delta_{\nu \sigma} - \delta_{\mu \sigma}
\delta_{\nu \rho} \big) \ . 
\label{18b}
\ee
The elements of $A$ and of $H^{\rm GOE}$ are uncorrelated. The real
dimensionless parameter $\alpha$ describes the strength of the
violation of time--reversal invariance. For $\alpha = 0$ we deal with
the GOE and for $\alpha = N^{1/2}$, with the GUE.

For tests of the violation of time--reversal invariance in nuclei, the
central feature of the GUE is level repulsion at small distances
(scaled spacing $s \ll 1)$. In the case of the GOE, level repulsion
leads to a linear dependence of the NNS distribution for small $s$,
see the Wigner surmise~(\ref{7}). In contradistinction, the NNS
distribution for the GUE increases quadratically with $s$ for small
$s$. A test for a violation of time--reversal invariance in nuclei is,
therefore, based on a detailed examination of the NNS distribution at
small spacings (French {\it et al.}, 1985). As explained below
Eq.~(\ref{2f}) and, in a different context, in Section~\ref{isos}, the
mixing of levels (and, thus, the spacing distribution) are sensitive
to very small mixing matrix elements which are of the order of the GOE
mean level spacing $d = \pi \lambda / N$. For the perturbation
$N^{-1/2} \alpha i A$ in Eqs.~(\ref{18}, \ref{18b}), the
root--mean--square matrix element has the value $\alpha \lambda / N$.
This is comparable to $d$ for $\alpha \approx 1$ which explains the
choice of factors $N^{1/2}$ in Eq.~(\ref{18}). As in
Section~\ref{isos}, the mixing parameter $N^{-1/2} \alpha$ vanishes
asymptotically as $1 / \sqrt{N}$. The NNS distribution is sensitive to
a violation of time--reversal symmetry when the relevant matrix
elements are of the order of the average nuclear level spacing. The
analysis was done~(French {\it et al.}, 1985) for the nuclear data
ensemble (see Section~\ref{npres}), where typical spacings $d$ are of
the order of $10$ keV, and yielded an upper bound of about $d / 10$
for the time--reversal non--invariant matrix element of the nuclear
Hamiltonian. From this result French {\it et al.} (1985) inferred an
upper bound of about $1$ percent for the time--reversal non--invariant
part of the nucleon--nucleon interaction.

\section{Chaos in Nuclear Models}
\label{models}

Sections~\ref{RMT} and \ref{applRMT} were based almost entirely on
concepts of RMT and made very little use of the wealth of information
on the dynamical behavior of nuclei. We fill this gap in the present
Section. We discuss the two leading nuclear--structure models which
describe phenomenologically the dynamics of nuclei: The shell model
(which mostly applies to spherical nuclei), and the collective model
(which mostly applies to nuclei with surface deformations). We give a
brief introduction to both models which, in their simplest form, are
fully integrable and thus give rise to regular motion. We present
evidence that both models also allow for chaotic motion. We then turn
to a number of specific applications of RMT which incorporate
dynamical aspects.

\subsection{Spherical Shell Model}

\subsubsection{The Nuclear Shell Model}
\label{shell}

In the elementary version of the nuclear shell model nucleons move
independently under the influence of a common mean field. Attempts to
introduce a mean field into nuclear theory date back to the
1930s. These attempts were unsuccessful at the time, partly because
they lacked an essential ingredient (the strong spin--orbit coupling)
and partly because of the success of Bohr's idea of the compound
nucleus which depicted nuclei as systems of strongly interacting
particles. The introduction in 1949 of a central single--particle
potential with strong spin--orbit coupling by Haxel {\it et al.}
(1949)  and Mayer (1949)  changed that situation.
That model was very successful in the description of nuclear
properties in the ground--state domain and shifted attention away from
the compound--nucleus picture. It gave rise to a burst of
spectroscopic activity which lasted for many years and thoroughly
validated the model.

The sequence of single--particle levels of the nuclear shell model is
shown in Fig.~\ref{fig18}. The scheme is fundamentally that of the
harmonic oscillator in three dimensions with excitation energies $n
\hbar \omega$ and $n = 0,1,2,\ldots$. The integer $n$ defines the
major shells. Individual levels are denoted by $(n + 1)$, by the
single--particle angular momentum $\ell$ in spectroscopic notation,
and by the single--particle total spin $j$ obtained by
vector--coupling the angular momentum operator $\vec{\ell}$ and the
spin $\vec{s}$. (Here we do not distinguish neutrons and protons and
take isospin as a good quantum number. The picture requires some
modification for medium--weight and heavy nuclei which we do not
address). The degeneracy of the single--particle states in each major
shell,  which is characteristic of the harmonic oscillator, is lifted
because the single--particle potential does not have the shape of a
harmonic oscillator, and because of the presence of a strong
spin--orbit coupling that pushes  the states with highest spin in each
major shell down into the next--lower major shell (except for the
lowest shells where the spin--orbit interaction is not strong
enough). Thus each major shell contains a number of subshells each of
which is characterized by the quantum numbers $(n + 1, \ell, j)$.
Within a given major shell the index $n$ is redundant and will often
be omitted.

\begin{figure}[h]
\vspace{5 mm}
\includegraphics[width=\linewidth]{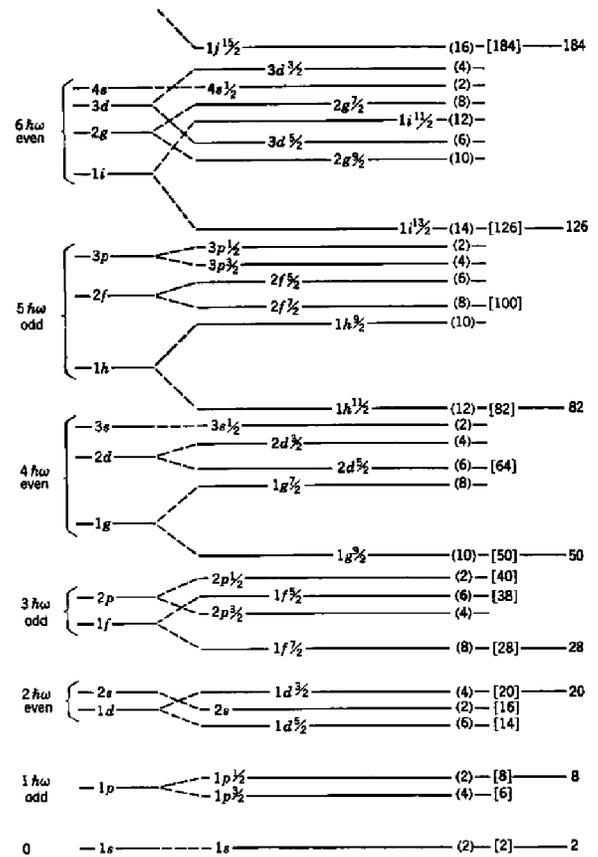}
\vspace{3 mm}
\caption{Level sequence in the nuclear shell model. From
Mayer and Jensen (1955).}
\label{fig18}
\end{figure}

For a nucleus with mass number $A$, the $A$ nucleons fill the lowest
shells in accord with the exclusion principle. The completely filled
shells are considered as inert, and the spectroscopic properties of
the low--lying states result from the $m$ ``valence nucleons'' which
partly fill the last shell (the ``valence shell''). This scheme
accounts for the strong binding energies of nuclei with closed shells,
and for the ground--state properties of nuclei differing from
closed--shell ones by the addition or removal of a single nucleon. For
nuclei with more than one valence nucleon (or more than one hole in
the valence shell), an extension of the model is called for because in
such cases strong degeneracies occur in the model which are not
observed in reality. For instance, two valence nucleons in the
$1d_{5/2}$--subshell of the $2s 1d$--shell can be coupled to total
isospin $T = 0$ or $T = 1$. For $T = 0$ ($T = 1$), the possible states
have odd (even) total spin values $J$ ranging from $J = 0$ to $J =
5$. All of these states are degenerate in the elementary shell model.

To remedy that situation, the elementary shell model is viewed as a
mean--field theory which takes account of most (but not all) of the
nucleon--nucleon interaction. The remaining ``residual interaction''
must be included to obtain quantitative agreement with data. From the
point of view of many--body theory, the residual interaction is an
effective interaction. The residual interaction is usually assumed to
be a two--body interaction (although there is evidence (Pieper and
Wiringa, 2001)
that three--body forces are needed in some cases to obtain good fits
to the data), to be time--reversal invariant, and to conserve spin and
parity. In the spirit of the shell model it is also assumed that the
residual interaction is weak (it does not significantly mix the
many--body states belonging to different major shells), so that it
effectively acts only amongst the $m$ valence nucleons in the valence
shell.

The input for this model (the full shell model or, in brief, the shell
model) consists of the single--particle energies $\ve_{\ell j}$ in the
valence shell, and of the matrix elements of the residual interaction
in that shell. The residual two--body interaction $V_{\rm res}$ is
completely characterized by a finite number of two--body matrix
elements. The two--body states $| j_1 j_2 s t \sigma \tau \rangle$ are
obtained by coupling any two single--particle states $(\ell_1 j_1)$
and $(\ell_2 j_2)$ in the valence shell to total spin $s$ and total
isospin $t$ with $z$--components $\sigma$ and $\tau$. The
antisymmetrized two--body matrix elements of $V_{\rm res}$ have the
form $\langle j_3 j_4 s t | V_{\rm res} | j_1 j_2 s t \rangle$. Our
notation implies conservation of spin and isospin and takes account of
the fact that the values of the matrix elements do not depend on
$\sigma$ and $\tau$. Conservation of parity imposes an additional
constraint not explicitly displayed in our notation. For brevity we
will refer to these matrix elements by the symbol $v_\alpha$. The
index $\alpha$ enumerates all allowed and distinct (i.e., not
connected by symmetry) two--body matrix elements in the valence
shell. For the $2s1d$--shell, the range of $\alpha$ is 63 while for
the $2p1f$--shell, it is 195.

In second quantization, the shell--model Hamiltonian governing the $m$
valence nucleons is then given by
\ba
{\cal H} &=& \sum_{\ell j} \ve_{\ell j} \sum_{\sigma \tau} a^{\dag}_{j
\ell \sigma \tau} a^{}_{j \ell \sigma \tau} \nonumber \\
&& \qquad + \frac{1}{4} \sum_{j_1 j_2 j_3 j_4 s t} \langle j_3 j_4 s t\
| V_{\rm res} | j_1 j_2 s t \rangle \nonumber \\
&& \qquad \qquad \times \sum_{\sigma \tau} A^{\dag}_{j_3 j_4 s t
\sigma \tau} A^{}_{j_1 j_2 s t \sigma \tau} \ .
\label{19}
\ea
Here $a^{\dag}_{j \ell \sigma \tau}$ creates a nucleon in a state
$(\ell j)$ with spin $z$--component $\sigma$ and isospin
$z$--component $\tau$ while $A^{\dag}_{j_3 j_4 s t \sigma \tau}$
creates a pair of nucleons in the state $| j_1 j_2 s t \sigma \tau
\rangle$. The operators $A^{\dag}$ are straightforwardly obtained by
vector--coupling products of two operators $a^{\dag}_{j \ell \sigma
\tau}$ and are not given explicitly.

The eigenvalues and eigenfunctions of ${\cal H}$ are identified with
the low--lying states of nuclei. For instance, nuclei pertaining to
the $2s1d$--shell (the ``$sd$--shell nuclei'') have mass numbers $17
\leq A \leq 39$, and the number of valence nucleons has the range $1
\leq m \leq 23$. Mass numbers $A = 16$ and $A = 40$ correspond to the
closed--shell nuclei $^{16}$O (the $1s$-- and the $1p$--shells are
filled) and $^{40}$Ca (the $2s1d$--shell is filled, too). Likewise
there are $2f1p$--shell nuclei etc. (We disregard the $1p$--shell
nuclei as they yield too few spectroscopic data for a meaningful
statistical analysis). According to the shell model, the Hamiltonian
${\cal H}$ determines the spectral properties of the low--lying states
of all nuclei pertaining to the same shell, at least in
principle. This claim is subject to a number of provisos. (i) To get
good fits to the spectra of all nuclei in a major shell it may be
necessary to allow for a weak dependence of the parameters $ \ve_{\ell
j}$ and $v_\alpha$ on the number $m$ of valence nucleons.  (ii)
Non--valence--shell states may be pushed down by the residual
interaction into the domain of low excitation energies (``intruder
states'') and require special treatment. Such states are obtained, for
instance, by lifting one or several nucleons from the valence shell
into the next higher major shell, or from the inert core into the
valence shell, or both. (iii) In its upper part, the spectrum of
${\cal H}$ cannot be expected to correspond to reality because the
much more numerous non--valence--shell states dominate the actual
spectrum and mix with the states in the valence shell. In the studies
of chaos reported below, one disregards this fact and confines
attention to the valence shell.  This is done in order to obtain a
manageable numerical problem. There are strong reasons to believe that
the results are universal and, thus, also hold when the mixing between
major shells is taken into account (see, for instance, Ormand and
Broglia, 1992).   (iv) In the form of Eq.~(\ref{19}), the shell model
applies to spherical nuclei (although it can be extended to weakly
deformed nuclei). Chaos in deformed nuclei is treated in
Section~\ref{collmod}.

For purposes of orientation, we cite a few numbers taken from Bohr and
Mottelson (1969) and Zelevinsky {\it et al.} (1996). The spacing
between major shells is approximately given by the
harmonic--oscillator energy $\hbar \omega \approx 40 \ A^{-1/3}$ MeV.
The spin--orbit interaction is $\approx - 20 \ (\vec{l} \cdot \vec{s})
\ A^{-2/3}$ MeV. The spacings of adjacent single--particle energies
may be as large as a couple of MeV. In the $sd$--shell, for instance,
the empirical values in $^{17}$O are $\ve_{d \ 5/2} = - 4.15$ MeV,
$\ve_{s \ 1/2} = - 3.28$ MeV, $\ve_{d \ 3/2} = 0.93$ MeV. With these
figures and for $m = 12$ valence nucleons, the range of the elementary
$sd$--shell spectrum would be about $42$ MeV. (These figures are not
completely representative since the $\ve_{\ell j}$s themselves are
actually used as fit parameters). The diagonal matrix elements of the
residual interaction may be as large as $1$ or $2$ MeV in magnitude.
This removes some of the degeneracies of the elementary shell model
referred to above and stretches the spectrum further. The non--diagonal
elements typically amount to several hundred keV in magnitude. These
numbers suggest that the residual interaction is able to mix the
states in different subshells of the valence shell, while the spectral
properties at low excitation energy are essentially determined by
valence--shell states (the admixtures of non--valence--shell states
are negligible).

Not all parameters of the shell model can be determined equally well
by a fit to spectroscopic data. That topic is discussed further in
Sections~\ref{chaossm} and \ref{another} below. To overcome that
difficulty, one uses nuclear many--body theory (i.e., variants of the
Bethe--Brueckner--Goldstone expansion) to calculate the two--body
matrix elements of $V_{\rm res}$. The results serve as starting values
for a fit to the data. In the fit, the two--body matrix elements
themselves or some parameters on which they depend, are
varied (Brown and Wildenthal, 1988; Honma {\it et al.}, 2002). There is evidence that the necessary
corrections mainly account for three--body forces (Caurier, 2005).

Since ${\cal H}$ conserves spin, isospin, and parity, the
eigenfunctions of ${\cal H}$ are simultaneously eigenfunctions of
total spin $J$, total isospin $T$, and parity $\Pi$. One way to obtain
the eigenfunctions of ${\cal H}$ with given $J,T,\Pi$ for $m$ valence
nucleons consists in constructing the matrix of ${\cal H}$ in a basis
of $m$--body states carrying these quantum numbers and in
diagonalizing that matrix. (This is not always the most efficient
procedure numerically, but is used here for the sake of the
argument). The $m$--body states needed for this procedure are obtained
from the elementary shell model by constructing all Slater
determinants of $m$ valence nucleons in the valence shell. These
determinants form classes, each class being defined by the set
$\{m_{\ell j}\}$ where $m_{\ell j}$ is the number of nucleons
occupying the subshell $(\ell j)$. The set $\{ m_{\ell j} \}$
obviously forms a partition of $m$ so that $\sum_{\ell j} m_{\ell j} =
m$ where the sum runs over the subshells of the valence shell. The
Slater determinants are antisymmetric by construction, but typically
are not eigenstates of $J$ and $T$. The $m$--body states with good
$J$, $T$, and $\Pi$ and fixed $\{m_{\ell j}\}$ are found as linear
combinations of the determinants in class $\{m_{\ell j}\}$ and are
denoted by $| J T \Pi \mu \rangle$ in the sequel. (We suppress the
magnetic quantum numbers and the class index $m_{\ell j}$). The states
$| J T \Pi \mu \rangle$ span a Hilbert space of finite dimension
${\cal D}(J, T, \Pi )$, which defines the range of the running index
$\mu$. In the $2s1d$--shell and depending on the quantum numbers $(J,
T)$, ${\cal D}$ ranges from a few to about $7000$ in the middle of the
shell ($m = 12$), while ${\cal D}$ attains considerably larger maximum
values already in the $2p1f$--shell. This is why the $2s1d$--shell has
been the preferred object for theoretical studies of chaos in nuclei.
The actual construction of the states $| J T \Pi \mu \rangle$ is
cumbersome, involves angular--momentum algebra, and is not given
here~(De Shalit and Talmi, 1963). In that basis, the matrix elements
of ${\cal H}$ have the form
\ba
&& {\cal H}_{\mu \nu}(J T \Pi) \ {\buildrel {\rm def} \over =} \ \langle
J T \Pi \mu | {\cal H} | J T \Pi \nu \rangle = \delta_{\mu \nu}
\sum_{\ell j} \ve_{\ell j} m_{\ell j} \nonumber \\
&& \qquad + \sum_\alpha v_\alpha C_{\mu \nu}(\alpha; J T \Pi ) \ .
\label{20}
\ea
The first term on the right--hand side of the last of Eqs.~(\ref{20})
is obvious. The form of the second term follows from Eq.~(\ref{19})
except that we have grouped together all matrix elements which are
connected by symmetry. Save for this operation, the coefficients $
C_{\mu \nu}(\alpha; J T \Pi )$ are matrix elements of the operator
$\sum_{m \tau} A^{\dag}_{j_3 j_4 s t m \tau} A^{}_{j_1 j_2 s t m
\tau}$ taken between the states $| J T \Pi \mu \rangle$ and $| J T \Pi
\nu \rangle$. By construction, the matrix ${\cal H}_{\mu \nu}$ is real
and symmetric.

The form~(\ref{20}) displays explicitly the dependence of the matrix
${\cal H}_{\mu \nu}(J T \Pi)$ on the input parameters $\ve_{\ell j}$
and $v_\alpha$ of the shell model. We emphasize that the coefficients
$C_{\mu \nu}(\alpha; J T \Pi )$ (which for fixed $\alpha$, $J$, $T$,
and $\Pi$ form a real and symmetric matrix in Hilbert space) are
determined entirely by the valence shell we are working in, by the
coupling scheme we have used to construct the states $| J T \Pi \mu
\rangle$, and by the two--body operator labeled $\alpha$ of which the
matrix elements are taken. Except for a set of unitary transformations
connecting the coupling scheme we have chosen with any other one, the
matrices $C_{\mu \nu}(\alpha)$ are uniquely determined. These matrices
reflect the symmetries and invariances of the elementary shell model
and are independent of the residual interaction actually considered.
In other words: Going from one residual interaction with matrix
elements $v_\alpha$ to another one with matrix elements $v'_\alpha$,
all we need to do is to replace the coefficients $v_\alpha$ in
Eq.~(\ref{20}) by the coefficients $v'_\alpha$, the matrices $C_{\mu
\nu}(\alpha)$ remaining the same. We emphasize this simple fact
because some properties of the shell--model Hamiltonian $H_{\mu \nu}(J
T \Pi)$ are determined by the matrices $C_{\mu \nu}(\alpha)$ alone and
are thus generic (i.e., largely independent of the choice of the
residual interaction).

It was mentioned before and we emphasize again that the shell model
accounts successfully for a vast amount of spectroscopic data in the
ground--state domain of spherical nuclei. This is not the place to go
into any details. Suffice it to say that some basic features of
$V_{\rm res}$ are well established: Pairs of nucleons coupled to
isospin $T = 1$ have a strong and attractive interaction (``pairing
force''). That force leads to spin--zero ground states in even--even
nuclei and favors the seniority coupling scheme~(De Shalit and Talmi,
1963). In the particle--hole channel, the diagonal elements of $V_{\rm
res}$ for pairs of nucleons with isospin $T = 0$ (neutron--proton
pairs) also show strong attraction, especially for large angular
momenta (``quadrupole--quadrupole interaction'') and favor nuclear
deformations, especially for nuclei far (in mass number) from
closed--shell nuclei~(see Bohr and Mottelson, 1969). For the
$sd$--shell, Wildenthal (1984) and Brown and Wildenthal
(1988) have established the standard parameters of the residual
interaction (``Brown--Wildenthal interaction'') by fitting the $66$
parameters ($63$ two--body matrix elements and $3$ single--particle
energies) to more than 400 pieces of data. The Brown--Wildenthal
interaction may optionally include the Coulomb interaction between
protons.

\subsubsection{Chaos in the Shell Model}
\label{chaossm}

Ever since the shell model was established, a strange dichotomy
pervaded nuclear--structure theory. On the one hand, the shell model
was extremely successful in describing the properties of low--lying
states in many nuclei, and the collective models in their various
forms afforded a description of those nuclei which were not accessible
to the shell model. Fits to data revealed the basic properties of the
residual interaction. Ever more sophisticated measurements widened the
data basis. Theoretical efforts were directed at both the
determination of $V_{\rm res}$ from the interaction between free
nucleons via many--body theory, and at the technology to calculate
nuclear properties from the shell--model Hamiltonian, including those
of deformed nuclei, and thus at understanding collective models. The
hope seemed justified that one day one would arrive at a complete
understanding of the structure of atomic nuclei. In the 1970s and
1980s, most work in nuclear--structure theory was devoted to that
vision.

On the other hand, during the same period the evidence for the
applicability of RMT to neutron and proton resonance data grew and was
definitively established in 1982, see Section~\ref{npres}. But aside
from fundamental symmetries, RMT lacks any dynamical structure
whatsoever. What did the success of RMT imply for the shell model? Is
there an excitation energy (somewhere below neutron threshold but
above the energies where shell--model calculations were so successful)
beyond which the shell model fails to work and chaos takes over? Or is
chaos possible (and perhaps even generic) in the framework of the
shell model itself?

In spite of a strong growth of chaos--related research in other fields
of physics in the 1980s and early 1990s, these questions did not
receive much attention at the time by the nuclear--structure
community. The main reason was probably the lack of statistically
significant experimental data. In addition, a strong commitment on the
part of the community to understand nuclei on a fundamental level
perhaps deflected attention away from the issue of chaos. Only a few
theoretical papers addressed that issue. Early work by Whitehead {\it
et al.} (1978), Verbaarschot and Brussard (1979), Brown and Bertsch
(1984) and Dias {\it et al.} (1989) addressed the validity of the
Porter--Thomas distribution for shell--model eigenfunctions, see the
comments below. The relation between spectral fluctuations and the
shell model was addressed by Weidenm\"{u]ller} (1985).  Meredith {\it
et al.} (1988) and Meredith (1993) studied the Lipkin--Meshkov--Glick
model: ${\cal M}$ Fermions occupy a system with three non--degenerate
subshells each containing ${\cal M}$ degenerate single--particle
states. In each subshell, the single--particle states do not carry any
further quantum numbers. The two--body interaction acts only between
particles in different subshells. All non--zero matrix elements are
identical. The Hilbert space has finite dimension. Symmetries reduce
the problem to manageable size even when ${\cal M}$ is large. The
system possesses a classical limit which is attained for coherent
states when ${\cal M}
\to \infty$. Classical chaos can thus be studied in its dependence on
the strength of the two--body interaction. As that strength is varied,
a close correspondence is numerically established between the
transition from regular to chaotic motion and that from Poisson to GOE
statistics for the spectral fluctuation measures of the quantum
system. The Bohigas--Giannoni--Schmit conjecture was verified for the
first time for an interacting many--body system (albeit in the
framework of a toy model without characteristic ingredients of the
shell model such as conserved quantum numbers). In the 1990s, several
papers ({\AA}berg, 1990; Alhassid {\it et al}, 1990; Alhassid and
Whelan (1991); Martinez-Pinedo {\it et al.} (1997)) addressed chaos in
the collective model. This topic is reviewed in Section~\ref{collmod}.

Ormand and Broglia (1992) reported a study of quantum chaos in the
shell model for the $sd$--shell. That work displayed several features
which were also discussed in the paper by Zelevinsky {\it et al.}
(1996). We focus on that later, much more extensive paper. The
authors undertook a thorough and systematic theoretical study of chaos
in the $sd$--shell. They used the well--established Brown--Wildenthal
(1988) form of the residual interaction, which is free of random
elements. In most of their calculations Coulomb effects were
neglected. They focused attention mainly on nuclei in the middle of
the shell ($m = 12$) where the dimensions ${\cal D}(J, T)$ of the
shell--model matrices are manageable and yet sufficiently large for a
statistical analysis. (We drop the label $\Pi$ since all states in the
$sd$--shell have positive parity). The authors calculated spectra and
eigenfunctions of the shell--model Hamiltonian numerically and
compared the results both with GOE predictions and with thermal
averages. The latter were considered because chaos was suspected by
Percival (1973) to cause the many--body eigenfunctions all ``to look
the same'', so that concepts such as mean occupation numbers for
single--particle states and the Fermi--Dirac distribution may make
sense for observables averaged locally over energy.

The authors used ergodicity (Section~\ref{ergod}) to compare GOE
ensemble averages with running averages over the calculated spectra.
To obtain statistically significant results, they used all eigenvalues
and/or eigenfunctions pertaining to fixed $m$ and to a pair $(J, T)$
of quantum numbers. As pointed out before (Section~\ref{shell}),
calculations using only states from the valence shell, yield results
which are unrealistic in the upper part of the spectrum where
non--valence--shell states actually dominate. Therefore the work of
Zelevinsky {\it et al.} (1996) must be considered a case study of
quantum chaos within the shell model in a restricted Hilbert space,
rather than a realistic calculation of nuclear properties. While we
summarize the results of this work in the present Section, we defer a
detailed analysis of how chaos is generated in the shell model to
Section~\ref{tbre}.

The average level density (calculated from the actual spectrum by
smoothing) differs from that of the GOE and has approximately Gaussian
shape. That result was anticipated many years ago by Mon and French
(1975), see also Section~\ref{embe}. The nearly Gaussian shape is an
artifact, of course, and due to the restriction to a single major
shell; the actual nuclear level density grows nearly exponentially
with excitation energy. Results of primary interest in the present
context relate to quantum chaos and concern the local spectral
fluctuation measures. After unfolding of the spectra, the NNS
distribution and the $\Delta_3$--statistic were found to agree well
with GOE predictions in the middle of the shell and for sufficiently
large values of the matrix dimension ${\cal} D(J, T)$. For $m = 12$
and the states with $J =2, T = 0$ (${\cal D}(2, 0) = 3276$) this is
shown in Figs.~\ref{fig19} and \ref{fig20}. It is of interest to
follow the onset of chaos as the strength of $V_{\rm res}$ is
varied. For $V_{\rm res} = 0$ the NNS distribution is found to be
close to Poissonian as expected. As the strength is increased, the
distribution approaches the Wigner distribution and, within
statistics, becomes indistinguishable from that distribution already
for strength values amounting to $20$ \% of the actual one. For
smaller values of $m$ ($^{24}$Mg with $m = 8$) the NNS distribution at
full strength value is still intermediate between the Poisson and the
Wigner distribution, even for sizable values of $D(J, T)$ (D(0, 0) =
1161). Here, inclusion of the Coulomb interaction reduces the
probability of very small spacings. The $\Delta_3$--statistic in
Fig.~\ref{fig20} agrees with the GOE prediction except for large
values of $L$ where it rises above that prediction. That seems to be a
systematic trend and is not fully understood at present. It perhaps
signals a weakening of GOE--type correlations between eigenvalues with
spacings of more than $7$ MeV or so and might suggest that strong
mixing is restricted to unperturbed shell--model configurations with
energies within an energy interval of $10$ to $20$ MeV. We return to
this point below. (A later evaluation of $\Delta_3(L)$ for the $3^-$
states showed good agreement with the GOE prediction op to $L \approx
3000$, however, see Zelevinsky (2007).)

\begin{figure}[h]
\vspace{5 mm}
\includegraphics[width=\linewidth]{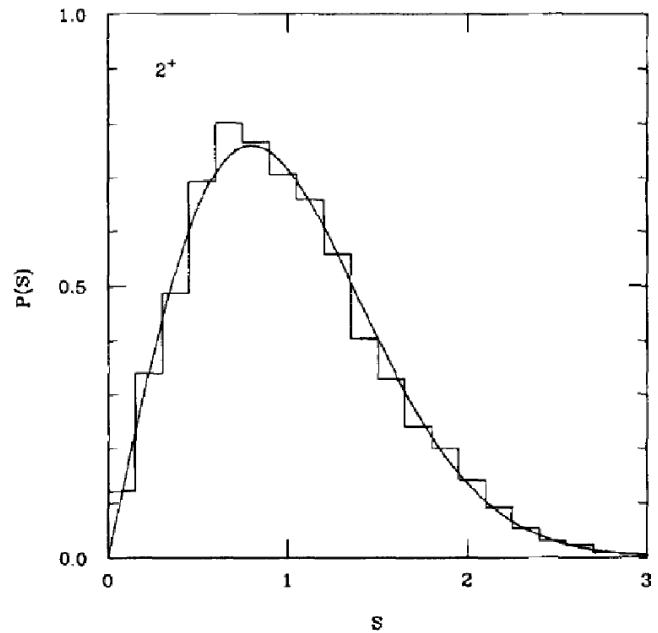}
\vspace{3 mm}
\caption{NNS distribution (histogram) and Wigner surmise (solid line)
for the states with $m = 12$ and $J = 2, T = 0$ in the $sd$--shell.
From Zelevinsky {\it et al.} (1996).}
\label{fig19}
\end{figure} 

\begin{figure}[h]
\vspace{5 mm}
\includegraphics[width=0.4\textwidth]{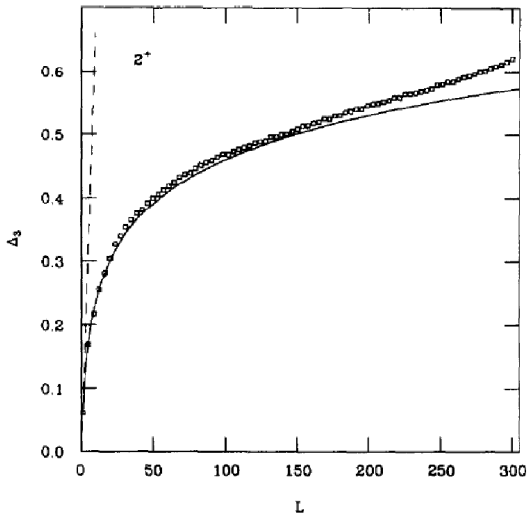}
\vspace{3 mm}
\caption{The $\Delta_3$--statistic (dots) and the GOE prediction
(solid line) for the states with $m = 12$ and $J = 2, T = 0$ in the
$sd$--shell. From Zelevinsky {\it et al.} (1996).}
\label{fig20}
\end{figure}

Extensive shell--model calculations have also been done for nuclei in
the $2p1f$--shell, especially by the Strasbourg--Madrid collaboration
(Caurier {\it et al.}, 2005). We are not aware, however, of an
equally thorough analysis of the results with respect to spectral
fluctuation measures and other indicators for quantum chaos as done
for the $sd$--shell by Zelevinsky {\it et al.} (1996). Kota (2001) has
analyzed the calculated spectrum of $4^+$ states in $^{48}$Ca
(Martinez-Pinedo {\it et al.}, 1997; Caurier {\it et al.}, 1999.)
Using the $1355$ states located in the middle of the spectrum (out of
a total of $1755$ levels), he finds good agreement with the Wigner
surmise for the NNS distribution, and with the GOE prediction for the
$\Delta_3$--statistic. The latter is reported only for $L \leq 60$.
Electromagnetic transition intensities and moments for $A \approx 60$
nuclei as calculated by Hamoudi {\it et al.} (2002) showed good
agreement with the Porter--Thomas distribution.

Investigation of the distribution of eigenfunctions offers additional
insight and shows the limitations of spectral fluctuation measures in
establishing chaos. Here the Porter--Thomas distribution is the
standard measure. In early calculations (Whitehead {\it et al.}, 1978;
Verbaarschot and Brussard, 1979; Brown and Bertsch, 1984) for the
$sd$--shell, it was found that the actual distribution of widths
differed from the Porter--Thomas form: There was an overabundance of
the largest and the smallest widths at the expense of those with
values closer to the median. This was ascribed to incomplete mixing of
the configurations especially in the wings of the unperturbed
spectrum: Too little mixing leads to admixtures with too small
amplitudes and leaves the original states too pure. The issue was
followed up in more detail by Zelevinsky {\it et al.} (1996). We
denote by $W_{k \mu}$ the square of the amplitude with which an
unperturbed shell--model configuration labeled $\mu$ is admixed into
an eigenstate of the shell--model Hamiltonian. The index $k$ labels
the eigenstates such that the associated eigenvalues $E_k$ increase
monotonically with $k$. The measure used was the ``information
entropy'' $S_k$, which is defined as
\be
S_k = - \sum_\mu W_{k \mu} \ln W_{k \mu} \ .
\label{21}
\ee
For complete mixing and large matrix dimension ${\cal D} \gg 1$,
wave--function normalization makes us expect $W_{k \mu} \approx 1 /
{\cal D}$ independently of $k$ and $\mu$: On average, all
configurations are equally mixed into every eigenstate. This
corresponds to a maximum value of $S_k = \ln {\cal D}$. That
naive estimate does not take into account the Porter--Thomas
distribution for the $W_{k \mu}$s. Doing so yields $S_k = \ln
(0.48 {\cal D} )$ for the maximum value of $S_k$ (Zelevinsky {\it et
  al.}, 1996). We
refer to that value as to the GOE limit. Incomplete mixing is bound to
reduce that value.

The measure~(\ref{21}) has the advantage that it allows for a detailed
study of configurational mixing versus the eigenfunction index $k$ and
thus offers more insight than afforded by the Porter--Thomas
distribution. This advantage has to be weighed against the shortcoming
that the measure depends on the representation chosen for calculating
the $W_{k \mu}$s. (Indeed, $S_k$ is not invariant under orthogonal
transformations of the basis of states $\{ \mu \}$, while the
Porter--Thomas distribution applies to every projection of the
eigenvectors of the GOE onto a fixed vector in Hilbert space). The
``natural'' representation chosen by Zelevinsky {\it et al.} (1996)
for $S_k$ is defined by the unperturbed shell--model
configurations. The implications and limitations of that choice are
discussed by Zelevinsky {\it et al.} (1996).

Fig.~\ref{fig21} shows the exponential of $S_k$ versus $k$ for the
same 3276 $J = 2, T = 0$ states as used in Figs.~\ref{fig19} and
\ref{fig20}. The GOE limit (1578) is almost reached in the middle of
the spectrum. In the wings, the mixing of states is fairly incomplete.
That pattern is generic and reinforces earlier findings (Whitehead
{\it et al},   
1978; Verbaarschot and Brussard, 1979; Brown and Bertsch, 1984; Dias
{\it et al.}, 1989).

\begin{figure}[h]
\vspace{5 mm}
\includegraphics[width=\linewidth]{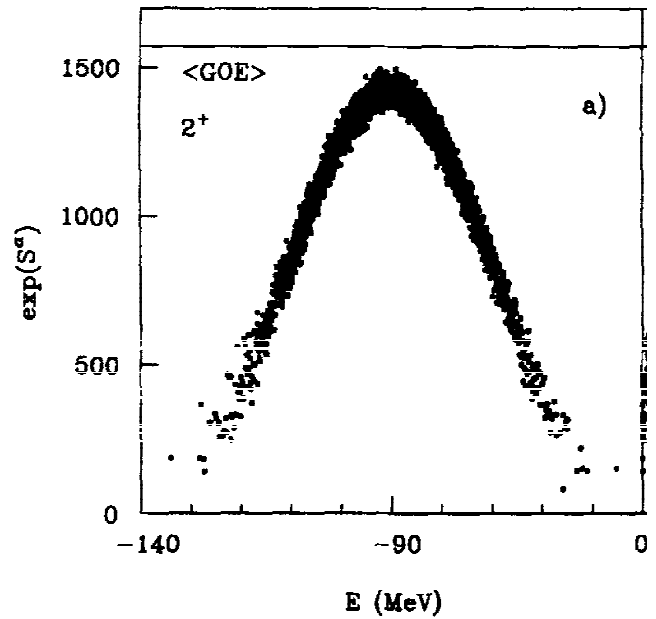}
\vspace{3 mm}
\caption{The exponential of the information entropy~(\ref{21}) versus
the energy of the state for the states with $m = 12$, $J = 2$ and
$T = 0$ in the $sd$--shell (dots). The GOE limit of 1578 is indicated
by the horizontal line. From Zelevinsky {\it et al.} (1996).}
\label{fig21}
\end{figure}

The eigenfunctions of the shell--model Hamiltonian are orthogonal and
normalized. In the GOE, these conditions impose weak correlations
among the expansion coefficients of the eigenfunctions in an arbitrary
basis which disappear for $N \to \infty$. The correlations for the
expansion coefficients of the shell--model Hamiltonian in the basis of
unperturbed shell--model configurations are found to be somewhat
larger than those of a GOE with the same dimension, especially in the
tails of the spectrum and for the $10$ or $20$ nearest eigenstates.

In Section~\ref{door} we introduced the concept of the strength
function for a doorway state. In analogy to Eq.~(\ref{13}), the
strength function for an unperturbed shell--model configuration $\mu$
with respect to the exact eigenstates of the shell--model Hamiltonian
is defined by $\sum_k W_{k \mu} \delta(E - E_k)$. Since no ensemble
averaging is involved in that definition, it is useful to average over
a group of neighboring (in energy) shell--model configurations to get
a smooth function. If the eigenstates were complete mixtures of
shell--model configurations, the smoothed strength function would be
constant (independent of energy $E$). In actual fact, the strength
functions are peaked with a full width at half maximum of about $20$
MeV even for shell--model configurations $k$ in the middle of the
unperturbed spectrum. The shape of the strength function is Gaussian
near its peak, but only falls off with an exponential tail in the
wings. Such exponential decay has been studied by Lewenkopf and
Zelevinsky (1994) and Frazier {\it et al.} (1996).

Another measure which shows that the eigenfunctions of the shell-model
Hamiltonian are not perfect mixtures of the unperturbed shell--model
configurations, is provided by the pairing force, an attractive
interaction between like nucleons. That force corresponds to a
particular linear combination of the interaction operators
$\sum_{\sigma \tau} A^{\dag}_{j_3 j_4 s t \sigma \tau} A^{}_{j_1 j_2 s
t \sigma \tau}$ appearing in Eq.~(\ref{19}). For a completely mixed
system, a plot of the expectation values of that linear combination
(taken with respect to the eigenstates of the shell model Hamiltonian)
versus $\mu$ is expected to fluctuate about a constant mean value. In
actual fact, the plot shows a systematic enhancement (with respect to
the mean value) of about $70$ \% in the ground--state domain at the
expense of a corresponding suppression at the upper end of the
spectrum, with small fluctuations.

The complete mixing of the eigenfunctions due to chaos may be similar
to thermalization. As a test, Zelevinsky {\it et al.} (1996) 
calculated the occupation numbers of the three single--particle states
$d_{5/2}$, $s_{1/2}$, and $d_{3/2}$ in the exact eigenstates, and
plotted the result versus $\mu$. The data have small fluctuations.
With the help of a properly defined temperature, the results can be
very well fitted with the Fermi distribution. The fit parameters are
``effective'' single--particle energies. These differ by only a few
$100$ keV from the input parameters (the single--particle energies
$\ve_{\ell j}$ of the shell model). This corroborates the picture of
thermalization.

In summary: If chaos is measured in terms of the usual spectral
fluctuation measures (NNS distribution and $\Delta_3$--statistic),
there is evidence that the residual interaction mixes the unperturbed
shell--model configurations sufficiently strongly to produce
chaos. This is true for nuclei with sufficiently many valence nucleons
and in the middle of the spectrum. Chaos is diminished for nuclei with
a smaller number of valence nucleons (or of holes in the valence
shell). Closer inspection of the eigenvectors offers a more subtle
picture. Even for nuclei in the middle of the shell, the mixing of the
unperturbed shell--model configurations is not complete, especially in
the tails of the spectrum. Small correlations between eigenvector
components exist beyond the ones imposed by orthogonality and
normalization. The strength functions of the unperturbed shell--model
configurations are not constant, but are peaked with a width of twenty
MeV or so. The expectation values of the operator of the pairing force
are enhanced in the ground--state domain. These deviations from GOE
properties may be related to the fact that the $\Delta_3$--statistic
shows an upward bend for large values of $L$. All this shows that
within a single major shell, chaos is not fully attained for realistic
strengths of the interaction, although it would be difficult to detect
such deviations experimentally (except for the behavior of $\Delta_3$).
In spite of the deviations, thermodynamic concepts apply, and the
single--particle occupation numbers follow the Fermi distribution
function.

We return to the questions posed at the beginning of this Section.
From the evidence presented, it seems likely that calculations that
would allow for the presence of non--valence--shell states would show
that as the excitation energy increases, nuclear levels attain an ever
greater similarity to GOE eigenstates. The approach to the GOE limit
is probably somewhat faster for nuclei in the middle between two
closed shells than for nuclei near closed shells. Chaos thus seems a
natural ingredient of the shell model, and not an exclusive
alternative to regular motion as seen in the ground--state domain. The
results on the strength function for the unperturbed shell--model
configurations and on the $\Delta_3$--statistic suggest that at
excitation energy $E$ the strong mixing (which is characteristic of
quantum chaos) involves only those unperturbed shell--model
configurations which are located in an energy interval centered at $E$
and $10$ to $20$ MeV wide.

The very success of the shell model, i.e., the ability of the model to
account for many basic features of nuclei, implies that many--body
states pertaining to different major shells are mixed only weakly. In
that sense the residual interaction is weak: It removes the numerous
degeneracies of the single--particle model in a manner that causes
nearly complete mixing of states within a major shell. However, the
residual interaction is not strong enough to destroy the overall shell
structure defined by the existence of major shells. In that sense,
nuclei are not fully chaotic systems. This point of view has been
emphasized by Bunakov (1999). 

Chaos does not preclude the existence of regular features. These are
seen in the ground--state domain where the mixing of unperturbed
shell--model configurations is incomplete, and in the presence of
collective modes of excitation which act as doorway states and are
seen as giant resonances. Chaos is helpful because it allows for the
description of average properties of excited nuclei in terms of
concepts of equilibrium statistical mechanics. It remains to show by
which mechanism chaos originates in the shell model, and to clarify
whether chaos is a generic feature of the shell model or depends upon
specific properties of the residual interaction.  We return to these
questions in Section~\ref{tbre}.

An analysis similar to the one in $sd$--shell nuclei described above
was carried out for the Ce atom by Flambaum {\it et al.} (1994) with very
similar results. This supports the view that chaos is a generic
property of self--bound many--body systems.

\subsubsection{Limits of Validity of RMT in Nuclei}

No real physical system can be expected to possess spectral
fluctuation properties that coincide exactly with RMT predictions.
Indeed, RMT is based upon a purely stochastic approach, and we must
expect that at some point, system--specific features dominate spectral
properties. Which then are the limitations of RMT in nuclei?

The answer to that type of question is known in systems with few
degrees of freedom (Berry, 1985) and in disordered systems (Imry,
2002).  For chaotic systems with few degrees of freedom, the
limitations of RMT are connected to the shortest periodic orbit in the
system. With $\tau_{\rm min}$ the period of the shortest periodic
orbit, $\Delta E = \hbar / \tau_{\rm min}$ defines the maximum energy
interval within which RMT predictions can be expected to hold. In
disordered systems, the period $\tau_{\rm min}$ is replaced by the
diffusion time $\tau_{\rm diff}$, and the characteristic energy
interval is given by the Thouless energy $E_c =
\hbar / \tau_{\rm diff}$. The dimensionless Thouless conductance $g$
is the ratio of either of these intervals and the average level
spacing and gives the number of levels over which RMT predictions are
expected to hold.

In nuclei, the situation is not completely clear, and two different
schools of thought exist. Bohigas and Leboeuf (2002) used a
mean--field approach. It is argued that the mean--field motion is
partly chaotic. The shortest periodic orbit at the Fermi energy is
used to estimate the characteristic energy interval as $\Delta E =
77.5 A^{- 1/3}$ MeV. The approach was worked out further by Olofsson
{\it et al.} (2006). However, the mean--field approach addresses
single--particle properties only. It is relevant for one--body chaos
but is not clearly related to chaos in a many--body system: Without
two-- or many--body interaction the eigenvalues of the many--body
system are sums of single--particle energies and have a Poisson
distribution irrespective of whether the single--particle motion is
regular or chaotic.

A different view was taken, for instance, by Bunakov (1999) and
Molinari and Weidenm\"{u}ller (2006), who argue that the
independent--particle model gives rise to regular motion, while the
residual interaction causes mixing of shell--model configurations and,
thus, chaos. The characteristic energy interval over which
configurations are strongly mixed (and the range of energies over
which RMT predictions apply) is then given by the spreading width. In
Section~\ref{chaossm} it was shown that in shell--model calculations,
the spreading width is found to be of the order of $10$ MeV. While
that number is not substantially different from the estimate obtained
within the chaotic mean--field scenario, the origins of both estimates
are clearly very different. A test of these predictions is not
possible using experimental data. As pointed out before, sufficiently
long sequences of levels with identical quantum numbers are not
known. The shell--model calculations reported on in
Section~\ref{chaossm} seem to lend substance to the second view.

\subsection{Collective Models}
\label{collmod}

\subsubsection{The Collective Model of Bohr and Mottelson}
\label{coll}

Nuclei can undergo shape deformations. This fact became obvious with
the discovery of nuclear fission in 1939: A very heavy nucleus splits
spontaneously into two fragments of about equal mass. The energy
liberated in that process could be roughly understood on the basis of
the liquid--drop model. In that model, nuclei are described as charged
droplets of an incompressible fluid held together by surface tension.
The latter mimics the attractive nuclear force.

A dynamical theory of surface deformations was developed by Bohr (1951
and 1952) and Bohr and Mottelson (1952) in the early 1950s. In an
expansion of the shape of the nuclear surface in terms of spherical
harmonics $Y_{\ell \mu}$, attention is focused on the lowest (in
$\ell$) non--trivial terms. These are the quadrupole terms $Y_{2 \mu}$
(the term with $\ell = 0$ is ruled out because of volume conservation
and the terms with $\ell = 1$ describe the motion of the center of
mass). For small values of the five expansion parameters $\alpha_{2
\mu}$ with $\mu = -2, -1, 0, +1, +2$, the surface has the shape of an
ellipsoid.  The principal axes of that ellipsoid define an intrinsic
(``body--fixed'') coordinate system. The five expansion parameters
$\alpha_{2 \mu}$ can be transformed into the three Euler angles which
specify the location of the body--fixed system with respect to the
laboratory system, and two parameters (commonly called $\beta$ and
$\gamma$) which specify the nuclear shape in the body--fixed
system. The potential energy $V(\beta, \gamma)$ defines the static
energy of quadrupolar nuclear shape deformations. A dynamical theory
is obtained by considering the parameters $\alpha_{2 \mu}$ (or their
transforms) as dynamical variables which obey bosonic commutation
relations. The resulting Hamiltonian (``Bohr Hamiltonian'') has a
number of parameters (the ``masses'' connected with the kinetic energy
terms and the parameters specifying the potential energy $V(\beta,
\gamma)$). Depending on the values of these parameters, the theory
predicts {\it inter alias} rotational motion and vibrational motion of
nuclei. In even--even nuclei, rotational motion manifests itself in
the occurrence of rotational bands (sequences of levels with spins $J
= 0,2,4,\ldots$). The excitation energies (above the band head) of the
states with spin $J$ in the band are proportional to $J(J+1)$. The
eigenfunctions are obtained by projecting the deformed intrinsic state
onto fixed angular momentum $J$. This is done using the Wigner ${\cal
D}$ functions and integrating over Euler angles. The electromagnetic
transitions within the band are electric quadrupole ($E2$) and are
strongly enhanced over simple single--particle (i.e., shell--model)
estimates. The transition matrix elements are proportional to the
static quadrupole moment of the intrinsic deformed state. In
Fig.~\ref{fig22}, two such rotational bands are displayed. Vibrational
motion manifests itself in (nearly) harmonic vibrations of the surface
about its equilibrium shape and is characterized by an
harmonic--oscillator--like spectrum. The electromagnetic transition
matrix elements are also enhanced over single--particle estimates but
not as much as in the rotational case. This ``collective model'' (so
named because many nucleons partake in an orderly way in the motion)
of Bohr and Mottelson has been extremely successful in accounting for
many spectroscopic data (Bohr and Mottelson, 1969).
 In cases of pure rotational and
pure vibrational motion we expect, of course, a predominance of
regular features. This view is supported by the empirical evidence
reviewed in Sections~\ref{analow} and \ref{anahss}.

\begin{figure}[h]
\vspace{5 mm}
\includegraphics[width=\linewidth]{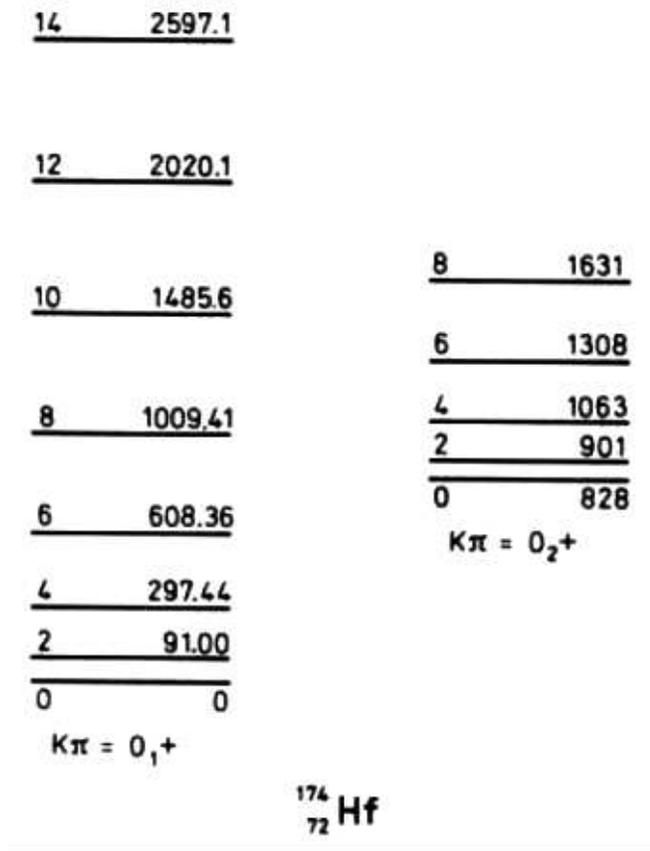}
\vspace{3 mm}
\caption{Two rotational bands in $^{174}$Hf. Adapted from Bohr and
Mottelson (1969) Vol. 2, p. 168.}
\label{fig22}
\end{figure}

\subsubsection{Onset of Chaos in Rapidly Rotating Nuclei}
\label{chaosyrast}

Studies of chaos in the collective model ({\AA}berg, 1990; Matsuo {\it
et al.}, 1997) have addressed the onset of chaos above the yrast line
for states of high spin. (The yrast line is defined in
Section~\ref{tests}. As a function of $J$, it is given by the energy
of the lowest level with spin $J$).  We review first the later, more
extensive paper by Matsuo {\it et al.} (1997) and then the earlier
work by {\AA}berg (1990). In both papers, very similar techniques were
used and similar results were obtained. The starting point is a
generalized shell model. Chaos in these approaches is due to the
residual interaction which mixes the basic shell--model
configurations. The approach is thus similar to that of
Section~\ref{shell}. However, the shell--model used differs from the
spherical shell model of Section~\ref{shell}. This ``Nilsson model''
takes into account the fact that rotational motion in medium--weight
and heavy nuclei is due to deformations of the nuclear shape. Thus the
single--particle Hamiltonian $H_{\rm Nilsson}$ of the Nilsson model
contains, in addition to the kinetic energy and a spin--orbit coupling
term, a non--spherical, elliptically deformed single--particle
potential. The parameters of that potential are obtained by fits to
the data. (We mention in passing that deformed single--particle
potentials with a pure quadrupole deformation do not give rise to
chaotic single--particle motion. An additional octupole deformation is
needed. Even then, chaos arises only in the oblate case (Arvieu {\it
et al.}, 1987; Heiss {\it et al.}, 1994).

The obvious difficulty with this approach is that by construction,
$H_{\rm Nilsson}$ is not rotationally invariant. This is not a problem
in principle. We recall that according to the collective model (see
Section~\ref{coll}) each rotational band is due to a deformed
``intrinsic state''. The wave function of each band member (which has
definite spin $J$) is obtained by projecting the intrinsic state onto
spin $J$. In the same sense, the many--body eigenstates of $H_{\rm
Nilsson}$ (Slater determinants) are viewed as microscopic realizations
of intrinsic states. Projection onto states of good total spin will
generate from each such Slater determinant the members of a rotational
band.

However, in practice projection of Slater determinants is cumbersome, 
and an approximation (the ``cranking model'') is used. It is
postulated that in the laboratory system, the deformed
single--particle potential rotates (is ``cranked'') about some axis
with fixed frequency $\omega$. That axis must be perpendicular to the
symmetry axis of the potential (since in quantum mechanics, rotation
about a symmetry axis is not possible). Under a coordinate
transformation from the laboratory to the body--fixed frame of
reference, the rotating deformed single--particle potential becomes a
static deformed potential. As in classical mechanics, that coordinate
transformation induces an additional term in the Hamiltonian, and the
single--particle Hamiltonian of the cranking model is
\be
H_{\rm cranking} = H_{\rm Nilsson} - \vec{\omega} \cdot \vec{j} \ .
\label{24}
\ee
Here $\vec{\omega}$ points in the direction of the axis of rotation,
and $\vec{j}$ is the total spin of the nucleon. The $z$--direction of
the body--fixed system is commonly assumed to coincide with the
symmetry axis, and the direction of $\vec{\omega}$ is assumed to
coincide with the $x$--axis, so that the last term in Eq.~(\ref{24})
(comprising both Coriolis and centrifugal forces) takes the form
$\omega j_x$.

The many--body solutions of the cranking Hamiltonian are Slater
determinants. Let $| 0 \rangle$ be the vacuum state, and let
$a^\dag_i$ be the creation operator for the cranked single--particle
state labeled $|i(\omega)\rangle$. Then, $|\mu \rangle = \prod_i
a^\dag_i | 0 \rangle$ is a Slater determinant of cranked states (all
taken at the same frequency $\omega$). The label $\mu$ represents a
set of occupied orbitals. Different choices of $\mu$ correspond to the
ground state and to the excited states of the cranking model. It is
assumed that the deformed potential is the same for all of these
states. This assumption is realistic up to excitation energies of
several MeV for nuclei for which the potential energy of deformation
displays a deep and stable minimum. The states $| \mu \rangle$ depend
on the cranking frequency $\omega$. Taken as functions of $\omega$,
the single--particle energies of the cranking model display avoided
crossings. At such crossings, the adiabatic single--particle wave
functions change abruptly. To obtain states $| \mu \rangle$ which
depend smoothly on $\omega$, a diabatic single--particle basis is used
by {\AA}berg (1990) and Matsuo {\it et al.} (1997).

To generate states of (approximate) total spin $J$ from a given state
$| \mu \rangle$, the consistency condition $\langle J_x
\rangle(\omega) = J$ is used to determine $\omega$. Here $J_x$ is the
$x$--component of the total spin operator (the sum of the spins of all
nucleons). In other words, the cranking frequency $\omega$ is adjusted
so that the system rotates on average with the desired spin $J$. In
Matsuo {\it et al.} (1997) the average $\langle J_x \rangle$ is the
thermal average over all cranked single--particle states taken with a
temperature $T = 0.4$ MeV. This temperature corresponds to the
excitation energies of interest (about $2$ MeV above yrast). The
states with different spins generated in this way form a rotational
band, with the state $| \mu \rangle$ playing the role of the intrinsic
state. This is seen by using a Taylor expansion of the expectation
value $\langle \mu | H_{\rm cranking} | \mu \rangle$ in powers of
$\omega$. That same expansion is used to transform the intrinsic
energies back to the laboratory system.

The states of fixed $J$ generated in this fashion are orthogonal. They
are mixed by the residual interaction. In Matsuo {\it et al.}, (1997) a
two--body interaction of the surface delta type was used (the
interaction is confined to an infinitely thin layer of the nuclear
surface). The interaction strength was determined by previous studies
of rotational nuclei. The resulting Hamiltonian contains the
eigenvalues of the cranking model in the laboratory system as diagonal
elements and the matrix elements of the residual interaction. To
obtain a manageable problem, only the lowest $1000$ states $| \mu
\rangle$ were used. The resulting lowest $300$ eigenstates of the
total Hamiltonian were found to be rather stable against that
truncation. These are used in the statistical analysis.  They cover a
region of excitation energy up to about $2.4$ MeV above yrast.

After unfolding, the spectra were binned. The NNS distribution and the
$\Delta_3$--statistic show a gradual transition from near Poissonian
behavior in the lowest bin to near GOE behavior in the highest one.
For the NNS distribution, this is indicated by the Brody parameter
(see Section~\ref{gen}), which increases monotonically with excitation
energy $U$ and reaches values close to unity at the upper end of the
spectrum. The dependence of the Brody parameter on $U$ is the same for
all spin values studied. The $\Delta_3$--statistic agrees with the GOE
value only up to a maximum value of $L$. That value increases with
increasing $U$. Even for the bin containing the $50$ states with
highest excitation energies, however, this maximum value is as small
as $6$. This shows that in the cranking model, spectral stiffness is a
local phenomenon. We recall that similar features (although for much
larger values of $L$) are found in the spherical shell model, see
Section~\ref{chaossm}.

Essentially the same model was used in {\AA}berg (1990). However,
rather than studying the mixing of rotational bands at fixed
interaction strength as a function of excitation energy, the
excitation energy was held fixed and the strength $\Delta$ of the
interaction was varied. This procedure simulates the increase of level
density with increasing excitation energy. The results for the
$\Delta_3$--statistic were analyzed by writing $\Delta_3$ as the sum
of the GOE expression~(\ref{9}) (taken at a scaled length $qL$ with $0
< q < 1$) and of a term linear in $L$ (as for the Poisson
distribution) with relative weight $(1 - q)$. The fit parameter $q$
was determined as a function of the strength $\Delta$ of the two--body
interaction. The result is shown in Fig.~\ref{fig23}. The
fragmentation of collective strength was also investigated. The
cranking frequency depends upon $J$ and so do, therefore, the
single--particle energies of the cranking model. As a result, states
with different values of $J$ are mixed differently by the residual
interaction. This implies that the collective $E2$ transition strength
(which for unmixed rotational bands only connects states within the
same band) becomes fragmented. As a measure of the distribution of the
reduced matrix elements for $E2$ decay connecting a mother state with
spin $J$ and the daughter states with spin $J - 2$, the standard
deviation is used. An average value $\overline{\sigma}_{E2}$ for the
standard deviation is obtained by averaging over $50$ mother states.
Figure~\ref{fig23} also shows $\overline{\sigma}_{E2}$ versus
$\Delta$, the strength of the two--body interaction. Also shown is
$1/2 \ \Gamma_{\rm rot}$ ({\AA}berg, 1990; Matsuo {\it et al.},
1997), the half width of the average distribution function of the
matrix elements. That function changes from Gaussian to Breit--Wigner
form as $\Delta$ increases.  This fact explains why as a function of
$\Delta$, $1/2 \ \Gamma_{\rm rot}$ displays a maximum. We observe that
as functions of $\Delta$, $q$ and $\overline{\sigma}_{E2}$ behave very
similarly. This shows that the fragmentation of the $E2$ transition
strength is an indirect measure of chaos.

\begin{figure}[h]
\vspace{5 mm}
\includegraphics[width=\linewidth]{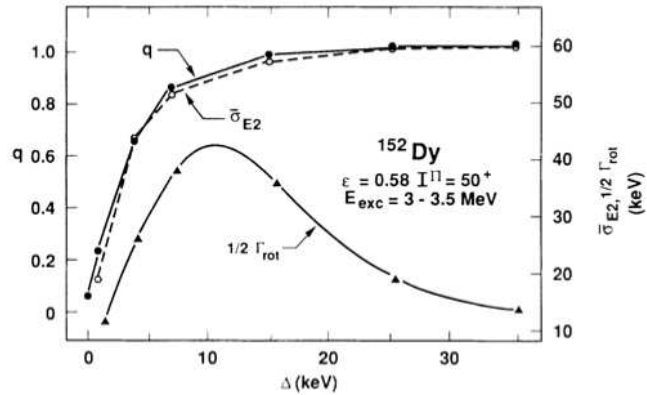}
\vspace{3 mm}
\caption{Mixing parameter $q$ for the $\Delta_3$--statistic (see text)
(left--hand scale), average standard deviation $\overline{\sigma}_{E2}$
of the distribution of reduced $E2$ matrix elements (right--hand scale),
and half width at half maximum $1/2 \Gamma_{\rm rot}$ (right--hand
scale) for the same distribution, all versus $\Delta$, the strength of
the two--body interaction which mixes the rotational bands. From
{\AA}berg (1990).}
\label{fig23}
\end{figure}

The theoretical ideas of {\AA}berg (1990) and Matsuo {\it et al.}
(1997) were applied to data by Stephens {\it et al.} (2005).  A beam
of $^{48}$Ca ions of 215 MeV hit a target of $^{124}$Sn. The resulting
isotopes of Yb decay by gamma emission.  Pairs of gamma quanta were
measured in coincidence. Numerical simulations along the lines
described above were compared with the measured spectra. The fits were
used to determine the ratio of the strength $\Delta$ of the
nucleon--nucleon interaction versus the mean level spacing. The ratio
covers the range from $0.15$ (nearly fully ordered) to $1.5$ (nearly
fully chaotic), in an energy interval which is consistent with the
theoretical work.

Some authors ({\AA}berg, 1992; Mottelson, 1992) have proposed that at
excitation energies of a few MeV above the yrast line, the Coriolis
term in Eq.~(\ref{24}) may be rather weak. Then rotational bands would
exist with completely mixed wave functions of the GOE
type. Nevertheless, the collective $E2$ decay would take place
entirely within each band.

On the basis of a representation using both quasiparticles and
phonons, order, chaos, and the order--to--chaos transition are
discussed by Soloviev (1995).

\subsubsection{Chaos in the Interacting Boson Model}
\label{chaosibm}

The shell model is generally considered the fundamental
phenomenological nuclear--structure model (Caurier {\it et al.},
2005). However, applications of the model have been restricted in
practice to nuclei for which the number of valence nucleons is not too
large. This is the case for all nuclei in the $1p$--shell and the
$2s1d$--shell and, more recently, for most nuclei in the
$2p1f$--shell. For yet heavier nuclei (and thus shells beyond the
$2p1f$--shell), the number of valence nucleons for nuclei near the
middle of the shells is simply too large, and a shell--model
calculation is prohibitively difficult. At the same time, it is in
these mass regions that the collective model finds its most successful
application. This fact has prompted many authors to look for a
derivation of the collective model from the shell model by introducing
suitable collective variables. We mention, in particular, the idea of
using a boson expansion within the shell model to obtain a simplified
description with built--in collective features. In this approach,
pairs of nucleons form boson--like entities. None of these approaches
was truly successful in providing a derivation of the Bohr Hamiltonian
from the shell model within controlled approximations.  However, in
the midst of these efforts a phenomenological approach emerged which
has become eminently successful, the interacting boson model (IBM) of
Arima and Iachello (1975). The IBM postulates the existence of
$s$--bosons and $d$--bosons. These bosons may be thought of as
representing pairs of nucleons coupled to spin $0$ and $2$,
respectively. Alternatively, the $d$--bosons may be viewed as the five
quanta of quadrupole surface deformations. The $s$--boson is then an
artifice which is used to simplify the mathematics. For each nucleus,
the total number $N$ of bosons is fixed in the IBM. In the limit $N
\to \infty$, the solutions of the boson Hamiltonian approach those of
the collective model (Dieperinck {\it et al.}, 1980). The connection
between the IBM and the shell model is discussed by Iachello and Talmi
(1987). The IBM at large is reviewed by Iachello and Arima (1987).

In the original form of the IBM (often referred to as IBM 1) no
distinction is made between nucleon pairs formed of protons and pairs
formed of neutrons. This is the form of the IBM for which an analysis
with regard to chaotic properties has been performed (Alhassid {\it et
al.}, 1990; Alhassid and Whelan, 1991).  To introduce the model, we
defined $s^\dag$ and $d^\dag_\mu$ as the creation operators for the
$s$--boson and the five $d$--bosons, respectively. While $s^\dag$ is a
scalar, the $d^\dag_\mu$ transform under rotations as the components
of an irreducible tensor of rank $2$. The modified annihilation
operators $\tilde{s} = s$ and $\tilde{d}^{}_\mu = (-)^\mu d^{}_\mu$
have the same transformation properties. The symbol $(d^\dag \times
\tilde{d}^{})^{(k)}_\kappa$ denotes the irreducible tensor of rank $k$
with spherical components $\kappa$ obtained by vector--coupling
$d^\dag$ and $\tilde{d}$, and likewise for $s^\dag$ and $\tilde{s}$. A
quadrupole operator $Q(\chi)$ is defined by $Q(\chi) = (d^\dag \times
\tilde{s} + s^\dag \times \tilde{d})^{(2)} + \chi (d^\dag \times
\tilde{d})^{(2)}$ where $\chi$ is a real parameter. Coupling $Q(\chi)$
with itself to an irreducible tensor of rank zero generates the scalar
$Q(\chi) \cdot Q(\chi)$. In the spherical representation, the three
components of the angular--momentum operator $\vec{L}$ are defined by
$\sqrt{10} (d^\dag \times \tilde{d})^{(1)}_\kappa$, with $\kappa = -1,
0, +1$. With $\hat{n}_d$ the number operator for $d$--bosons, the IBM
Hamiltonian studied in Alhassid and Whelan (1991) is
\be
H = E_0 + c_0 \hat{n}_d + c_2 Q(\chi) \cdot Q(\chi) + c_1 \vec{L}^2 \ .
\label{22}
\ee
The Hamiltonian conserves the total number $N$ of bosons and depends
on the four parameters $c_0, c_1, c_2$, and $\chi$. (The energy $E_0$
is irrelevant as it only shifts the overall spectrum). Except for an
overall multiplicative scaling factor, only three parameters are
relevant for the structure of the spectrum. Aside from the energy of
the $s$--boson (which is put equal to zero) and the energy of the
$d$--bosons (given by $c_0$ with $c_0 \geq 0$ because the energy of
the $d$--bosons is generically found not to be below that of the
$s$--boson), $H$ contains two scalar boson--boson interaction terms.
The parameter $\chi$ appearing in $Q$ is empirically restricted by the
limiting symmetries described below. Moreover, there is an isomorphism
which maps $\chi$ onto $- \chi$. Therefore, $\chi$ has the range $-
\sqrt{7}/2 \leq \chi \leq 0$. The parameters $c_0$ and $c_2$ have
opposite signs (the quadrupole--quadrupole is attractive). The total
angular momentum is a good quantum number; the value of the parameter
$c_1$ only defines the positions of states with different angular
momenta with respect to one another and is irrelevant when chaos is
investigated. Within the model, quadrupole transition matrix elements
are calculated with the help of the same quadrupole operator $Q(\chi)$
that also appears in the Hamiltonian.

The IBM is an algebraic model and therefore easy to solve. Fitting
data is also easy because of the small number of parameters in
Eq.~(\ref{22}) and in other versions of the model. The fits obtained
are very good and cover a wide range of nuclei within a major
shell. The parameters change slowly with mass number. The connection
between the IBM and the shell model is understood reasonably well
although it has not been possible yet to derive the IBM parameters
from the shell model. All this suggests that the IBM encapsulates
real properties of nuclei and explains the success and popularity of
the model. Studies of chaos have been restricted to the
form~(\ref{22}) of the model which applies to even--even nuclei. It is
possible, however, to extend both the Bohr--Mottelson model and the
interacting boson model to nuclei with an odd number of protons and/or
neutrons.

Tests of chaos in the IBM face the same difficulty as do tests of
chaos in the nuclear shell model using only a single major shell: The
IBM is well established as a tool to describe low--lying states. At
higher excitation energies we expect to find states which cannot be
modelled by the IBM. In tests of GOE predictions one needs large data
sets and typically uses the entire spectrum of states of the IBM
pertaining to fixed quantum numbers, thereby exceeding the domain of
applicability of the IBM to real nuclei.

The IBM has the advantage that domains (in the space of parameters) of
full integrability coincide with symmetries of the Hamiltonian. Three
such symmetries exist. They are found using group--theoretical
arguments. Let $| 0 \rangle$ denote the vacuum state. The six
single--boson states $s^\dag | 0 \rangle$ and $d^\dag_\mu | 0 \rangle$
span a six--dimensional space. Under the general unitary
transformations in six dimensions, i.e., under the group $U(6)$, that
space transforms onto itself. The generators of the Lie algebra of
$U(6)$ are the operators $b^{\dag}_i b_j$ where $b_i$ stands for one
of the boson operators $s$ or $d_\mu$. Since $H$ is constructed from
scalar quantities involving these operators, $H$ has non--vanishing
matrix elements only between pairs of states that belong to the same
irreducible representation of $U(6)$.  Equivalently, the eigenstates
of $H$ form bases of irreducible representations of $U(6)$. We deal
with bosons and thus require totally symmetric eigenstates. These
belong to the fully symmetric irreducible representations which are
characterized by the integer $N$, the total boson number. The three
symmetries of the Hamiltonian correspond to the three chains of
subgroups of $U(6)$ given by
\ba
U(6) \supset U(5) \supset O(5) \supset O(3) \ ; \ (I) \nonumber \\
U(6) \supset SU(3) \supset O(3) \ ; \ (II) \nonumber \\
U(6) \supset O(6) \supset O(5) \supset O(3) \ . \ (III)
\label{23}
\ea
Case $(I)$ is realized for $c_2 = 0$, case $(II)$ for $c_0 = 0$ and
$\chi = - \sqrt{7}/2$, and case $(III)$ for $c_0 = 0$ and $\chi = 0$.
Each of the three cases corresponds to a simple dynamical situation.
In case $(I)$ we deal with vibrational nuclei, in case $(II)$ with
rotational nuclei, and in case $(III)$ with $\gamma$--unstable nuclei
(instability with respect to the equilibrium value of $\gamma$). Full
integrability of the Hamiltonian suggests classical regularity and,
in the quantum regime, Poisson statistics for the eigenvalues.

Alhassid {\it et al.} (1990) and Alhassid and Whelan (1991) checked
these expectations. Using coherent states and the limit $N \to
\infty$, one reaches the classical limit (similarly as in Meredith
{\it et al.}, 1988.) The behavior of the trajectories in phase space
was studied for several values of the total angular momentum. For
fixed $\vec{L}^2$ the parameter $c_1$ is redundant. Moreover, the
Hamiltonian can be rescaled. This leaves two parameters, $\eta$
(defined in terms of the negative ratio of $c_0$ and $N c_2$, with the
factor $N$ originating in the classical limit of the quantum
Hamiltonian~(\ref{22})) with $0 \leq \eta \leq 1$ and $\chi$ with $-
\sqrt{7}/2 \leq \chi \leq 0$. The results are shown in
Fig.~\ref{fig24}. The three symmetries in relation~(\ref{22})
correspond to the three corners of the triangles shown: Case $(I)$
corresponds to $\eta = 1$, case $(II)$ to $\eta = 0$ and $\chi = -
\sqrt{7}/2$, and case $(III)$ to $\eta = 0$ and $\chi = 0$. The three
corners are surrounded by regions of no or weak chaos as
expected. Chaos is strongest for small angular momenta and in the
center and lower part of the triangle. The authors emphasize the
existence of a narrow strip of nearly regular motion connecting the
upper and the left--hand corners of the triangle.

\begin{figure}[h]
\vspace{5 mm}
\includegraphics[width=\linewidth]{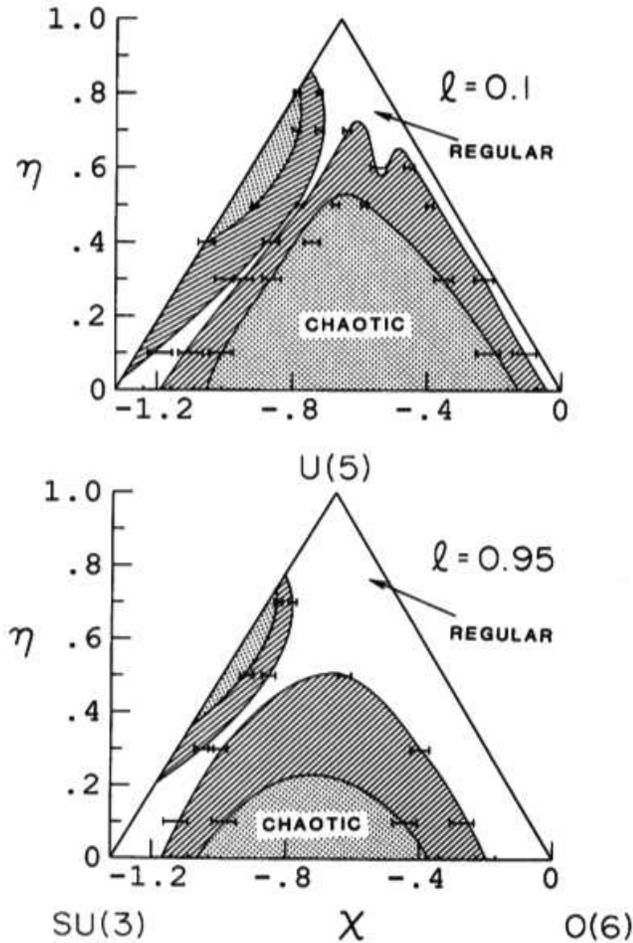}
\vspace{3 mm}
\caption{Domains of regular or chaotic motion in the
$\eta-\chi$--plane for two values of $\ell$, the angular momentum per
boson. The accessible parameter space has the shape of a triangle.
Three domains in parameter space are separated by solid lines: Domains
with nearly regular motion (fraction of chaotic phase--space volume
$\leq 0.3$; white areas); domains with nearly chaotic motion (fraction
of chaotic phase--space volume $> 0.7$; dotted areas); and the regions
in between. From Alhassid and 
Whelan  (1991).}
\label{fig24}
\end{figure}

The quantum spectra were analyzed with the help of the NNS
distribution and of the $\Delta_3$--statistic. The Porter--Thomas
distribution was checked using $E2$ transition probabilities. As a
result, the Bohigas--Giannoni--Schmit conjecture (see
Section~\ref{RMTchaos}) was confirmed once again, this time for a
system of interacting bosons.  More specifically, all three
fluctuation measures are close to the GOE values in the domains of
parameter space where the classical motion is close to fully chaotic,
and show significant deviations and a tendency toward Poissonian
behavior in the domains where the classical motion is almost
regular. To obtain good statistics, the results shown were for $N =
25$ bosons and $J = 2$ states. This is close to the angular momentum
value $\ell = 0.1$ shown in the upper part of Fig.~\ref{fig24}. The
authors stress that for smaller, more realistic values of $N$ similar
effects are seen. These IBM studies of chaos were extended to higher
energies and spins by including broken--pair degrees of
freedom~(Vretenar and Alhassid, 1992). The regular behavior predicted
for nuclei in the ``arc of regularity'' which separates vibrational
and rotational motion (see Fig.~\ref{fig24}), was recently confirmed
experimentally (Jolie {\it et al.}, 2004).

In summary, collective models also display quantum chaos. Regular
features dominate near the yrast line and at and near symmetries of
the Hamiltonian. Chaos is strongest in regions of parameter space that
are far removed from such symmetries.  This is consistent with the
analyses of data reviewed in Sections~\ref{analow} and \ref{anahss}.

\subsection{Special Issues}

In Sections~\ref{chaossm}, \ref{chaosyrast} and \ref{chaosibm} we have
shown that the dynamics of nuclei often produce spectral fluctuations
of the GOE type and thus quantum chaos. In the present Section, the
emphasis is different. We address a number of complex situations where
we cannot test whether RMT applies. Rather, we postulate that it does
and use RMT as a tool to model the physical system. This leads to
results which are used in analyzing data.

\subsubsection{Decay out of a Superdeformed Band}
\label{sdb}

The potential energy of deformation $V(\beta, \gamma)$ introduced in
Section~\ref{coll} differs from the deformation energy of the
liquid--drop model. The difference is due to the shell--structure of
nuclei. Nuclei with a completely filled major shell for neutrons
and/or protons (see Fig.~\ref{fig18}) are particularly stable. But
major shells well separated from each other in energy exist not only
for spherical nuclei. As the nucleus is being deformed, the energies
of the single--particle states change, and new major shells emerge for
certain values of the deformation parameters. This statement applies
quite generally for single--particle potentials including those with a
spin--orbit force. More generally, at fixed excitation energy the
density of single--particle states of the shell model shows
considerable fluctuations as a function of deformation. The same is
true of the sum of the energies of all occupied single--particle
states and, thus, of the deformation potential $V(\beta, \gamma)$.
Beyond the smooth dependence on deformation parameters due to the
liquid--drop model, $V(\beta, \gamma)$ therefore displays maxima and
minima that reflect the deformation dependence of the filled
single--particle states of the shell model. For those values of the
deformation parameters and for those mass numbers where a major shell
is filled, the nuclear binding energy is expected to have a maximum
and, consequently, the potential energy of deformation $V(\beta,
\gamma)$ is expected to have a minimum.

Strutinsky (1966, 1967, 1968) realized that shell closures and, more
generally, maxima and minima of $V(\beta, \gamma)$ are generic
phenomena that are not limited to small deformations. His
``shell--correction method'' is based on the semiclassical
periodic--orbit theory for independent--particle motion and allows the
calculation of a correction to the deformation energy of the
liquid--drop model. The correction is the difference between the
ground--state energy calculated from the actual single--particle level
density (with states filled up to the Fermi energy), and that obtained
from a smoothed level density. The method of calculation and many of
its results are summarized by Brack {\it et al.} (1972) and by Brack
and Bhaduri (1997).  A particular prediction was that in certain
nuclei and as a function of deformation, $V(\beta,
\gamma)$ would display a second pronounced minimum (in addition to the
absolute minimum defining the nuclear ground state). This is
schematically shown in Figure~\ref{fig25}.

\begin{figure}[h]
\vspace{5 mm}
\includegraphics[width=\linewidth]{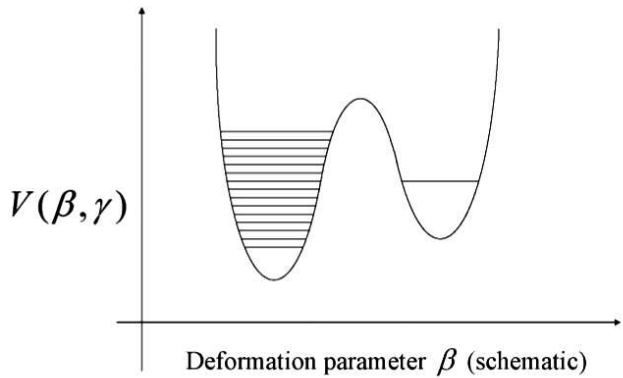}
\vspace{3 mm}
\caption{The deformation energy $V(\beta, \gamma)$ for fixed total
spin $J_0$ versus deformation (schematic). In some nuclei, $V(\beta,
\gamma)$ may display a second minimum at large deformation. The level
shown in the second minimum, a member of the SD band with spin $J_0$,
may decay either via gamma emission to the next lower level in the SD
band (not shown) or, via  tunneling, mix with 
 the ND states with spin
$J_0$ located in the first minimum.  The latter decay dominantly by
statistical $E$1 emission.}
\label{fig25}
\end{figure}

Following earlier evidence, the discovery of ``superdeformed''
rotational bands in $^{152}$Dy (Twin {\it et al.}, 1986) and other
medium--weight nuclei confirmed the existence of the second
minimum. The moment of inertia of a very strongly deformed nucleus is
larger than usual. A rotational band corresponding to an intrinsic
state located in the second minimum can therefore be identified by its
particularly large moment of inertia and the resulting narrow spacings
of its members.  The intensities of the $E2$ gamma transitions within
a SD band show a remarkable feature: The intraband $E2$ transitions
follow the band down with practically constant intensity. At some
point, the intraband transition intensity starts to drop sharply. It
either disappears abruptly or is much reduced in the next intraband
transition(s) and then disappears. This phenomenon is referred to as
the decay out of a SD band (Ragnarsson and {\AA}berg, 1986; Herskind
{\it et al.}, 1987).  It is attributed to the mixing of the SD
state(s) and the normally deformed (ND) states of the same spin and
parity located in the first minimum, see Fig.~\ref{fig25}.
Calculations using the shell--correction method show that the barrier
separating the first and the second minimum of the deformation
potential depends on and decreases with decreasing spin $J$. Decay out
of the SD band sets in at a spin value $J_0$ for which penetration
through the barrier is competitive with the intraband $E2$ decay.
Theoretical efforts aim at a quantitaive description of that process.
We describe here one of the theoretical approaches which is based on
the use of random--matrix theory.

The first minimum of the deformation potential is typically a few MeV
deeper than the second minimum. Therefore, the ND states populated by
decay out of the SD band have a typical excitation energy of $3$ to
$4$ MeV above yrast. These states are believed to decay largely via
statistical emission of $E1$ gamma quanta. Such decay would only
contribute to the background in the coincidence spectrometer used to
detect the rotational bands. This view is consistent with the fact
that no signal for the decay of the ND states has been found. In view
of the total lack of spectral information on the ND states, a
statistical model is used in the analysis of the data. It is assumed
that the spectral fluctuation properties of the ND states can be
modeled by the GOE.  This is the view taken by Vigezzi {\it et al.}
(1990a, 1990b) and Shimizu {\it et al.} (1992), and in most subsequent
work on the subject; see, however, D{\o}ssing {\it et al.} (2004). It
is the aim of these works to gain information on the barrier
separating the first and the second minimum through the analysis of
the decay data.

The quantity of central interest is the probability $I_{\rm out}$ for
decay out of the SD band. A plausible formula for $I_{\rm out}$ was
obtained by  Vigezzi {\it et al} (1990a, 1990b) and  
Shimizu {\it et al.} (1992) as follows. When one disregards
the coupling of both the SD states and the ND states to the
electromagnetic field, the resulting model for the Hamiltonian has the
form of Eq.~(\ref{12}), with $H_{0 \mu}$ the matrix elements for
penetration of the barrier separating the SD state $| 0 \rangle$ and
the $N$ ND states $| \mu \rangle$ with $\mu = 1, \ldots, N$. This is
in fact a doorway model for the SD state. The difference from the
usual doorway situation is that because of the barrier, the spreading
width $\Gamma^\downarrow$ is small compared to or at best of the order
of the mean level spacing $d$ of the ND states. Let $| m \rangle$ with
$m = 1, 2, \ldots, N + 1$ denote the eigenstates of $H$ and $c_m =
\langle 0 | m \rangle$ the amplitudes with which the SD state $| 0
\rangle$ is admixed into the eigenstates $| m \rangle$. It is assumed
that $E2$ decay out of the next--higher state in the SD band populates
the state $| m \rangle$ with probability $|c_m|^2$, and that the
widths for decay of the state $| m \rangle$ back into the SD band or
by statistical emission of $E1$ gamma rays are given by $|c_m|^2
\Gamma_S$ and by $(1 - |c_m|^2) \Gamma_N$, respectively. Here
$\Gamma_S$ is the width for electromagnetic decay within the SD band,
and $\Gamma_N$ is the common total decay width for $E1$ emission of
the ND states (it has the same value for every state $| \mu \rangle$).
We then have  (Vigezzi {\it et al.}, 1990a,  1990b; Shimizu {\it et
  al.}, 1992)
\be
I_{\rm out} = \sum_m |c_m|^2 \frac{ (1 - |c_m|^2) \Gamma_N}{ (1 -
|c_m|^2) \Gamma_N + |c_m|^2 \Gamma_S} \ .
\label{25}
\ee
To compare with the measured decay probability out of a SD band, the
ensemble average of this expression was numerically simulated by
putting $E_0$ in the middle of the GOE spectrum, by diagonalizing a
large number of random matrices of the form~(\ref{12}) and by using
the eigenfunctions to calculate the coefficients $c_m$. Repeating the
procedure for different strengths of the barrier penetration matrix
elements, one determines the actual strength by a fit to the data.

The arguments used to write down Eq.~(\ref{25}) imply a perturbative
treatment of the coupling to the electromagnetic field. As a result,
$I_{\rm out}$ depends on two dimensionless parameters, the ratio
$\Gamma_N / \Gamma_S$ and the ratio $\Gamma^\downarrow / d$ which
determines the coefficients $c_m$. Thus, $I_{\rm out}$ is independent
of the fine--structure constant. This is physically implausible
because once the SD state is populated, the competition is between the
intraband decay and the population of the ND states and thus between
the electromagnetic and the strong interaction. More importantly, the
analyses of data using the approach of Vigezzi {\it et al} (1990a,
1990b) and Shimizu {\it et al.} (1992) yielded values of
$\Gamma^\downarrow$ that were about two orders of magnitude smaller
than $\Gamma_N$, putting a perturbative treatment of $\Gamma_N$ into
question. In Weidenm\"{u}ller {\it et al.} (1998) and Gu and
Weidenm\"{u}ller (1999) the same statistical model~(\ref{12}) was used
in a non--perturbative way. The amplitudes for intraband decay of the
SD state and for $E1$ decay of the ND states were added as diagonal
terms on the right--hand side of Eq.~(\ref{12}). An expression for the
amplitude for decay out of the SD band was derived using that modified
Hamiltonian. The ensemble average of that amplitude was calculated
analytically. The resulting expression for the average probability for
decay out of the Sd band involves a threefold integral and depends on
the dimensionless parameters $\Gamma_N / d$ and $\Gamma_S /
\Gamma^\downarrow$. The result was compared with that of
Vigezzi {\it et al.}
(1990a, 1990b)and  Shimizu {\it et al.}  (1992)
 and the limits of validity of the
latter approach were determined. Analytical approximation formulas to
the exact result were given to simplify the data analysis. Many data
have been analyzed using either approach, see, for instance,
Kruecken {\it et al.} (2001).
 and references therein. A simplified treatment was
developed  by Stafford and Barrett (1999). 

A few figures taken from Kruecken (2001) may serve as examples for the
results obtained. Superdeformed bands occur not only for mass numbers
around $150$ (where they were first discovered) but also, for
instance, in the lead region ($A \approx 200$). A case in point is the
first SD band in $^{194}$Hg. For spin $J = 12 \hbar$, one finds
$\Gamma_S = 0.097$ meV, $\Gamma_N = 4.8$ meV, and $d = 16.3$ eV. The
analysis using Vigezzi (1990a, 1990b) yields $\Gamma^\downarrow = 37$
meV and that using Gu and Weidenm\"{u}ller (1999) yields
$\Gamma^\downarrow = 25$ meV. Calculating the rms barrier penetration
matrix element $v$ from $\Gamma^\downarrow$ and using results for
several values of $J$, one obtains a dependence on $J$ roughly in
agreement with the theoretical expectation $v \propto \exp ( - \alpha
J )$.

Decay out of a SD band continues to receive considerable attention,
see, for instance, Sargeant {\it et al.} (2005) and references
therein. Our aim here was to show how RMT is basically used in
analyzing data.

\subsubsection{Double Giant Dipole Resonance}
\label{secondGDR}

We recall the discussion of the Giant Dipole Resonance (GDR) at the
beginning of Section~\ref{door}. Action of the dipole operator on the
nuclear ground state induces the dipole mode. That mode can be viewed
as an oscillation of the center--of--mass of the protons against that
of the neutrons. This simple intuitive picture is exact in the
framework of two extremely opposite models of nuclear structure, the
harmonic--oscillator independent--particle model, and a collective
model using neutron and proton fluids. Therefore, the picture is
expected to have general validity. If the oscillation is approximately
harmonic, a repeated $E1$ excitation of the GDR should be possible and
should lead to the double giant dipole resonance (DGDR). The DGDR does
indeed exist and was first observed in a number of nuclei in the
1990s. For a review, see 
Bertulani and Ponomarev (1999).   Compared to predictions of
the harmonic picture, the measured cross sections for excitation of
the DGDR were found to be larger by factors ranging from $1.3$ to
$2$. The widths of the DGDR were found to be about $1.4$ times larger
than those of the GDR. This last fact is in keeping with some but not
all theoretical estimates. These findings have attracted much
theoretical attention. Anharmonicities of the Hamiltonian and
non--linearities of the external field were studied as possible causes
for the discrepancies.

Here we focus on an explanation originally due to Carlson (1999a,
1999b) which uses the Brink--Axel hypothesis. The Brink--Axel
hypothesis postulates the existence of a giant dipole resonance built
not only on the nuclear ground state, but also on every excited
nuclear state as well (Brink, 1955; Axel, 1962).  With the above
picture for the GDR as an oscillation of the center--of--mass of the
protons against that of the neutrons, the hypothesis is very
plausible. The hypothesis suggests an enhancement of the excitation
cross section of the DGDR by the following mechanism. The GDR is a
doorway state and mixes with the background states having the same
spins and parities as the GDR. The mixing time is simply estimated as
$\hbar / \Gamma^{\downarrow}$. If that mixing time is small or at most
of the order of the time it takes to excite the DGDR from the GDR,
then each of the background states admixed to the GDR mode may get
excited into its own GDR. The contributions from the excitation of the
background states would add to that of the GDR and lead to an
enhancement of the cross section. With a GOE model for the background
states as in Eq.~(\ref{12}), the average intensity for excitation of
the DGDR in a collision between two heavy ions can be worked out using
controlled approximations (Gu and Weidenm\"{u}ller, 2001). For
realistic values of the parameters, it is found that the contribution
of the background states is significant, and that the results of the
calculations agree well with data on the reaction $^{208}$Pb +
$^{208}$Pb at $640$ MeV/nucleon. The DGDR is, thus, another case where
RMT is successfully used to understand data. The use of RMT vindicates
the Brink--Axel hypothesis.

\subsubsection{Damping of Collective Modes and Friction}

Collective motion is due to the coherent motion of many nucleons.
Examples are the rotational or vibrational motion of nuclei
(Section~\ref{coll}) or the giant dipole resonance, an oscillation of
the center of mass of the neutrons against that of the protons
(Sections~\ref{door} and \ref{secondGDR}). Other examples are
encountered in heavy--ion collisions and in induced nuclear fission.
In the first case, the grazing collision of two heavy nuclei at
energies of a few MeV per nucleon (at the Coulomb barrier) leads to a
process known as deep inelastic scattering (N\"{o}renberg and
Weidenm\"{u}ller, 1980; Weidenm\"{u}ller, 1980): The kinetic energy of
relative motion and the associated angular momentum are partly
transformed into intrinsic excitation energy and spin of both
fragments. At the same time, nucleons are transferred between the two
reaction partners. As a result, the reaction mainly produces pairs of
fragments in highly excited states with masses similar to those of the
incident nuclei, but with considerably smaller kinetic energies.
Similar processes occur at higher incident energies (Cassing and
N\"{o}renberg, 1985).  The excitation energies of either fragment
cannot be measured precisely. A phenomenological description of the
process focuses attention on the collective degrees of freedom, i.e.,
on the relative coordinates of both fragments. These move under the
influence of a conservative potential (due to the Coulomb force and to
the overlap of both nuclei) and of dissipative forces. The latter
account for the transfer of energy and angular momentum into intrinsic
degrees of freedom as well as for nucleon transfer between both
reaction partners. In the simplest approximation, the relative motion
is described classically and the dissipative forces are represented by
a friction constant (Wilczynski, 1973).  Induced fission is a somewhat
similar process: Here the shape deformation leading to fission is
identified as the collective degree of freedom. For fission to happen,
that degree of freedom must overcome the fission barrier. Between the
top of the fission barrier and the scission point, the fission degree
of freedom gains kinetic energy. That energy is partly transformed
into intrinsic excitation energy of the fissioning system. The process
can again be described phenomenologically as involving dissipative
forces acting on the fission degree of freedom (Grang\'{e} and
Weidenm\"{u}ller, 1980).  The number of neutrons emitted prior to
fission serves as a measure of the strength of that dissipation
(Gavron {\it et al.}, 1986). In nuclear physics, dissipative processes
typically arise in the context of nuclear reactions. On the other
hand, the theoretical treatment usually does not refer to scattering
processes. That is why we deal we dissipation here and not in Part 2
of this review.

The description of dissipative forces due to the interaction of a
quantum system with its environment is a standard topic in
nonequilibrium statistical mechanics. If the $f$ collective degree(s)
of freedom $q_\nu$ with $\nu = 1,2,\ldots,f$ can be treated
classically, the phenomenological description of the collective motion
may use an equation of the type
\be
\frac{{\rm d}}{{\rm d}t} \frac{\partial {\cal L}}{\partial {\dot
q}_\nu} - \frac{\partial {\cal L}}{\partial q_\nu} = F_\nu(q, {\dot q})
\ .
\label{26}
\ee
Here ${\cal L}(q_\nu, {\dot q}_\nu)$ is the classical Lagrangian, $t$
is the time, and $F_\nu$ are the friction forces. To get the most
general description, stochastic Langevin forces must be added to the
friction forces. An equivalent description is in terms of a
Fokker--Planck equation. If the classical approximation is not
justified, Eq.~(\ref{26}) must be replaced by its quantum analog which
typically involves the density matrix. Often it is justified to
characterize the friction forces by a friction constant and the
Langevin forces by a diffusion constant. We refer to these constants
as to transport coefficients. Calculating transport coefficients is
the aim of a microscopic approach to these processes.

The canonical approach is that of Caldeira and Legett (1983), who
described the environment (the ``heat bath'') as an infinite set of
harmonic oscillators. The nuclear case differs from the standard
situation in statistical mechanics in several respects. First, the
intrinsic degrees of freedom cannot be viewed as an infinitely
extended heat bath. The total energy of the system (intrinsic plus
collective degrees of freedom) is conserved, and the energy content of
either subsystem is roughly the same. In other words, the nucleus is a
mesoscopic system for which the thermodynamic limit is not appropriate
(although some statistical description may still apply). Second, the
coupling between the collective and the intrinsic degrees of freedom
is typically very strong. The energy content of the interaction part
of the Hamiltonian is not small compared to that of either
subsystem. As a consequence, the usual Markov approximation often does
not apply (Brink {\it et al.}, 1978). These facts must be taken into
account as one works out the transport coefficients appearing in
Eq.~(\ref{26}) or in the quantum analog of that equation from a
microscopic approach.

The microscopic approach itself also encounters some difficulties.
First, the definition of the collective degree(s) of freedom in terms
of the coordinates of the participating nucleons poses problems.
Second, in nuclei dissipative processes typically involve the transfer
of several MeV or even several ten MeV from the collective to the
intrinsic degrees of freedom, perhaps together with the transfer of
several ten units of angular momentum.  The level densities in the
fragments at the resulting excitation energies are very high. A
detailed microscopic description of the intrinsic degrees of freedom
is therefore not possible. Rather, one employs stochastic models of
the random--matrix type. In such models, the coupling matrix elements
between collective and intrinsic degrees of freedom are
Gaussian--distributed random variables (in accord with the
Porter--Thomas distribution in Eq.~(\ref{6})). The mean coupling
strength and the density of intrinsic states are estimated using
simple nuclear models. As a result, the force acting on the collective
degree(s) of freedom is random, and the transport coefficients are
defined as ensemble averages involving that random force.

Calculating the transport coefficients as ensemble averages poses
physical questions and technical difficulties quite different from
those encountered in calculating the average of the strength function
$\sum_\tau | \langle 0 | \tau \rangle |^2 \delta(E - E_\tau)$ in
Section~\ref{door}. That calculation leads to Eq.~(\ref{13}). In a
time--dependent picture, Eq.~(\ref{13}) shows that the amplitude of
the collective mode decays exponentially due to mixing with the
background states. The decay width is given by the spreading width of
Eq.~(\ref{13a}). (To see that it is the {\it amplitude} of the
collective mode which undergoes exponential decay we note that the
strength function can be expressed identically in terms of the Green's
function as $- (1 / \pi) \Im \langle 0 | (E^+ - H)^{-1} | 0 \rangle$).
In the case of transport coefficients, it is necessary to average {\it
probabilities} (squares of amplitudes). Quantum mechanics implies that
total sum of all occupation probabilities is conserved in time.
Therefore, transport coefficients describe the redistribution in time
of occupation probabilities caused by dissipative forces. Friction,
for instance, implies that collective states at lower energy become
occupied at the expense of states at higher energy. Averaging
expressions that depend on squares of amplitudes is technically harder
than averaging amplitudes.

The description of dissipative processes in heavy--ion reactions and
of fission in terms of RMT has received much theoretical attention.
Aside from the reviews (N\"{o}renberg and Weidenm\"{u}ller, 1980;
Weidenm\"{u}ller, 1980) and in addition to the papers cited above, we
mention the work of Agassi {\it et al.} (1977) and of Hofmann and
collaborators (1977, 2001). In a very general context, the interaction
of collective and microscopic degrees of freedom has been studied in a
series of papers by Bulgac, Kusnezov, and collaborators (see Bulgac
and Kusnezov (1996) and references therein). There are other
approaches to nuclear dissipation which do not use RMT but emphasize
the chaotic aspects of single--particle motion, see Blocki {\it et
al.} (1995) and references therein.

\subsubsection{Fluctuations of Binding Energies}

The masses of atomic nuclei are keys to the understanding of many
physical and astrophysical processes. For this reason it is important
to construct reliable theoretical models for the values of the nuclear
masses or, equivalently, for the binding energy $B(A)$ as function of
mass number $A$. (We suppress for simplicity the additional dependence
of $B$ on neutron number $N$ or proton number $Z$). This function is
also needed to predict the masses of (as yet) unknown nuclei.

The standard approach to a global modeling of the function $B(A)$
starts out from the liquid--drop model of the nucleus and considers in
addition shell corrections (see Section~\ref{sdb}), as well as
corrections due to the pairing force (see Section~\ref{chaossm}). The
latter lead to an odd--even staggering of $B(A)$. The resulting
``semi--empirical mass formula'' contains about $30$ parameters and is
fitted to a large number of data. Years of painstaking work have
culminated in a best fit (M\"oller {\it et al.}, 1995) that reproduces
the data points very well but not exactly. The overall difference
(root--mean--square value taken in a limited window of mass values) is
of the order of $0.5$ MeV and decreases with increasing $A$. Other
approaches (Samyn {\it et al.}, 2004; Duflo and Zuker, 1995) have led
to similar differences. The figure of $0.5$ MeV is obviously very
small and of the order of $5 \times 10^{-4}$ in comparison with the
total binding energy which for medium--weight and heavy nuclei is of
the order of GeV. Nevertheless, that small value has attracted
considerable attention.

Bohigas and Leboeuf (2002) have suggested that there are two types of
contributions to the shell correction for the nuclear binding
energy. The first one is due to the regular motion of nucleons in the
mean field and is taken into account in terms of the Strutinsky shell
correction method or, in the fits just mentioned, in terms of the
semiempirical mass formula. The second one is due to the (partly)
chaotic motion of nucleons within the nucleus. In Bohigas and LeBoeuf
(2002) that part is also evaluated within the mean--field
approximation. The authors argue, however, that their final result for
the fluctuations of the chaotic part is of much more general validity,
and may be interpreted as arising from the residual interactions.

With $\rho(E)$ the density of single--particle states, the shell
correction to the binding energy has the form $\int {\rm d} E \ E
\rho(E)$. Fluctuations of the binding energy are characterized by the
variance of that expression, the average being taken over a window of
mass numbers and indicated by angular brackets. That variance can be
expressed identically in terms of the form factor $K(\tau)$ as
$\langle B^2(A) \rangle - (\langle B(A) \rangle)^2 = (\hbar^2 / (2
\pi^2)) \int_0^\infty {\rm d} \tau K(\tau) / \tau^4$. The form factor
is essentially the Fourier transform of the density--density
autocorrelation function. For the chaotic contribution to the
variance, $K(\tau)$ is approximated by the random--matrix result
$K(\tau) = 2 \tau$. This expression applies for values of $\tau$ below
the Heisenberg time, and for $\tau \geq \tau_{\rm min}$ where
$\tau_{\rm min}$ is the period of the shortest periodic orbit in the
system while $K(\tau) = 0$ for $\tau < \tau_{\rm min}$. The value of
$\tau_{\rm min}$ is estimated using free motion at the Fermi velocity
over a distance typically given by the nuclear radius. All this yields
for the root--mean--square value of the fluctuation $\sigma(A)$ of the
binding energy the result $\sigma(A) = 2.78 A^{- 1/3}$ MeV. That
result is in reasonable agreement with the deviations of the data from
the semiempirical mass formula.

The interpretation proposed by Bohigas and Leboeuf (2002) of the
deviations of measured binding energies from the semiempirical mass
formula as being due to chaotic motion has caused a lively discussion
that cannot be reviewed here. It implies that throughout the valley of
stability, the binding energies of all nuclei are correlated (Molinari
and Weidenm\"{u}ller, (2006); Olofsson {\it et al.} (2006)). Such
correlations, while expected on dynamical grounds, are surprising from
a statistical viewpoint.

\section{Random--matrix Models inspired by Nuclear--Structure Concepts}
\label{concepts}

In Section~\ref{models} is was shown that in the framework of standard
nuclear models such as the shell model or the cranking model, chaos is
a generic property of nuclei. However, these models differ very much
from the GOE, the standard ensemble to model chaotic systems. To
motivate the study of the random--matrix ensembles treated in the
present Section, we elucidate the difference between these nuclear
models and the GOE, more precisely: Between these models and a generic
realization of the GOE. In doing so, we focus attention on the
shell--model Hamiltonian~(\ref{19}); what will be said applies {\it
mutatis mutandis} likewise to the Hamiltonian of the cranking model in
Section~\ref{chaosyrast}. For the sake of simplicity we disregard
total spin and isospin. A more complete discussion for the $sd$--shell
including these quantum numbers may be found in Section~\ref{tbre}.

Without spin and isospin, the many-body eigenstates of the
single--particle part of the shell--model Hamiltonian~(\ref{19}) are
Slater determinants. These are multiply degenerate. The degeneracies
are lifted by the residual interaction $V_{\rm res}$. The way this
happens depends on the rank $k$ of $V_{\rm res}$. If $V_{\rm res}$ is
a pure two--body interaction ($k = 2$), then $V_{\rm res}$ has
non--vanishing matrix elements only between those pairs of Slater
determinants which differ in the occupation numbers of at most two
single--particle states. In the Hilbert space spanned by Slater
determinants, the matrix of $V_{\rm res}$ then has many zeros. If
$V_{\rm res}$ also contains three--body forces ($k = 3$), the number
of zeros is reduced because $V_{\rm res}$ now connects all Slater
determinants which differ in the occupation numbers of at most three
single--particle states. The systematic appearance of zeros in the
matrix representation of $V_{\rm res}$ is completely removed only when
$V_{\rm res}$ contains $m$--body interactions where $m$ is the number
of valence nucleons. For a generic realization of the GOE, on the
other hand, zeros do not appear systematically. This is why it is
sometimes said that the GOE contains interactions of arbitrary rank.

When the first evidence for the validity of a RMT approach to nuclei
became available in the 1960s, the question arose how that difference
between the GOE and a realistic residual interaction would affect
nuclear properties. Would the spectral statistics be the same? Later,
other nuclear properties (like the average level density) came into
the focus of RMT. How would these depend on the rank of $V_{\rm res}$?
To answer these questions, several non--canonical random--matrix
ensembles were introduced. French and Wong (1970) and Bohigas and
Flores (1971) defined and studied what became known as the two--body
random ensemble (see Section~\ref{tbre}). A few years later, Mon and
French (1975) defined the embedded $k$--body random ensembles where
the rank of the residual interaction is a free parameter (see
Section~\ref{embe}). Quite recently, constrained ensembles (where
certain matrix elements of the GOE are suppressed) were introduced in
Papenbrock {\it et al.} (2006) (see Section~\ref{const}). The present
Section is devoted to a review of these ensembles. The analytical
treatment of the first two ensembles is much more difficult than that
of the canonical ensembles of RMT because they lack the orthogonal
invariance of the GOE. Therefore, a complete analytical theory of
these ensembles does not exist. The analytical treatment of the
constrained ensembles is yet in its infancy.

\subsection{Embedded Ensembles}
\label{embe}

For the $k$--body embedded ensembles of random matrices (introduced by
Mon and French, 1975) we confine ourselves to a brief account of the
main features of the orthogonal case and refer to the reviews (Brody
{\it et al.}, 1981; Kota, 2001; Benet and Weidenm\"{u}ller, 2003) for
further details.

The embedded ensembles dispose of all the complexities due to the
couplings of angular momentum and spin but retain the symmetries
imposed by the exclusion principle. One considers $m$ fermions in $l
\geq m$ degenerate single--particle states (which carry no further
quantum numbers) that interact via a random $k$--body interaction with
$k \leq m$. To obtain an understanding of the transition from the
two--body ensemble to the GOE, the parameter $k$ is allowed to range
from $k = 1$ to $k = m$, although $k = 2$ is the most interesting
(i.e., realistic) case. The case $k = 1$ is integrable and, therefore,
somewhat exceptional.

Labeling the single--particle states with a running index $j = 1,
\ldots, l$, we introduce the usual fermionic creation operators
$a^{\dag}_j$ and define the creation operators for $k$ fermions by
\be
\psi^{\dag}_{k, \alpha} = \prod_{s = 1}^k a^{\dag}_{j_s}
\label{27}
\ee
where $\alpha$ stands for the set $j_1 < j_2 < \ldots < j_k$. The
corresponding annihilation operators are defined by $\psi^{}_{k,
\alpha} = (\psi^{\dag}_{k, \alpha})^{\dag}$. The random $k$--body
interaction is
\be
V_k = \sum_{\alpha, \gamma} v_{k; \alpha \gamma} \psi^{\dag}_{k,
\alpha} \psi^{}_{k, \gamma} \ . 
\label{28}
\ee
The coefficients $v_{k; \alpha \gamma}$ are Gaussian--distributed
random variables with zero mean values and a common second moment.
The matrix elements $v_{k; \alpha \gamma}$ are real symmetric with
a second moment $\overline{v_{k; \alpha \gamma} v_{k; \alpha'
\gamma'}} = v^2 (\delta_{\alpha \alpha'} \delta_{\gamma \gamma'} +
\delta_{\alpha \gamma'} \delta_{\gamma \alpha'})$. The Kronecker
deltas stand for the string $\delta_{j_1 j'_1} \delta_{j_2 j'_2}
\ldots \delta_{j_k j'_k}$ etc. The interaction $V_k$ lifts the
degeneracy of the many--body states; the parameter $v^2$ sets the
scale for the spectrum. Without loss of generality we may put $v^2 =
1$.

The $k$--body embedded ensembles are then defined in terms of the
three parameters $k, l, m$ with $k \leq m \leq l$. The ensembles are
referred to as EGOE($k$) and jointly as EGE($k$). In canonical RMT,
universal results are obtained in the limit of infinite matrix
dimension. For EGE($k$), the same limit is obtained by taking $l \to
\infty$. For fixed $k$ this can be done by imposing constraints on the
ratio $m / l$. Brody {\it et al.} (1981) define the dilute limit
as $l \to \infty, m \to \infty, m / l \to 0$.

Central questions in the theory of the embedded ensembles are: (i)
What is the shape of the spectral density? (ii) What are the spectral
fluctuation properties? (iii) Are these properties universal (i.e.,
independent of the assumed Gaussian distribution)? (iv) Are the
spectra ergodic? (v) Can the embedded ensembles be usefully applied to
real nuclei? If so, what are their predictions? Ideally, questions (i)
to (iv) should be answered in the limit $l \to \infty$.

Partial answers to these questions have been obtained with a variety
of methods. The moments method (Mon and French, 1975) evaluates
moments of $V_k$ in the limit $l \to \infty$. The distribution of
Hamiltonian matrix elements being Gaussian, products of Hamiltonian
matrices are averaged by Wick contraction involving all pairs of
matrices. In the dilute limit $k \ll m \ll l$, $m / l \to 0$ for $l
\to \infty$, only pairs of neighboring matrices are taken into
account (``binary correlation approximation''). Numerical methods give
some insight although the extrapolation to large matrix dimension may
pose problems. The study of the second moments of the many--body
matrix elements of $V_k$ reveals a duality symmetry between EGOE$(k)$
and EGOE($m - k$).

For the spectral density, there is a gradual transition from the
semicircular shape (which is attained for $k = m$) to the Gaussian
shape (which applies for $k \ll m$). The transition sets in at $2 k =
m$. Less is known for the level statistics. It is clear that for $k =
m$ the spectral fluctuations are Wigner--Dyson--like and that for $k =
1$ they are Poissonian. The cases $1 < k < m$ have been much debated
without firm analytical conclusion. The numerical evidence points
toward Wigner--Dyson statistics for $k \geq 2$. Universality has not
been addressed. Ergodicity has been proved for some observables in the
limit $l \to \infty$, but the non--ergodic contributions disappear
very slowly with increasing $l$.

We turn to the applications of EGOE($2$) to nuclei. These are based
upon the binary correlation approximation (Kota, 2001). Moreover, it
is stipulated that the suppression of spin and isospin quantum numbers
does not limit the predictive power of EGOE($k$) in nuclei. With that
assumption, the EGOE($2$) result for the spectral density explains why
shell--model calculations with a large number of valence nucleons
yield approximately Gaussian spectra. But a Gaussian spectrum results
also for $k = 1$ irrespective of $m$, so that spectra of approximately
Gaussian shape are generically expected. EGOE($k$) also predicts the
distribution of transition strengths. For an operator ${\cal O}$
causing a transition from energy $E_i$ to energy $E_f$, the transition
strength distribution is defined as the ensemble average of the trace
of ${\cal O}^{\dag} \delta(V_2 - E_f) {\cal O} \delta(V_2 -
E_i)$. This expression equals the square of the transition matrix
element multiplied with the densities of the initial and final
states. The average can be worked out and yields a bivariate Gaussian
distribution. The comparison with shell--model calculations in Kota
(2001) shows good agreement. Similarly, transition strength sums can
be worked out in closed form. The same is true for the average
occupation numbers of single--particle states which were referred to
already in Section~\ref{chaossm} and which also agree well with
results of the shell model. The method can be extended to cases where
the Hamiltonian is the sum of a single--particle operator and
$V_2$. It is then possible, for instance, to predict the onset of
chaos versus the strength of $V_2$, again in good agreement with
numerical calculations.

In summary we see that EGOE($2$) yields a number of results that are
in good agreement with the shell model. That means, in turn, that
EGOE($2$) is capable of making predictions that can reliably be used
when shell--model calculations are not available. All this is in stark
contrast to the GOE and is possible only because EGOE($2$) takes
account of an essential aspect of the shell model. On the other hand,
it is very difficult to obtain analytical results for EGOE($k$) which
are not based upon the binary correlation approximation. Perhaps most
importantly, there still is no definitive analytical result on the
spectral fluctuation properties of EGOE($k$).

The embedded ensembles have recently been generalized to cover
particles with spin (see Kota, 2007).

\subsection{Two--body Random Ensemble}
\label{tbre}

The two--body random ensemble (TBRE) addresses the questions: Which
nuclear properties obtained by diagonalizing the shell--model
Hamiltonian~(\ref{20}) are generic (i.e., hold for most two--body
interactions), and which are specific properties of a given
interaction? And how does the residual interaction mix the states so
as to produce chaos in nuclear spectra?  To this end, the TBRE uses
the actual form~(\ref{20}) of the shell--model Hamiltonian but
replaces the matrix elements $v_\alpha$ of the actual two--body
interaction by Gaussian--distributed real random variables with zero
mean value and a common second moment (multiplied by the factor $2$
for the diagonal elements, see Eq.~(\ref{30})). In contrast to the
embedded two--body random ensemble (see Section~\ref{embe}), the TBRE
obviously does take into account spin and isospin quantum numbers. It
is time--reversal invariant. Moreover, the TBRE explores the
properties of the residual two--body interaction uniformly in the
space spanned by the variables $v_\alpha$. Statements derived for the
TBRE apply to almost all two--body interactions with the exception of
a set of measure zero.

The TBRE was introduced by French and Wong (1970) and Bohigas and
Flores (1971). The authors were mainly interested in the spectral
fluctuation properties of the TBRE.  The numerical results reported by
French and Wong (1970) and Bohigas and Flores (1971) showed that the
NNS distribution and the $\Delta_3$--statistic of the TBRE agree with
those of the GOE. These results were confirmed by later numerical
studies and showed that chaos is a generic property of the TBRE.
Interest in the TBRE was revived in 1998 by the work of Johnson {\it
et al.} (1998, 1999). These authors showed that in even--even nuclei,
the TBRE predicted ground states with spin zero and positive parity
much more frequently than corresponds to their statistical weight, in
spite of the fact that the matrix elements $v_\alpha$ are random. That
finding caused substantial theoretical activity (see the reviews by
Zelevinsky and Volya, 2004 and Zhao {\it et al.}, 2004) including
studies of bosonic systems interacting via a random two--body
interaction. Here we confine ourselves to the nuclear--physics aspects
of the TBRE. Papenbrock and Weidenm\"{u}ller (2004, 2005, 2006, 2007)
investigated the mechanism by which the TBRE mixes the shell--model
configurations.

The TBRE contains the non--degenerate single--particle energies (first
term on the right--hand side of Eq.~(\ref{20})) as non--random
parameters. The mixing of shell--model configurations in the TBRE
depends on the mean strength of the two--body matrix elements measured
in units of the spacing of the single--particle levels. Complete
mixing occurs only when that ratio is large, see Section~\ref{chaossm}.
For simplicity and clarity, theoretical studies of the TBRE focusing
on chaos often assume that the single--particle energies are
degenerate (in which case they can be put equal to zero by a shift of
the energy scale). We follow that custom here and study the mixing of
shell--model configurations in a ``pure'' TBRE. It has to be borne in
mind, however, that the non--degeneracy of the single--particle levels
tend to weaken the mixing found in that model.

\subsubsection{Comparison GOE -- TBRE}
\label{compGOETBRE}

Before discussing specific properties of the TBRE, it is instructive
to compare the TBRE with the GOE. In the GOE and with $N >> 1$ the
matrix dimension, the number of independent random variables is $N (N
+ 1) / 2$ and, thus, large compared to $N$. In the TBRE, on the other
hand, the number ${\cal N}$ of independent random variables is
typically small compared to the dimension ${\cal D}(J, T, \Pi)$ of the
Hamiltonian matrix. We recall that in the $sd$--shell, we have ${\cal
N} = 63$ while typically ${\cal D} \approx 10^3$ in the middle of the
shell and for low values of $J$. In the $pf$--shell, the corresponding
figures are ${\cal N} = 195$ and ${\cal D} \approx 10^4$ to $10^5$.
The complete mixing of the basis states and the ensuing validity of
Wigner--Dyson statistics for the spectrum cannot be achieved by such a
small number of random variables alone. In an essential way it is also
due to the matrices $C_{\mu \nu}(J T \Pi; \alpha)$ appearing in
Eq.~(\ref{20}). Therefore, studies of the TBRE must focus on the
structure of these matrices, see Section~\ref{struc}.

The GOE is mathematically accessible and formally attractive because
it is orthogonally invariant, universal, and ergodic (see
Section~\ref{propGOE}). By construction, the TBRE is much more
realistic than the GOE (if one includes the non--degenerate
single--particle states of the shell model in the Hamiltonian) but
probably lacks all these properties. It is not orthogonally invariant
because the matrices $C_{\mu \nu}(J T \Pi; \alpha)$ are fixed by the
shell model. An orthogonal transformation would change the chosen
representation of these matrices in Hilbert space but would leave
every realization of the TBRE unchanged. It is not clear whether the
TBRE is universal (i.e., yields results which do not depend on the
assumed Gaussian distribution of the matrix elements $v_\alpha$). We
are not aware of any paper addressing that question. The TBRE is not
ergodic because the limit of infinite matrix dimension cannot be taken
in a meaningful way (except for the case of a single $j$--shell where
$j \to \infty$ is a meaningful limit that has not been explored
yet). In spite of these shortcomings, the TBRE has attractive
features, see below. In Section~\ref{propGOE} it was pointed out that
GOE spectra carry no information content. The TBRE produces spectra
with Wigner--Dyson level statistics. At the same time, the TBRE does
carry information content because the number of random variables is
small compared to typical matrix dimensions.  Ideally it takes ${\cal
N}$ data points to completely determine the values of the random
variables in the TBRE; that number is typically small compared to the
number of eigenvalues pertaining to fixed values of $J$, $T$, and
$\Pi$. Again, this shows the important role played by the matrices
$C_{\mu \nu}(J T \Pi; \alpha)$ in the TBRE. These matrices are fixed
by the geometry of the shell model itself. At the same time, they are
obviously very important for the strong mixing of the shell--model
configurations. The choice of the residual interaction only determines
the particular linear combination of the $C$'s that forms the
shell--model Hamiltonian ${\cal H}_{\mu \nu}(J T \Pi)$ in
Eq.~(\ref{20}).

Properly speaking, the TBRE is not a single ensemble but a set of
ensembles. Indeed, in every shell a given set of matrix elements
$\{v_\alpha\}$ determines for all values of $A$ pertaining to that
shell the Hamiltonian matrices ${\cal H}_{\mu \nu}(J T \Pi)$ for all
values of $J$, $T$, and $\Pi$. Taking the $v_\alpha$ as random
variables implies that all these matrices become Gaussian
random--matrix ensembles. Since all these ensembles depend upon the
same set $\{v_\alpha\}$ of random variables, they are correlated. Such
correlations are the hallmark of the TBRE. Correlations of this type
do not occur naturally within the GOE.

\subsubsection{Structure of the Matrices $C_{\mu \nu}(J T \Pi;
\alpha)$}
\label{struc}

The discussion in Section~\ref{compGOETBRE} has revealed the central
role played by the matrices $C_{\mu \nu}(J T \Pi; \alpha)$ in the
TBRE. As emphasized in Section~\ref{shell}, these matrices are
completely determined by the coupling scheme chosen to construct the
many--body states $|J T \Pi \mu \rangle$. They depend upon
vector--coupling coefficients and on coefficients of fractional
parentage and, thus, are given in terms of group--theoretical
concepts. (The coefficients of fractional parentage give the
decomposition of a state $|J T \Pi \mu \rangle$ for $m$ nucleons in
terms of the same states for $(m-1)$ nucleons). The matrices $C_{\mu
\nu}(J T \Pi; \alpha)$ are determined by the intrinsic symmetries of
the shell model and are known completely. At the same time, these
matrices are highly complex, and there is no analytical theory yet to
describe their structure. We are confined to describing briefly some
of their properties and refer for details to Papenbrock and
Weidenm{\"u}ller (2007).

To see how the non--diagonal elements of $V_{\rm res}$ mix the
unperturbed configurations, we use the coupling scheme described above
Eq.~(\ref{20}) and classify the many--body states in terms of the
partitions $\{ m_{\ell j} \}$ of $m$. For $m = 12$ there are $41$ such
partitions in the $sd$--shell. Each partition consists of a string of
three non--negative integers which give the number of nucleons in each
subshell. In that basis, the matrix elements of $V_{\rm res}$ attain
block structure. Each block is labeled by a pair of partitions $\{
m_{\ell j} \}$. The blocks come in four classes: (i) The partition $\{
m_{\ell j} \}$ is unchanged (diagonal blocks arising from those
two--body matrix elements which do not change partition although they
may change the actual many--body basis state); (ii) the partition $\{
m_{\ell j} \}$ is changed by adding unity to one of its elements at
the expense of another element (the residual interaction lifts a
single nucleon from one subshell to another); (iii) the partition $\{
m_{\ell j} \}$ is changed by adding unity to two of its elements or
two to one of its elements at the expense of one or two other elements
(the residual interaction lifts two nucleons into different
subshells); (iv) the partition $\{ m_{\ell j} \}$ is changed by moving
more than two nucleons into different subshells. These blocks are
empty because a two--body interaction cannot move more than two
nucleons.

The block structure of the matrices $C_{\mu \nu}(J T \Pi; \alpha)$
(also displayed in Zelevinsky {\it et al.}, 1996) is here shown in
Fig.~\ref{fig26} for $m = 12$ and for the states with $J = 0$, $T = 0$
of the $sd$--shell. Blocks in classes (i), (ii), (iii), and (iv) are
shown in black, light gray, dark gray, and white, respectively (color
online: red, green, blue, and white, respectively). It is obvious that
the structure of the matrix is fundamentally different from that of a
typical GOE matrix where all states are coupled with each other. That
fact implies that complete mixing of the shell--model configurations
can never be the result of a single matrix $C_{\mu \nu}(J T \Pi;
\alpha)$ alone. A linear combination of the type appearing in
Eq.~(\ref{20}) is definitely needed. It is also obvious that the block
structure shown in Fig.~\ref{fig26} is not restricted to the
$sd$--shell and is a generic property of the shell model.

\begin{figure}[h]
\vspace{5 mm}
\includegraphics[width=\linewidth]{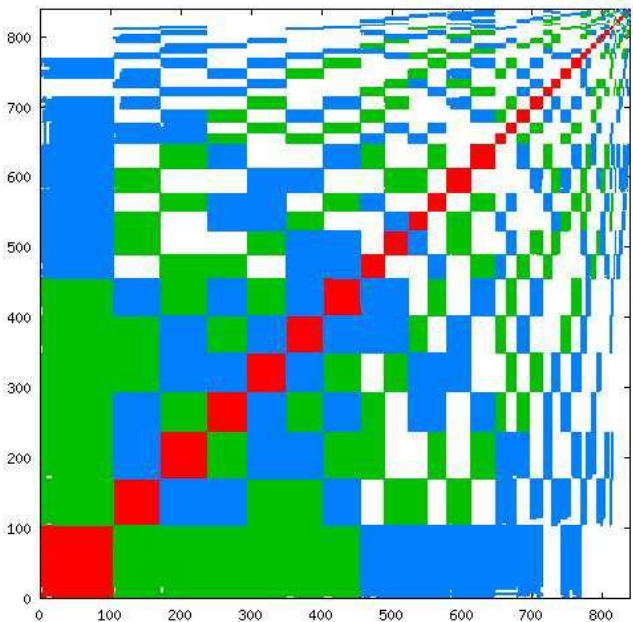}
\vspace{3 mm}
\caption{(Color online) Block structure of the matrix of the
shell--model Hamiltonian for the $sd$--shell. Blocks in black, light
gray, dark gray (color online: red, green, blue) indicate matrix
elements which change the partition by zero, one or two units,
respectively, as explained in the text. White areas do not carry
matrix elements. From Papenbrock and Weidenm\"{u}ller (2005).}
\label{fig26}
\end{figure} 

The structure of individual matrices $C_{\mu \nu}(J T \Pi; \alpha)$ is
governed by ``geometric chaos'' (Zelevinsky {\it et al.}, 1996). The
many--body states $| J T \Pi \mu \rangle$ may be constructed by
coupling first a pair of nucleons to given intermediate values of spin
and isospin, by then vector--coupling a third nucleon to the resulting
pair etc. etc. There are obviously very many different paths leading
to the same total spin $J$ and isospin $T$; their number determines
the dimension ${\cal D}(J, T, \Pi)$ of the resulting Hilbert
space. Each path corresponds to a different set of vector--coupling
coefficients. Supposing that the vector--coupling coefficients are
pseudorandom numbers, we expect that the states $|J T
\Pi \mu \rangle$ also behave randomly. Several tests (Zelevinsky {\it
et al.}, 1996; Zelevinsky and Volya, 2004) support this idea of
``geometric chaos''. It is, therefore, reasonable to expect that the
elements of the matrices $C_{\mu \nu}(J T \Pi; \alpha)$ behave
randomly in those blocks where they do not vanish identically. This
view is obviously not restricted to the $sd$--shell but applies
likewise to every major shell. Further insight into the mixing
properties of the matrices $C_{\mu \nu}(J T \Pi; \alpha)$ is obtained
by counting the number of matrices which yield non--vanishing
contributions to a matrix element of the shell--model Hamiltonian
${\cal H}_{\mu \nu}(J T \Pi)$ and by calculating the inverse
participation ratios for these matrices (Papenbrock and
Weidenm\"{u}ller, 2005). All this evidence points to strong mixing of
shell--model configurations by the matrices $C_{\mu
\nu}(J T \Pi; \alpha)$.

Although we are very far from a complete theory of the TBRE, the
evidence presented makes it plausible that chaos is a generic property
of the shell model. The matrices $C_{\mu \nu}(J T \Pi; \alpha)$ are
both, the fundamental building blocks of the TBRE, and the agents for
complete mixing of the shell--model configurations.

\subsubsection{Another Representation of the Shell--Model Hamiltonian}
\label{another}

Can we gain further insight into the structure of the matrices $C_{\mu
\nu}(J T \Pi; \alpha)$? Is it possible to define a quantitative
measure for the information content of shell--model spectra? Why are
spin zero ground states dominant in the TBRE? These questions are
answered by transforming the interaction part of the shell--model
Hamiltonian ${\cal H}_{\mu \nu}(J T \Pi)$ in Eq.~(\ref{20}). This is
done (Papenbrock and Weidenm\"uller, 2004) by diagonalizing for each
set of quantum numbers $\{J T \Pi \}$ the real, symmetric and
positive--semidefinite matrix $S_{\alpha \beta} = {\cal D}^{-1}(J T
\Pi) {\rm Trace} [ C(J T \Pi, \alpha) C(J T \Pi; \beta) ]$. We denote
by $s^2_\alpha \geq 0$ the eigenvalues of $S_{\alpha \beta}$, by
$s_\alpha \geq 0$ the roots of these eigenvalues, and by ${\cal
O}_{\alpha \beta}$ the eigenvector belonging to the eigenvalue
$s^2_\alpha$. We define the new random variables $w_\alpha =
\sum_\beta {\cal O}_{\alpha \beta} v_\alpha$ and, for $s_\alpha > 0$,
the matrices $B_{\mu \nu}(J T \Pi; \alpha) = (1 / s_\alpha) \sum_\beta
{\cal O}_{\alpha \beta} C_{\mu \nu}(J T \Pi; \beta)$. By construction,
the matrices $B_{\mu \nu}(J T \Pi; \alpha)$ are orthonormal with
respect to the trace, ${\cal D}^{-1}(J T \Pi) {\rm Trace} [ B(J T \Pi,
\alpha) B(J T \Pi; \beta) ] = \delta_{\alpha \beta}$. The shell--model
Hamiltonian~(\ref{20}) takes the form
\be
{\cal H}_{\mu \nu}(J T \Pi) = \delta_{\mu \nu} \sum_{\ell j} \ve_{\ell
j} m_{\ell j} + \sum_\alpha w_\alpha s_\alpha B_{\mu \nu}(J T \Pi;
\alpha) \ .
\label{29}
\ee
In the last sum over $\alpha$, only terms pertaining to non--zero
eigenvalues $s^2_\alpha$ appear.

The form~(\ref{29}) of the shell--model Hamiltonian is instructive for
several reasons. First, by adding further matrices we could enlarge
the set $\{ B(J T \Pi; \alpha) \}$ of orthonormal matrices to a
complete set of ${\cal D}(J T \Pi) [ {\cal D}(J T \Pi) + 1 ] / 2$
orthonormal real symmetric matrices. The linear combination of all
these matrices with random coefficients would be equivalent to the
GOE. The number of matrices $B(J T \Pi)$ is very much smaller than
${\cal D}(J T \Pi) [ {\cal D}(J T \Pi) + 1 ] / 2$, however. This shows
once again that the TBRE is very different from the GOE; it is a
constrained ensemble in the sense of Section~\ref{const}. Second,
being obtained from the $v_\alpha$ by an orthogonal transformation,
the new random variables $w_\alpha$ have the same Gaussian
distribution as the former. However, not all $w_\alpha$ but only those
pertaining to non--vanishing eigenvalues $s^2_\alpha$ do appear in the
sum in Eq.~(\ref{29}). At least one (and often several) eigenvalues
$s^2_\alpha$ always vanish. That implies that one or several linear
combinations of the $v_\alpha$ can never be measured (nor do they
affect the shell--model spectrum). Third, those $w_\alpha$ that do
appear in Eq.~(\ref{29}) are multiplied with the root factors
$s_\alpha$. The matrices $B(J T \Pi)$ are orthonormal.  Therefore, the
difficulty of determining the $w_\alpha$ in Eq.~(\ref{29}) from data
increases with increasing smallness of the factors $s_\alpha$. These
root factors are derived from the matrix $S_{\alpha \beta}$ and
reflect intrinsic properties of the shell model.

In Papenbrock and Weidenm\"uller (2004) the distribution of the roots
$s_\alpha$ has been worked out for the case of a single $j$--shell and
for two nuclei in the $sd$--shell. In Fig.~\ref{fig27} we show
$s_\alpha(J)$ versus $J$ for $n = 6$ identical nucleons in the $j =
19/2$--shell. Very similar results have been found in all other cases
(different nucleon numbers in the same shell and two nuclei in the
$sd$--shell). All roots are smooth functions of total spin $J$. One
root is substantially bigger than the rest. The associated two--body
operator is the monopole operator and essentially determines the
centroid of the shell--model spectrum. These features can be
understood semi--analytically (Papenbrock and Weidenm\"uller, 2004).
Identically vanishing eigenvalues are related to conserved quantum
numbers. For instance, the matrix representation of the two--body
operator $\vec{J}^2 - J(J+1)$ (with $\vec{J}$ the total spin operator)
can be written as a linear combination of the matrices $C(J T \Pi;
\alpha)$; that linear combination vanishes identically, and similarly
for total isospin $T$.

\begin{figure}[h]
\vspace{5 mm}
\includegraphics[width=\linewidth]{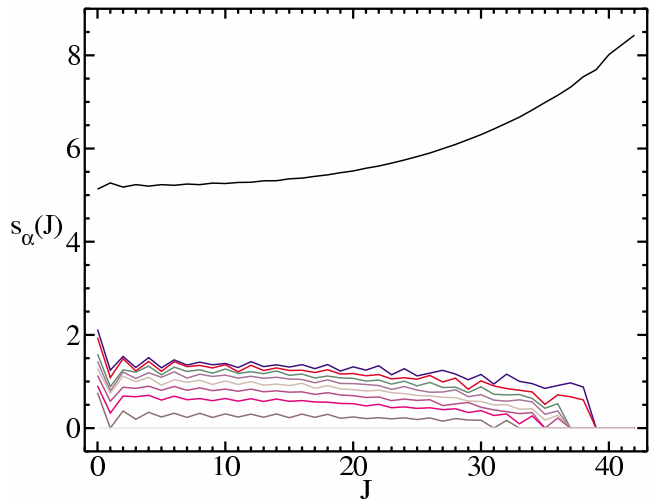}
\vspace{3 mm}
\caption{(Color online) The square roots of the eigenvalues of the
matrix $S_{\alpha \beta}(J)$ (see text) for the $j = 19/2$--shell
with 6 identical nucleons versus total spin $J$. From Papenbrock and
Weidenm\"{u}ller (2004).}
\label{fig27}
\end{figure}

\subsubsection{Preponderance of Spin Zero Ground States in the TBRE}

The discovery by Johnson {\it et al.} (1998) of the preponderance of
spin zero ground states used a specific version of the TBRE. That
version favors neither a particle--particle nor a particle--hole
representation, and the authors refer to it as to the random
quasiparticle ensemble. Following Johnson {\it et al.} (1998), we
denote the two--body matrix elements $\langle j_3 j_4 s t | V_{\rm
res} | j_1 j_2 s t \rangle$ appearing in Eq.~(\ref{19}) by $V_{\rho'
\rho}$ where $\rho$ and $\rho'$ label the two--body states $|j_1 j_2 s
t \rangle$ and $|j_3 j_4 s t \rangle$. The $V_{\rho' \rho}$ have zero
mean values and second moments given by
\be
\overline{V_{\rho' \rho} V_{\sigma' \sigma}} = \frac{v^2}{(2s+1)(2t+1)}
[ \delta_{\rho \sigma} \delta_{\rho' \sigma'} + \delta_{\rho \sigma'}
\delta_{\rho' \sigma} ] \ .
\label{30}
\ee

Here $v^2$ is a constant, i.e., independent of all the quantum numbers.
The factor $1 / (2s+1)(2t+1)$ guarantees that this a random
quasiparticle ensemble. The single--particle energies $\ve_{\ell j}$
in Eq.~(\ref{20}) are neglected.

In Johnson {\it et al.} (1998) it was found that for several nuclei in
the $sd$--shell, the probability of finding a ground state with spin
zero lies between $2/3$ and $3/4$ although the total fraction of spin
zero states is less than 10 percent. This result is known as the
preponderance of spin zero ground states. It holds for the random
interaction~(\ref{30}). That interaction does not possess a strong
pairing force (the agent usually held responsible for the
preponderance of spin zero ground states in actual nuclei). Other
regularities were also observed (Johnson {\it et al.}, 1998, 1999;
Zhao {\it et al.}, 2004b). We confine ourselves to the preponderance
of spin zero ground states. A large number of theoretical papers is
devoted to this phenomenon. We confine ourselves to the two successful
explanations that have been offered for the phenomenon and refer the
reader to the reviews: Zelevinsky and Volya (2004); Zhao {\it et al.} 
(2004a) for further references.

The method used by Zhao and Arima (2001) and refined in later papers
(Zhao {\it et al.}, 2002, 2004a) is based upon a simple counting
procedure. The authors put one of the ${\cal N}$ different matrix
elements of the residual interaction equal to $(-1)$ and all others to
zero and calculate the spectrum. The procedure is repeated ${\cal N}$
times, each time with a different non--zero matrix element. Let ${\cal
N}_J$ be the number of times the ground state is found to have spin
$J$. We obviously have $\sum_J {\cal N}_J = {\cal N}$. The probability
of finding a spin zero ground state is then estimated as ${\cal N}_0 /
{\cal N}$. Comparing the results with an average over many
diagonalizations of the TBRE, the authors find good agreement for a
number of cases (four to six fermions in single $j$--shells and two
$j$--shell systems, boson systems).

The analysis of Papenbrock and Weidenm\"{u}ller (2004) uses a
two--step argument. (For simplicity we replace the quantum numbers $J,
T, \Pi$ by the single symbol $J$). First, the square of the width
$\sigma_J$ of the spectrum of levels with quantum number $J$ is
defined as the normalized variance of the shell--model
Hamiltonian. Since shell--model spectra have approximately Gaussian
shape (Mon and French, 1975), $\sigma_J$ has a direct physical
interpretation. From Eq.~(\ref{29}) we obtain $\sigma^2_J =
\sum_\alpha w^2_\alpha(J) s^2_\alpha(J)$. The argument $J$ on $w$ and
on $s$ serves as a reminder that both quantities depend on $J$. That
dependence is weak, however, at least for the eigenvalues, see
Fig.~\ref{fig27}. More importantly, the spectral widths depend via the
$w_\alpha$ on the same random variables $v_\alpha$ and are, therefore,
strongly correlated: They tend to be all large or all small for a
given realization of the $v_\alpha$. The variances are biggest when
many terms contribute almost uniformly to the sum over $\alpha$;
Fig.~\ref{fig28} shows that this is the case for
$\sigma_0$. Altogether the correlations favor either $\sigma_0$ or the
spectral width of the largest spin. Second, the quantity of interest
is $R_J$, the largest (or smallest) eigenvalue of the spectrum of
levels with spin $J$. (There is no distinction between the two because
the $v_\alpha$ have random signs). Numerical calculations show that
the linear relation $R_J = r_J \sigma_J$ holds with $r_J$ practically
constant (i.e., independent of the realization of the $v_\alpha$). For
the states with $J = 0$ this is shown for 6 identical nucleons in the
$j = 19/2$ shell in the inset in Fig.~\ref{fig28}. The dependence of
$r_J$ on $J$ is largely determined by ${\cal D}(J)$, the dimension of
Hilbert space. Typically, ${\cal D}(J)$ decreases with increasing $J$
and shows odd--even staggering. Both features are present in $r_J$,
see Fig.~\ref{fig28}. These tendencies of $r_J$ suppress the
competition of the spectral width belonging to the largest spin and
favor $R_0$ over all other $R_J$, see Fig.~\ref{fig28}. The
explanation carries over to other cases including nuclei with an odd
number of nucleons. In the nuclei $^{20}$Ne and $^{24}$Mg, it leads to
a semiquantitative agreement with numerical TBRE calculations
(Papenbrock {\it et al.}, 2006a).

\begin{figure}[h]
\vspace{5 mm}
\includegraphics[width=\linewidth]{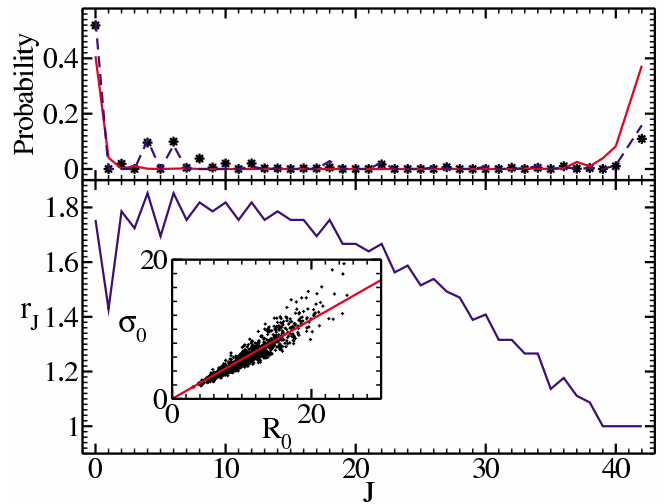}
\vspace{3 mm}
\caption{(Color online) 6 identical fermions in a $j = 19/2$ shell.
Inset: Spectral radius $R_0$ versus spectral width $\sigma_0$ (data
points) and the linear fit (line). Bottom: Scaling factor $r_J$
versus $J$. Top: Probability that the ground state has spin $J$
(points); probability that spin $J$ has the largest spectral width
(solid line); probability that the product $r_J \sigma_J$ is maximal
(dashed line). From Papenbrock and Weidenm\"{u}ller (2004).}
\label{fig28}
\end{figure}

The method of Papenbrock and Weidenm\"{u}ller (2004) was improved by
Yoshinaga {\it et al.} (2006). The authors considered, for instance, a
single $j$--shell with identical nucleons. The energy $E(J)$ of the
lowest state of spin $J$ was written as $E(J) = {\cal D}(J)^{-1} {\rm
Trace} ( {\cal H} ) - \Phi_J \sigma_J$. This equation differs from
that of Papenbrock and Weidenm\"{u}ller (2004) by the inclusion of the
trace of ${\cal H}$ and by the fact that an analytical form for the
function $\Phi_J$ was proposed. The trace of ${\cal H}$ vanishes upon
taking the ensemble average but, for each realization, fluctuates
around zero. Inclusion of the trace in the equation for $E(J)$ removes
the scatter of the points around the best linear approximation to
$r_J$ shown in Fig.~\ref{fig28}. The function $\Phi_J$ was fitted and
given analytically as $\Phi_J = \sqrt{0.99 \ln {\cal D}(J) +
0.36}$. Very good agreement is obtained for $m = 4$ to $m = 6$
fermions in several single $j$--shells and systems with two
$j$--shells between TBRE results and the predictions based upon this
approach. The method works also for bosons.

\subsubsection{Correlations between Spectra with Different Quantum
Numbers}

As pointed out at the end of Section~\ref{compGOETBRE}, the TBRE
causes correlations between spectra carrying different quantum numbers
($A, J, T, \Pi$) but belonging to the same major shell. From the point
of view of the shell model, this is not surprising: Switching from one
realization of the TBRE to another is tantamount to using a different
residual interaction; such a different choice of $V_{\rm res}$ is
bound to affect the spectra of all nuclei in the shell. However, the
existence of such correlations is surprising from the point of view of
RMT and exceeds the traditional framework of the theory. Moreover, the
statistical analysis of nuclear data has always assumed the absence of
correlations between observables with different quantum numbers. That
assumption is put into question by the TBRE.

In Papenbrock and Weidenm\"{u}ller (2006) correlations in $sd$--shell
nuclei were displayed between states in the same nucleus carrying
different quantum numbers and between states with identical quantum
numbers in different nuclei.  The normalized cross--correlation
functions had maxima of around $6 - 10$ per cent. Similar figures
result by calculating correlations in actual shell--model spectra by
averaging over $17$ $sd$--shell nuclei (Papenbrock and
Weidenm\"{u}ller, 2006).

\subsubsection{Summary}

The studies of the TBRE reviewed above suggest that chaos is a generic
property of the shell model. The strong mixing is due to the matrices
$C_{\mu \nu}(J T \Pi; \alpha)$ which, for every major shell, are
determined by the geometry of that shell, by the number of valence
nucleons, and by the quantum numbers $\{ J T \Pi \}$. A complete
theory of the TBRE would have to be based on the analysis of these
matrices; such an analysis is not available yet. As a consequence,
there is no analytical proof yet that TBRE spectra do obey
Wigner--Dyson statistics. TBRE and GOE are very different; in the TBRE
there exist spectral correlations which are totally absent in the GOE.

\subsection{Constrained Ensembles}
\label{const}

The spectral fluctuation properties of both, the embedded ensembles
and the TBRE, are known only through numerical simulations. Such
studies are necessarily confined to matrices of small dimensions $N$.
A generic answer (valid for any $N$) can only be obtained
analytically.  A renewed attempt at such an answer was made in
Papenbrock {\it et al.} (2006). The authors introduced and studied
constrained Gaussian random--matrix ensembles. We review here the
unitary case.

Starting point is a complete basis of $N^2$ orthonormal Hermitian
matrices $B_\alpha$ in the $N$--dimensional Hilbert space.
Orthonormality is defined in terms of the trace,
\be
\langle B_\alpha | B_\beta \rangle = {\rm Trace} ( B_\alpha B_\beta)
= \delta_{\alpha \beta} \ . 
\label{31}
\ee
We note that in contrast to Section~\ref{another} we do not include
the matrix dimension in the definition~(\ref{31}). Any Hermitian
matrix $H$ can be expanded in terms of that basis,
\be H = \sum_{\alpha = 1}^{N^2} h_\alpha B_\alpha \ .
\label{32}
\ee
Taking the $N^2$ complex coefficients $h_\alpha$ in Eq.~(\ref{32}) as
uncorrelated Gaussian--distributed random variables with zero mean
value and a common second moment, one obtains the GUE. A constrained
ensemble is obtained by requiring a subset $\{ h_q \}$ of the
expansion coefficients $h_\alpha$ in Eq.~(\ref{32}) to vanish
identically. (It was shown in Section~\ref{another} that the TBRE can
be viewed as a constrained ensemble, and it is easy to see that the
same statement holds for the embedded ensembles). The resulting
ensemble is not unitarily invariant, however. Unitary invariance is
restored by integrating the constraining condition $\prod_q \delta (
\langle B_q | H \rangle )$ over the unitary group, see
Section~\ref{GOE}. This yields for the probability density of the
matrices $H$
\ba
W(H) &\propto& \exp \bigg( - \frac{N}{2 \lambda^2} \langle H | H
\rangle \bigg) \nonumber \\
&& \qquad \times \int {\rm d}{\cal U} \bigg( \prod_q \delta ( \langle
U B_q U^\dag | H \rangle ) \bigg) \ . 
\label{33}
\ea
The first factor on the right--hand side is the same as in
Eq.~(\ref{2g}). The second factor represents the constraints. It is
unitarily invariant and, therefore, does not affect the eigenvector
distribution. The eigenvalue distribution of the constrained ensemble
differs from that on the right--hand side of Eq.~(\ref{2i}) by the
additional factor
\be
F(E_1, \ldots, E_N) = \int {\rm d}{\cal U} \bigg( \prod_q \delta (
\langle B_q | U E U^\dag \rangle ) \bigg) \ . 
\label{34}
\ee
Here $E$ is the diagonal matrix of eigenvalues.

In the GUE, quadratic level repulsion is a consequence of the
Vandermonde determinant $\prod_{\rho \leq \sigma} | E_\rho -
E_\sigma|$ appearing on the right--hand side of Eq.~(\ref{2i}). Level
repulsion is lifted by the constraints if and only if $F(E_1, \ldots,
E_N)$ is singular whenever two eigenvalues coincide. Using Fourier
transformation, one replaces each of the delta functions by an
integral over a plane wave. The function $F(E_1, \ldots, E_N)$ can
then be written as a Harish--Chandra Itzykson Zuber integral
(Harish--Chandra 1957, Itzykson and Zuber, 1980). Inspection of the
result of the integration yields a sufficient condition for $F(E_1,
\ldots, E_N)$ not to be singular. In its simplest form, that condition
reads
\be
N_Q < \frac{N(N-1)}{2} \ .
\label{35}
\ee
Here $N_Q$ is the number of constraints. It is quite remarkable that
the sufficient condition depends only on $N_Q$.

The result can be extended to the GOE and the GSE. Moreover, other
properties (spectral radius, distribution of matrix elements) of the
constrained ensembles can be investigated and are seen to differ from
those of the canonical ensembles. Relations can be established between
the ensemble defined by the constraints on the set $\{ h_q \}$ of
matrix elements, and the one defined by the complementary set (all
$h_\alpha$ but the set $\{ h_q \}$ are constrained). Unfortunately,
the condition~(\ref{35}) does not cover the physically interesting
cases of the TBRE and of the embedded ensemble with two--body
interactions. The spectral fluctuation properties of these ensembles
remain an open theoretical problem.

\section{Summary and Conclusions}
\label{concl}

Comparison of GOE predictions with data reviewed in
Section~\ref{applRMT} shows that there is good agreement on spectral
fluctuations not only near neutron threshold but also in the
ground--state domain. This is true also for the three nuclei with
complete level schemes. Extensions of the GOE are used to successfully
describe isospin mixing, and to test time--reversal invariance.
Deviations from GOE predictions are found up to several $100$ keV
above yrast and in case of symmetries. An enlargement of the data
basis would be highly desirable. There is hope that the next
generation of large--scale gamma detectors will contribute pertinent
information on the interesting energy interval $500$ keV to $1$ MeV
above yrast.

Studies of the spherical shell model, of the Nilsson model, and of the
interacting boson model reviewed in Section~\ref{models} show a joint
tendency towards strong mixing of the unperturbed configurations and
towards GOE fluctuation properties. This resolves the apparent
dichotomy between Bohr's picture of the compound nucleus described in
Section~\ref{RMT} and the independent--particle model. Typically, the
GOE limit is not fully attained, however. For the $sd$--shell, mixing
is strongest for nuclei in the middle of the shell but even here the
spreading width of the unperturbed shell--model configurations located
in the center of the spectrum is $20$ MeV or so and, thus, smaller
than the range of the spectrum.

Random--matrix ensembles patterned after nuclear--structure concepts
were reviewed in Section~\ref{concepts}. The embedded two--body random
ensemble possesses strong predictive power for average properties such
as the shape of the spectrum or the distribution of transition
strengths. This is true even though that ensemble disregards conserved
quantum numbers like spin or isospin. The two--body random ensemble is
closest in structure to the shell model. The strong mixing of
shell--model configurations is seen to be a generic feature of the
shell model. At the same time, the two--body ensemble possesses
features that go beyond the GOE such as the correlations between
spectra with different quantum numbers. It would be intriguing to find
such correlations experimentally.

Following the Bohigas--Giannoni--Schmit conjecture, we have identified
chaos in nuclei with the occurrence of spectral fluctuations of the
GOE type. Closer inspection has shown, however, that chaos in nuclei
differs from that in few--degrees--of--freedom systems. The evidence
presented shows that chaos is caused by the residual two--body
interaction of the shell model. Shell structure is a typical feature
of fermionic many--body systems. The residual interaction strongly
mixes the shell--model configurations within a major shell,  but leaves
the overall shell structure largely intact. Evidence for this
assertion comes not only from the success of shell--model calculations
in the ground--state domain,  but also from the success of the optical
model for elastic scattering and from the existence of distinct maxima
in the neutron strength function. (The last two features are mentioned
without explanation for the sake of the argument and will be treated
in part 2 of this review). The existence of regular features in nuclei
also attests to the incomplete mixing of the shell--model
configurations. We mention the regularities in the ground--state
domain caused by the pairing force, collective motion, and doorway
states.

The evidence presented in this review strongly supports the view that
chaos is a generic property of nuclei. At the same time, we are far
from having a complete theoretical understanding of chaos in nuclei.
First, we lack an overall many--body theory that would permit the
calculation of nuclear spectra from the nucleon--nucleon interaction
within controlled approximations (except for the lightest nuclei). In
contrast to few--degrees--of--freedom systems, there is also no
theoretical framework such as the semiclassical approximation which
would establish the connection between classical chaos and spectral
fluctuation properties of the RMT type. On a more mundane level, the
evidence presented above for chaos in the spherical shell model comes
mainly from the $sd$--shell. Although the arguments seem generic, it
would be gratifying to have similar evidence in other major shells.
Would the sizeable correlations between spectra carrying different
quantum numbers found in the two--body random ensemble persist with
increasing shell size? Likewise, it is desirable to attain a deeper
analytical understanding of how chaos arises within a major shell.
Does the difference between the GOE and the two--body random ensemble
entail other differences beyond those correlations?  Also, the
analysis of properties of the matrices $C_{\mu \nu}$ in Eq.~(\ref{20})
within a single $j$--shell would be of substantial interest, coupled,
if possible, with the proof of GOE spectral fluctuation properties for
$j \gg 1$.

Chaos limits the predictability of spectral properties in nuclei in
terms of simple models. For instance, the shell--model eigenfunctions
become mixed ever more strongly as the excitation energy increases,
see Section~\ref{chaossm}. Such strong mixing requires the presence of
many matrix elements of the two--body interaction and cannot be
modeled in simple terms. We are not aware of any attempts to formulate
and quantify that limitation, however.

The way it is defined in this review, chaos is a statistical property
of levels carrying identical quantum numbers. Regular dynamical
features in nuclei typically relate several states carrying different
quantum numbers. The coexistence of those two aspects of nuclear
motion deserves deeper analysis. Hopefully, it would also shed light 
on the dynamical properties of other fermionic many--body systems.

\section*{Acknowledgments}

We owe thanks to many colleagues with whom we have discussed topics
relating to chaos in nuclei over many years, and from whom we have
learned much. Special and cordial thanks are due to those who took the
trouble to read the first draft of this paper, and who made numerous
valuable and thoughtful comments and suggestions: S. {\AA}berg,
Y. Alhassid, O. Bohigas, T. Friedrich, G. Garvey, T. Guhr,
T. Papenbrock, A. Richter, T. H.  Seligman, V. Zelevinsky, Y. M. Zhao.
We would also like to thank R. Chankova, A. Chyzh, H. Friedrich, and
J. F. Shriner, Jr., for assistance with the figures and manuscript
preparation.  This work was supported in part by the U.S. Department
of Energy grant DE-FG02-97ER41042.


\begin{thebibliography}{99}

\bibitem{Abe90} {\AA}berg, S., 1990, Phys. Rev. Lett. {\bf 64}, 3119.

\bibitem{Abe92} {\AA}berg, S., 1992, Prog. Nucl. Part. Phys. {\bf 28},
11. 

\bibitem{Abu85} Abul-Magd A. Y., and  H. A. Weidenm\"{u}ller, 1985, 
Phys. Lett. {\bf B 162}, 223.

\bibitem{Abu04} Abul-Magd, A. Y., H. L. Harney, M. H. Simbel, and H.
A. Weidenm\"{u}ller, 2004 Phys. Lett. {\bf B 579}, 278.

\bibitem{Ada98} Adams, A. A., G. E. Mitchell, and J. F. Shriner, Jr.,
1998, Phys. Lett. {\bf B 422}, 13.

\bibitem{Aga77} Agassi, D., C. M. Ko, and H. A. Weidenm\"uller, 1977,
Ann. Phys. (N.Y.) {\bf 107}, 140. 

\bibitem{Agv03} Agvaanluvsan, U., G. E. Mitchell, J. F. Shriner,
Jr., and M. P. Pato, 2003, Nucl. Instrum. Methods Phys. Res. {\bf A
498}, 459.

\bibitem{Alh90} Alhassid, Y., A. Novoselsky, and N. Whelan, 1990,
Phys. Rev. Lett. {\bf 65}, 2971.

\bibitem{Alh91} Alhassid, Y., and N. Whelan, 1991, Phys. Rev. Lett.
{\bf 67}, 816.

\bibitem{Alh93} Alhassid, Y., and N. Whelan, 1993, Phys. Rev. Lett.
{\bf 70}, 572.

\bibitem{Alt95} Alt, H., H.-D. Gr{\"a}f, H. L. Harney, R. Hofferbest,
H. Lengeler, A. Richter, P. Schaardt, and H. A. Weidenm{\"u}ller,
1995, Phys. Rev. Lett. {\bf 74}, 62.

\bibitem{Alt98}H. Alt, C. I. Barbosa, H.-D. Gr\"{a}f, T. Guhr, H. L.
Harney, R. Hofferbert, H. Rehfeld, A. Richter, 1998, Phys. Rev. Lett.
{\bf 81}, 4847.

\bibitem{Alt97} Altland, A. and M. R. Zirnbauer, 1997, Phys. Rev.
{\bf B 55}, 1142.

\bibitem{Ari75} Arima, A., and F. Iachello, 1975, Phys. Rev. Lett.
{\bf 35}, 1069.

\bibitem{Arv87} Arvieu, R., F. Brut, J. Carbonell, and J. Touchard,
1987, Phys. Rev. {\bf A 35}, 2389.

\bibitem{Axe62} Axel, P., 1962, Phys. Rev. {\bf 126}, 671.

\bibitem{Bae92} Bae, M. S., T. Otsuka, T. Misusaki, and N. Fukunishi,
1992, Phys. Rev. Lett. {\bf 69}, 2349.

\bibitem{Bal68} Balian, R., 1968, Nuovo Cim. Soc. Ital. Fis.
{\bf B 57}, 183.

\bibitem{Bal70} Balian, R., and C. Bloch, 1970, Ann. Phys. (N.Y.)
{\bf 60}, 401.

\bibitem{Bal76} Baltes, H. P., and  E. R. Hilf, 1976, {\it Spectra
of Finite Systems} (Wissenschaftsverlag, Mannheim).

\bibitem{Bar00} Barbosa, C. I., T. Guhr, and H. L. Harney, 2000,
Phys. Rev. {\bf E 62}, 1936.

\bibitem{Ben03} Benet, L., and H. A. Weidenm\"uller, 2003, J. Phys.
A: Math. Gen. {\bf 36}, 3569.

\bibitem{Ber77} Berry, M. V., and M. Tabor, 1977, Proc. Roy. Soc.
London {\bf A 356}, 375.

\bibitem{Ber81} Berry, M. V., 1981, Ann. Phys. (N.Y.) {\bf 131}, 163.

\bibitem{Ber84} Berry, M. V., and  M. Robnik, 1984, J. Phys. A: Math.
Gen. {\bf 17}, 2413.

\bibitem{Ber85} Berry, M. V., 1985, Proc. Roy. Soc. London {\bf A
400}, 229.

\bibitem{Ber99} Bertulani, C. A., and V. Yu. Ponomarev, 1999, Phys.
Rep. {\bf 321}, 139.

\bibitem{Bil76} Bilpuch, E. G., A. M. Lane, J. D. Moses, and G. E. 
Mitchell, 1976, Phys. Rep. {\bf 28}, 145.

\bibitem{Blo95} Blocki, J., J. Skalski, and W. J. Swiatecki, 1995,
Nucl. Phys. {\bf A 594}, 137.

\bibitem{Boh71} Bohigas, O., and J. Flores, 1971, Phys. Lett. {\bf
B 34}, 261.

\bibitem{Boh83} Bohigas, O., R. U. Haq, and A. Pandey, 1983, in:
{\it Nuclear Data for Science and Technology}, edited by K. H.
B\"ockhoff (Reidel, Dordrecht), p. 809.

\bibitem{Boh84} Bohigas, O.,  M. J. Giannoni, and C. Schmit, 1984,
Phys. Rev. Lett. {\bf 52}, 1.

\bibitem{Boh84a} Bohigas, O., and M. J. Giannoni, 1984, {\it Lecture
Notes in Physics {\bf 209}} (Springer, Heidelberg).

\bibitem{Boh94} Bohigas, O., and H. A. Weidenm\"uller, 1988, Ann. Rev.
Nucl. Part. Science {\bf 38}, 421.

\bibitem{Boh02} Bohigas, O., and P. Leboeuf, 2002, Phys. Rev. Lett.
{\bf 88}, 092502.

\bibitem{Boh04} Bohigas, O., and M. P. Pato, 2004, Phys. Lett.
{\bf B 595}, 171.

\bibitem{Boh84b} Bohle, D., A. Richter, W. Steffen, A. E. L.
Dieperink, N. Lo Iudice, F. Palumbo, and O. Scholten, 1984, Phys.
Lett.  {\bf B 137}, 27.

\bibitem{Boh36} Bohr, N., 1936, Nature {\bf 137}, 344.

\bibitem{Boh51} Bohr, A., 1951, Phys. Rev. {\bf 81}, 134.

\bibitem{Boh52} Bohr, A., 1952, K. Dan. Vidensk. Selsk. Mat. Fys.
Medd. {\bf 26}, No. 14.

\bibitem{Boh53} Bohr, A., and B. R. Mottelson, 1952, K. Dan. Vidensk.
Selsk. Mat. Fys. Medd. {\bf 27}, No. 16.

\bibitem{Boh69} Bohr, A., and B. Mottelson, {\it Nuclear Structure},
W. A. Benjamin, New York, Vol. 1 (1969) and Vol. 2 (1975).

\bibitem{Bol68} Bollinger, L. M., and G. E. Thomas, 1968, Phys. Rev.
C {\bf 171}, 1293.

\bibitem{Bol70} Bollinger, L. M., and G. E. Thomas, 1970, Phys. Rev.
C {\bf 2}, 1951.

\bibitem{Bra72} Brack, M., J. Damgaard, A. S. Jensen, H. C. Pauli,
V. M. Strutinsky, and C. Y. Wong, 1972, Rev. Mod. Phys. {\bf 44},
320.

\bibitem{Bra97} Brack, M., and R. K. Bhaduri, 1997,
{\it Semiclassical Physics} (Addison Wesley, New York).

\bibitem{Bri55} Brink, D., 1955, D. Phil. Thesis, Oxford University,
unpublished.

\bibitem{Bri78} Brink, D., J. Neto, and H. A. Weidenm\"uller, 1978,
Phys. Lett. {\bf B 80}, 170.

\bibitem{Bro81} Brody, T. A., J. Flores, J. B. French, P. A. Mello,
A. Pandey, and S. S. M. Wong, 1981, Rev. Mod. Phys. {\bf 53}, 385.

\bibitem{Bro84} Brown, B. A., and G. Bertsch, 1984, Phys. Lett.
{\bf B 148}, 5. 

\bibitem{Bro88} Brown, B. A., and B. H. Wildenthal, 1988, Ann. Rev.
Nucl. Part. Sci. {\bf 38}, 29.

\bibitem{Bul96} Bulgac, A., and D. Kusnezov, 1996, Phys. Rev. {\bf
E 54}, 3468.

\bibitem{Bun99} Bunakov, V. E., 1999, Yadernaya Fisica {\bf 62}, 5 
[Phys. Atomic Nucl. {\bf 62}, 1 (1999)].

\bibitem{Cal83} Caldeira, A. O., and A. J. Leggett, 1983, Ann. Phys.
(N.Y.) {\bf 149}, 374.

\bibitem{Car99a} Carlson, B. V., L. F. Canto, S. Cruz--Barrios, M.
S. Hussein, and A. F. R. de Toledo Piza, 1999, Ann. Phys. (N.Y.)
{\bf 276}, 111.

\bibitem{Car99b} Carlson, B. V., M. S. Hussein, A. F. R. de Toledo
Piza, and L. F. Canto, 1999, Phys. Rev. {\bf C 60}, 014604.

\bibitem{Cas80} Casati, G., F. Valz--Gris, and I. Guarneri, 1980, 
Lett. Nuov. Cim. {\bf 28}, 279.

\bibitem{Cas85} Cassing, W., and W. N\"orenberg, 1985, Nucl. Phys.
{\bf A 433}, 467.  

\bibitem{Cau99} Caurier, E., G. Martinez--Pinedo, F. Nowacki, J.
Retamosa, and A. P. Zuker, 1999, Phys. Rev. {\bf C 59}, 2033.

\bibitem{Cau05} Caurier, E., G. Martinez--Pinedo, F. Nowacki, A.
Poves, and A. P. Zuker, 2005, Rev. Mod. Phys. {\bf 77}, 427.

\bibitem{Che46} Chevalley, C., 1946, {\it Theory of Lie Groups},
(Princeton University Press, Princeton).

\bibitem{Con90} Conway, J., 1990, {\it A Course in Functional
Analysis}, (Springer, New York).

\bibitem{Dem05} Dembowski, C., B. Dietz, T. Friedrich, H.-D.
Gr\"{a}f, H. L. Harney, A. Heine, M. Miski-Oglu, A. Richter, 2005,
Phys. Rev. {\bf E 71}, 046202.

\bibitem{Dep07} De Pace, A.,  A. Molinari, and H. A.
Weidenm\"uller, 2007, Ann. Phys. (N.Y.) {\bf 322}, 2446.

\bibitem{Des63} De Shalit, A., and I. Talmi, 1963, {\it Nuclear
Shell Theory} (Academic, New York).

\bibitem{Dia89} Dias, H., M. S. Hussein, N. A. de Oliveira, and B.
H. Wildenthal, 1989, J. Phys. G: Nucl. Part. {\bf 15}, L79.

\bibitem{Die80} Dieperink, A. E. L., O. Scholten, and F. Iachello,
1980, Phys. Rev. Lett. {\bf 44}, 1747.

\bibitem{Die06} Dietz, B., T. Guhr, H. L. Harney, and A. Richter,
2006, Phys. Rev. Lett. {\bf 96}, 254101.

\bibitem{Dos96} D{\o}ssing, T., B. Herskind, S. Leoni, A. Bracco,
R. A. Broglia, M. Matsuo, and E. Vigezzi, 1996, Phys. Rep. {\bf
268}, 1.

\bibitem{Dos04} D{\o}ssing, T., A. P. Lopez-Martens, T. L. Khoo,
T. Lauritsen, and S. {\AA}berg, 2004, in {\it The Labyrinth in Nuclear
  Structure}, edited by A. Braco and C. A. Kalfas, (AIP, Melville,
NY), p. 164.

\bibitem{Duf95} Duflo, J., and A. P. Zuker, 1995, Phys. Rev. {\bf
C 52}, R23.

\bibitem{Dys63} Dyson, F. J., and M. L. Mehta, 1963, J. Math. Phys.
{\bf 4}, 701.

\bibitem{Egi86} von Egidy, T., A. N. Behkami, and H. H. Schmidt,
1986, Nucl. Phys. {\bf A 454}, 109; ibid. 1989, {\bf A 481}, 189.

\bibitem{Ell96}  Ellegaard, C., T. Guhr, K. Lindeman, J. Nygard,
and M. Oxborrow, 1996, Phys. Rev. Lett. {\bf 77}, 4918 .

\bibitem{End00} Enders, J., T. Guhr, N. Huxel, P. von Neuman-Cosel,
C. Rangacharyulu, and A. Richter, 2000, Phys. Lett. {\bf B 486}, 273.

\bibitem{End04} Enders, J., T. Guhr, A. Heine, P. von Neuman-Cosel,
V. Yu. Ponomarev, A. Richter, and J. Wambach, 2004, Nucl. Phys.
{\bf A 741}, 3.

\bibitem{Fer34} Fermi, E., E. Amaldi, O. D'Agostino, F. Rasetti, and 
E. Segre, 1934, Proc. Roy. Soc. {\bf A 146}, 483; ibid. {\bf 149}
522 (1935).

\bibitem{Fla94} Flambaum, V. V., A. A. Gribakina, G. F. Gribakin,
and M. G. Kozlov, 1994, Phys. Rev. {\bf A 50}, 267.

\bibitem{Fra96} Frazier, N., B. A. Brown, and V. Zelevinsky, 1996,
Phys. Rev. {\bf C 54} 1665.

\bibitem{Fre70} French, J. B., and S. S. M. Wong, 1970, Phys. Lett.
{\bf B 33}, 449.

\bibitem{Fre85} French, J. B., V. K. B. Kota, A. Pandey, and S.
Tomsovic, 1985, Phys. Rev. Lett. {\bf 54}, 2313.

\bibitem{Gar64} Garg, J. B., J. Rainwater, J. S. Petersen, and W. W.
Havens, Jr., 1964, Phys. Rev. {\bf 134}, B985.

\bibitem{Gar97} Garrett, J. D., J. Q. Robinson, A. J. Foglia, and
H.-Q. Jin, 1997, Phys. Lett. {\bf B 392}, 24.

\bibitem{Gau61} Gaudin, M., 1961, Nucl. Phys. {\bf 25}, 447.

\bibitem{Gav86} A. Gavron, A. Gayer, J. Boissevain, H. C. Britt, J.
R. Nix, A. J. Sierk, P. Grangé, S. Hassani, H. A. Weidenm\"uller, J.
R. Beene, B. Cheynis, D. Drain, R. L. Ferguson, F. E. Obenshain, F.
Plasil, G. R. Young, G. A. Petitt and C. Butler, 1986, Phys. Lett.
{\bf B 176}, 312.

\bibitem{Gra80} Grang${\acute e}$, P., and H. A. Weidenm\"uller,
1980, Phys. Lett. {\bf B 96}, 26.

\bibitem{Gro00} Grossmann, C. A., M. A. LaBonte, G. E. Mitchell, J.
D. Shriner, J. F. Shriner, Jr., G. A. Vavrina, P. M. Wallace, 2000,
Phys. Rev. C {\bf 62}, 024323.

\bibitem{Gu99} Gu, J. Z., and H. A. Weidenm\"uller, 1999, Nucl. Phys.
{\bf A 660}, 197.

\bibitem{Gu01} Gu, J. Z., and H. A. Weidenm\"uller, 2001, Nucl. Phys.
{\bf A 690}, 382.

\bibitem{Guh90a} Guhr, T., and H. A. Weidenm\"uller, 1990a, Ann. Phys.
(N.Y.) {\bf 199}, 412.

\bibitem{Guh90b} Guhr, T., and H. A. Weidenm\"uller, 1990b, Chem. Phys.
{\bf 146}, 21.

\bibitem{Guh98} Guhr, T., A. M\"uller--Groeling, and H. A.
Weidenm\"uller, 1998, Phys. Rep. {\bf 299}, 189.

\bibitem{Gur57} Gurevich, I. I., and M. I. Pevsner, 1957, Nucl. Phys.
{\bf 2}, 575.

\bibitem{Gut95} Gutzwiller, M., 1990, {\it Chaos in Classical and
Quantum Mechanics} (Springer, Berlin).

\bibitem{Haa01} Haake, F., 2001, {\it Quantum Signatures of Chaos,
2nd edition} (Springer, Berlin).

\bibitem{Hac95} Hackenbroich, G., and H. A. Weidenm\"uller, 1995,
Phys. Rev. Lett. {\bf 74}, 4118.

\bibitem{Ham02} Hamoudi, A., R. G. Nazmitdinov, E. Shahaliev, and Y.
Alhassid, 2002, Phys. Rev. {\bf C 65}, 064311. 

\bibitem{Haq82} Haq, R. U., A. Pandey, and O. Bohigas, 1982, Phys.
Rev. Lett. {\bf 48},  1086.

\bibitem{Har57}Harish--Chandra, 1957, Am. J. Math. {\bf 79}, 87.

\bibitem{Har84} Harney, H. L., 1984, Z. Phys. {\bf A 316}, 177.

\bibitem{Har86} Harney, H. L., A. Richter, and H. A.
Weidenm{\"u}ller, 1986, Rev. Mod. Phys. {\bf 58}, 607.

\bibitem{Hax49} Haxel, O., J. H. D. Jensen, and H. E. Suess, 1949, 
Phys. Rev. {\bf 75}, 1766.

\bibitem{Hei94} Heiss, W. D., R. G. Nazmitdinov, and S. Radu, 1994,
Phys. Rev. Lett. {\bf 72}, 2351.

\bibitem{Her87} Herskind, B., B. Lauritzen, K. Schiffer, R. A.
Broglia, F. Barranco, M. Gallardo, J. Dudek, and E. Vigezzi, 1987,
Phys. Rev. Lett. {\bf 59}, 2416.

\bibitem{Heu07} Heusler, S., S. M\"uller, A. Altland, P. Braun, and
F. Haake, 2007, Phys. Rev. Lett. {\bf 98}, 044103.

\bibitem{Hof77} Hofmann, H., and P. J. Siemens, 1977, Nucl. Phys.
{\bf A 275}, 464.

\bibitem{Hof01} Hofmann, H., F. A. Ivanyuk, C. Rummel, and S. Yamaji,
2001, Phys. Rev. {\bf C 64}, 054316.

\bibitem{Hon02} Honma, M., T. Otsuka, B. A. Brown, and T. Mizusaki,
2002, Phys. Rev. {\bf C 65}, 061301.

\bibitem{Hus00} Hussein, M. S., and M. P. Pato, 2000, Phys. Rev. 
Lett. {\bf 84}, 3873.

\bibitem{Iac87a} Iachello, F., and I. Talmi, 1987, Rev. Mod. Phys.
{\bf 59}, 339.

\bibitem{Iac87b} Iachello, F., and A. Arima, 1987, {\it The
Interacting Boson Model} (Cambridge Press, Cambridge).

\bibitem{Imr02} Imry, Y., 2002 {\it Introduction to Mesoscopic
Physics, 2nd edition}, (Oxford  Press, Oxford). 

\bibitem{Itz80} Itzykson, C., and J. B. Zuber, 1980, J. Math. Phys.
{\bf 21}, 411.

\bibitem{Joh98} Johnson, C. W., G. F. Bertsch, and D. J. Dean, 1998,
Phys. Rev. Lett. {\bf 80}, 2749.

\bibitem{Joh99} Johnson, C. W., G. F. Bertsch, D. J. Dean, and I.
Talmi, 1999, Phys. Rev. {\bf C 61}, 014311.

\bibitem{Jol04} Jolie, J., R. F. Casten, P. Cejnar, S. Heinze, E. A.
McCutchan, and N. V. Zamfir, 2004, Phys. Rev. Lett. {\bf 93}, 132501.

\bibitem{Koe07} Koehler, P. E., J. L. Ullmann, T. A. Bredeweg, J. M.
O’Donnell, R. Reifarth, R. S. Rundberg, D. J. Vieira, and J. M. Wouters,
2007, Phys. Rev. {\bf C 76}, 025804.

\bibitem{Kot01} Kota, V. K. B., 2001, Phys. Rep. {\bf 347}, 223.

\bibitem{Kot07} Kota, V. V. B., {\bf 2007} J. Math. Phys. {\bf 48},
053304.

\bibitem{Kru01} Kruecken, R., A. Dewald, P. von Brentano, and H. A.
Weidenm\"uller, 2001, Phys. Rev. {\bf C 64}, 064301.

\bibitem{Lan58} Lane, A. M., and R. G. Thomas, 1958, Rev. Mod. Phys.
{\bf 30}, 257.

\bibitem{Leb06} Leboeuf, and P., J. Roccia, 2006, Phys. Rev. Lett.
{\bf 97}, 010401.

\bibitem{Lev86} Leviandier, L., M. Lombardi, R. Jost, and J. P. Pique,
1986, Phys. Rev. Lett. {\bf 56}, 2449.

\bibitem{Lew94} Lewenkopf, C., and V. Zelevinsky, 1994, Nucl. Phys. 
{\bf A 569}, 183c.

\bibitem{Lio72a} Liou, H.I., H. S. Camarda, S. Wynchank, M. Slagowitz,
G. Hacken, F. Rahn, and J. Rainwater, 1972, Phys. Rev. C {\bf 5}, 974.

\bibitem{Lio72b} Liou, H. I., H. S. Camarda, and F. Rahn, 1972, Phys.
Rev. C {\bf 5}, 1002.

\bibitem{Lom93} Lombardi, M., and  T. H. Seligman, 1993, Phys. Rev.
{\bf A 47}, 3571 (1993).

\bibitem{Lom94} Lombardi, M., O. Bohigas, and T. Seligman, 1994, Phys.
Lett. {\bf B 324}, 263.

\bibitem{Lyn68} J. E. Lynn, 1968, {\it The Theory of Neutron Resonances}
(Clarendon, Oxford).

\bibitem{Mah69} Mahaux, C., and H. A. Weidenm\"uller, 1969,
{\it Shell--Model Approach to Nuclear Reactions}  (North--Holland,
Amsterdam).

\bibitem{Mat97} Matsuo, M., T. D{\o}ssing, E. Vigezzi, and S.
{\AA}berg, 1997, Nucl. Phys. {\bf A 620}, 296.

\bibitem{Mar97} Martinez--Pinedo, G., A. P. Zuker, A. Poves, and E.
Caurier, 1997, Phys. Rev. {\bf C 55}, 187.

\bibitem{May49} Mayer, M. G., 1949, Phys. Rev. {\bf 75}, 1969.

\bibitem{May55} Mayer, M. G., and J. H. D. Jensen, 1955,
{\it Elementary Theory of Nuclear Shell Structure} (Wiley, New York).

\bibitem{Mcd79} McDonald, S. W., and A. N. Kaufman, 1979, Phys. Rev.
Lett. {\bf 42}, 1189.

\bibitem{Meh04} Mehta, M. L., 2004, {\it Random Matrices, 3rd edition}
(Academic Press, New York).

\bibitem{Mer88} Meredith, D. C., S. E. Koonin, and M. R. Zirnbauer,
  1988, Phys.
Rev. {\bf A 37}, 3499.

\bibitem{Mer93} Meredith, D. C., 1993, Phys. Rev. {\bf E 47}, 2405.

\bibitem{Mit85} Mitchell, G. E., E. G. Bilpuch, J. F. Shriner, Jr.,
and A. M. Lane, 1985, Phys. Rep. {\bf 117}, 1.

\bibitem{Mit01} Mitchell, G. E., and J. F. Shriner, Jr., 2001,
Physica Scripta {\bf T 90}, 105.

\bibitem{Mol95} M\"oller, P., J. R. Nix, W. D. Myers, and W. J.
Swiatecki, 1995, At. Data Nucl. Data Tables {\bf 59}, 185.

\bibitem{Mol06} Molinari, A., and H. A. Weidenm\"uller, 2006, Phys.
Lett. {\rm B 637}, 48.

\bibitem{Mon75} Mon, K. F., and J. B. French, 1975, Ann. Phys. (N.Y.)
{\bf 95}, 90.

\bibitem{Mot92} Mottelson, B. R., 1992, Nucl. Phys. {\bf A 557},
717c.

\bibitem{Nat36} Nature, 1936, {\bf 137}, 351.

\bibitem{Neu64} Neutron Cross Section, Sigma Center, Brookhaven
National Laboratory {\bf BNL 325}, Suppl. 2, Brookhaven, N. Y. (1964). 

\bibitem{Noe80} N\"orenberg, W., and H. A. Weidenm\"uller, 1980, 
{\it Introduction to the Theory of Heavy--Ion Reactions, 2nd edition}
(Springer, Berlin).

\bibitem{Olo06} Olofsson, H., {\AA}berg, O. Bohigas, and Leboeuf, 
2006, Phys. Rev. Lett. {\bf 96}, 042502.

\bibitem{Orm92} Ormand, W. E., and R. A. Broglia, 1992, Phys. Rev.
{\bf C 46}, 1710.

\bibitem{Pan81}Pandey, A., 1981, Ann. Phys. (N. Y) {\bf 134}, 110.

\bibitem{Pap04} Papenbrock, T., and H. A. Weidenm\"uller, 2004,
Phys. Rev. Lett. {\bf 93}, 132503.

\bibitem{Pap05} Papenbrock, T., and H. A. Weidenm\"uller, 2005,
Nucl. Phys. {\bf A 757}, 422.

\bibitem{Pap06} Papenbrock, T., Z. Pluhar, and H. A. Weidenm\"uller
2006, J. Phys. A: Math. Gen. {\bf 39}, 9709.

\bibitem{Pap06a} Papenbrock, T., and H. A. Weidenm\"uller, 2006,
Phys. Rev. {\bf C 73}, 014311.

\bibitem{Pap07} Papenbrock, T., and H. A. Weidenm\"uller, 2007,
Rev. Mod. Phys. {\bf 79}, 997.

\bibitem{Per73} Percival, I. C., 1973, J. Phys. {\bf B 6}, L229.

\bibitem{Pie01} Pieper, S. C., and R. B. Wiringa, 2001, Ann. Rev.
Nucl. Part. Sci. {\bf 51}, 53. 

\bibitem{Por65} Porter, C.E., 1965, {\it Statistical Theories of
Spectra: Fluctuations}, Academic Press, New York.

\bibitem{Rag86} Ragnarsson, I., and S. {\AA}berg, 1986, Phys. Lett.
{\bf B 180}, 191.

\bibitem{Ram91} Raman, S., T. A. Walkiewicz, S. Kahane, E. T.
Jurney, J. Sa, Z. Gacsi, J. L. Weil, K. Allart, G. Bonsignori,
and J. F. Shriner, Jr., 1991, Phys. Rev. {\bf C 43}, 521. 

\bibitem{Ric95} Richter, A., 1995, Prog. Part. Nucl. Phys. {\bf34},
213.

\bibitem{Roc07} Roccia, J., and P. Leboeuf, 2007, Phys. Rev. {\bf
C 76}, 014301.

\bibitem{Ros60} Rosenzweig, N., and C. E. Porter, 1960, Phys. Rev.
{\bf 120}, 1698.

\bibitem{Sak03} Sakhr, J., and N. D. Whelan, 2003, Phys. Rev. {\bf
E 67}, 066213.

\bibitem{Sam04} Samyn, M.,  S. Goriely, M. Bender, and J. M.
Pearson, 2004, Phys. Rev. {\bf C 70}, 04439.

\bibitem{Sar05} Sargeant, A. J., M. S. Hussein, and A. N. Wilson,
  2005, in {\it Nuclei and Mesoscopic Physics}, East Lansing,
edited by V. Zelevinsky, (AIP, Melville, NY), p. 46.

\bibitem{Shi92} Shimizu, Y. R., F. Barranco, R. A. Broglia, T.
D{\o}ssing, and E. Vigezzi, 1992, Phys. Lett. {\bf B 274}, 253.

\bibitem{Shr87} Shriner, Jr., J. F., G. E. Mitchell, and E. G.
Bilpuch, 1987, Phys. Rev. Lett. {\bf 59}, 435.

\bibitem{Shr89} Shriner, Jr., J. F., E. G. Bilpuch, and G. E.
Mitchell, 1989, Z. Phys. {\bf A 332}, 45.

\bibitem{Shr91} Shriner, Jr., J. F., G. E. Mitchell, and T. von
Egidy, 1991, Z. Phys. {\bf A 338}, 309.

\bibitem{Shr00} Shriner, Jr., J. F., C. A. Grossmann, and G. E.
Mitchell, 2000, Phys. Rev. {\bf C 62}, 054305.

\bibitem{Sol95} Soloviev, V. G., 1995, Nucl. Phys. {\bf A 586}, 265.

\bibitem{Som93} Sommermann, M., and H. A. Weidenm{\"u}ller, 1993, 
Europhys. Lett. {\bf 23}, 79.

\bibitem{Sta99} Stafford, C. A., and B. R. Barrett, 1999, Phys. Rev.
{\bf C 60}, 051305.

\bibitem{Ste05}Stephens, F. S., {\it et al.}, 2005, Phys. Rev. Lett.
{\bf 94}, 042501.

\bibitem{Str66} Strutinsky, V. M., 1966 Yad. Fiz. {\bf 3}, 614; Sov.
Journ. Nucl. Phys. {\bf 3}, 499 (1966).

\bibitem{Str67} Strutinsky, V. M., 1967, Nucl. Phys. {\bf A 95}, 420. 

\bibitem{Str68} Strutinsky, V. M., 1968, Nucl. Phys. {\bf A 122}, 1.

\bibitem{Twi86} Twin, P. J., B. M. Nyako, A. H. Nelson, J. Simpson, M.
  A. Bentley, H. W. Cranmer-Gordon, P. D. Forsyth, D. Howe,
  A. R. Mokhtar, J. D. Morrison, J. F. Sharpey-Schafer, G. Sletten, 
1986, Phys. Rev. Lett.
{\bf 57}, 811.

\bibitem{Ver79} Verbaarschot, J. J. M., and P. J. Brussard, 1979,
Phys. Lett. {\bf B 87}, 155.

\bibitem{Vig90a} Vigezzi, E., R. A. Broglia, and T. D{\o}ssing, 1990,
Nucl. Phys. A {\bf A 520}, 179c.

\bibitem{Vig90b} Vigezzi, E., R. A. Broglia, and T. D{\o}ssing, 1990,
Phys. Lett. {\bf B 249}, 163.

\bibitem{Vre92} Vretenar, D., and Y. Alhassid, 1992, Phys. Rev. {\bf
C 46}, 1334.

\bibitem{Wat81} Watson, III, W. A., E. G. Bilpuch, and G. E. Mitchell,
1981, Z. Phys. {\bf A 300}, 89.

\bibitem{Wei80} Weidenm\"uller, H. A., 1980, Progress in Particle and
Nuclear Physics {\bf 3}, 49.

\bibitem{Wei85} Weidenm\"uller, H. A., 1985, in: {\it Nuclear
Structure 1985}, edited by R. Broglia, G. B. Hagemann, B. Herskind
(Elsevier, Amsterdam) p. 213.

\bibitem{Wei93} Weidenm{\"u}ller, H. A., 1993, Phys. Rev. {\bf A 48},
1819.

\bibitem{Wei98} Weidenm\"uller, H. A., P. von Brentano, and B. R.
Barrett, 1998, Phys. Rev. Lett. {\bf 81}, 3603.

\bibitem{Whi78} Whitehead, R. R., A. Watt, D. Kelvin, and A. Conkie,
1978, Phys. Lett. {\bf B 76}, 149.

\bibitem{Wil73} Wilczynski, J., 1973, Phys. Lett. {\bf B 47}, 484.

\bibitem{Wil75} Wilson, W. M., E. G. Bilpuch, and G. E. Mitchell,
1975, Nucl. Phys. {\bf A 245}, 285.

\bibitem{Wil84} Wildenthal, B. H., 1984, Prog. Part. Nucl. Phys.
{\bf 11}, 5.

\bibitem{Yos93} Yoshinaga, N., N. Yoshida, T. Shigehara, and T.
Cheon, 1993, Phys. Rev. {\bf C 48}, R509.

\bibitem{Yos06} Yoshinaga, N., A. Arima, and Y. M. Zhao, 2006, Phys.
Rev. {\bf C 73}, 017303.

\bibitem{Zel96} Zelevinsky, V., B. A. Brown, N. Frazier, and M.
Horoi, 1996, Phys. Rep. {\bf 276}, 85.

\bibitem{Zel04} Zelevinsky, V., and A. Volya, 2004, Phys. Rep.
{\bf 391}, 311.

\bibitem{Zel07} Zelevinsky, V., 2007, private communication.

\bibitem{Zha01} Zhao, Y. M., and A. Arima, 2001, Phys. Rev. {\bf C
64}, 041301.

\bibitem{Zha02} Zhao, Y. M., A. Arima, and N. Yoshinaga, 2002, Phys.
Rev. {\bf C 66}, 034302.

\bibitem{Zha04} Zhao, Y. M., A. Arima, and N. Yoshinaga, 2004, Phys.
Rep. {\bf 400}, 1.

\bibitem{Zha04a}Zhao, Y. M., A. Arima, N. Shimizu, K. Ogawa, N.
Yoshinaga, and O. Scholten, 2004, Phys. Rev. {\bf C 70}, 054322.

\end{thebibliography}
\end{document}